\pdfoutput=1
\documentclass[PAPER, atlasdraft=false, texlive=2023, cernpreprint, UKenglish, texmf, orcidlogo]{atlasdoc}
 
\usepackage{atlaspackage}
\usepackage{atlasbiblatex}
 
\usepackage[hepprocess=true,hepparticle=true]{atlasphysics}
\usepackage{tabularx}
 
\graphicspath{{logos/}{figures/}}
 
\usepackage{ANA-HIGG-2019-09-PAPER-defs}
\usepackage{soul}
\usepackage{multirow}
\usepackage{arydshln}

\AtlasTitle{
Measurements of Higgs boson production cross-sections in the $H\to\tau^{+}\tau^{-}$ decay channel in $pp$ collisions at $\sqrt{s}=13\,\text{TeV}$ with the ATLAS detector
}
 
\AtlasAbstract{
Measurements of the production cross-sections of the Standard Model
(SM) Higgs boson ($H$) decaying into a pair of $\tau$-leptons are
presented.  The measurements use data collected with the ATLAS
detector from $pp$ collisions produced at the Large Hadron Collider at
a centre-of-mass energy of $\sqrt{s}=13\,\text{TeV}$, corresponding to
an integrated luminosity of $139\,\text{fb}^{-1}$.  Leptonic ($\tau\to\ell\nu_{\ell}\nu_{\tau}$) and
hadronic ($\tau\to\text{hadrons\,}\nu_{\tau}$) decays of the $\tau$-lepton
are considered.  All measurements account for the branching
ratio of $H\to\tau\tau$ and are performed with a requirement $|y_H|<2.5$, where $y_H$ is the true Higgs boson rapidity.
The cross-section of the $pp\to H\to\tau\tau$ process is measured to be
$2.94 \pm 0.21 \text{(stat)} ^{+\,0.37}_{-\,0.32} \text{(syst)}$~pb, in agreement with the SM prediction of \mbox{$3.17\pm0.09$\,pb}.
Inclusive cross-sections are determined separately for the four dominant production modes:
$2.65  \pm 0.41 \text{(stat)}  ^{+\,0.91}_{-\,0.67} \text{(syst)}$~pb for gluon--gluon fusion,
$0.197 \pm 0.028 \text{(stat)} ^{+\,0.032}_{-\,0.026} \text{(syst)}$~pb for vector-boson fusion,
$0.115 \pm 0.058 \text{(stat)} ^{+\,0.042}_{-\,0.040} \text{(syst)}$~pb for vector-boson associated production,
and $0.033 \pm 0.031 \text{(stat)} ^{+\,0.022}_{-\,0.017} \text{(syst)}$~pb for top-quark pair associated production.
Measurements in exclusive regions of the phase space, using the
simplified template cross-section framework, are also performed.  All
results are in agreement with the SM predictions.  }
\author{The ATLAS Collaboration}
 
\AtlasRefCode{HIGG-2019-09}

\PreprintIdNumber{CERN-EP-2021-217}
 
\journal{JHEP}
 
\AtlasJournalRef{JHEP 08 (2022) 175}
\AtlasDOI{10.1007/JHEP08(2022)175}

\hypersetup{pdftitle={ATLAS document},pdfauthor={The ATLAS Collaboration}}
 
\begin{document}
 
\maketitle
 
\tableofcontents
\clearpage
 
\section{Introduction}
\label{sec:intro}

A particle consistent with the Standard Model (SM) Higgs boson~\autocite{Englert:1964et,Higgs:1964ia,Higgs:1964pj,Guralnik:1964eu,Higgs:1966ev,Kibble:1967sv} was discovered in 2012 by the ATLAS and CMS collaborations~\autocite{HIGG-2012-27,CMS-HIG-12-028}  from the analysis of proton--proton ($pp$) collisions produced by the Large Hadron Collider (LHC)~\autocite{Evans:2008zzb}.
Since then, the analysis of data collected at centre-of-mass energies of \SI{7}{\TeV}, \SI{8}{\TeV} and \SI{13}{\TeV} in Runs~1 and 2 of the LHC\footnote{Run~1 signifies the LHC data-taking period in the years 2010--2012 and Run~2 the one in 2015--2018.} has led to the precise measurement of the Higgs boson mass, $m_H=\SI{125.09}{\GeV}$~\cite{HIGG-2014-14}, and to the observation and measurement of the four main production modes (gluon--gluon fusion, vector-boson fusion, and associated production with either a weak gauge boson or a pair of top quarks) and of several decay channels of the Higgs boson  predicted by the SM~\cite{HIGG-2016-21,HIGG-2016-22, HIGG-2016-25, HIGG-2016-33, HIGG-2017-11, HIGG-2016-07, CMS-HIG-17-025, CMS-HIG-16-040, CMS-HIG-16-041, CMS-HIG-16-042,HIGG-2015-07,HIGG-2017-07,CMS-HIG-16-043,HIGG-2018-13, CMS-HIG-17-035}.
 
The decay into a $\tau^{+}\tau^{-}$ pair\footnote{For simplicity, a $\tau^{+}\tau^{-}$ pair is denoted by $\tau\tau$ throughout the paper.} has the largest branching fraction of all leptonic Higgs boson decays (\SI{6.3}{\percent}~\autocite{Djouadi:1997yw, deFlorian:2016spz} for a mass of $m_H=\SI{125.09}{\GeV}$).
The large number of Higgs boson decays into $\tau\tau$ produced at the LHC ($\approx 500 \cdot10^3$ during Run 2) offers a unique opportunity to study the Yukawa mechanism in detail.
Measurements in this final state are, however, complicated at the experimental level, as the presence of two to four neutrinos\footnote{The number of neutrinos depends on the decay modes of the two $\tau$-leptons.} in the final state significantly degrades the resolution of the measured Higgs boson four-momentum, rendering the separation between the signal and the large background from \Ztt events difficult.
This effect can be mitigated through the dedicated study of the Higgs production modes where the event topology differs drastically from that of \Zjets events, the two most sensitive being the production of the Higgs boson through vector-boson fusion (VBF) and its production through gluon--gluon fusion (ggF) with Higgs boson produced with a large transverse momentum.
 
The first evidence of the $\tau\tau$ decay of the Higgs boson was obtained by the ATLAS~\cite{HIGG-2013-32} and CMS~\cite{CMS-HIG-13-004} collaborations using data collected at centre-of-mass energies of \SI{7}{\TeV} and \SI{8}{\TeV} during Run~1 of the LHC.
The combination~\cite{HIGG-2015-07} of these two results led to the first observation of the $\tau\tau$ decay of the Higgs boson.
More recent measurements in the $\Htautau$ decay channel are documented in Refs.~\cite{HIGG-2018-14,CMS-HIG-18-007,CMS-HIG-20-015}.

This paper presents measurements of the Higgs boson decaying into a $\tau\tau$ pair with the ATLAS detector, using the full Run~2 LHC dataset.
The $pp\to\Htautau$ process is measured inclusively, in the four dominant production modes simultaneously, and as a function of key properties of the event.
This is achieved with an optimised categorisation of the collected events.
Three $\tau\tau$ final states are targeted: two hadronically decaying $\tau$-leptons (\tauhad, where the tau decays into hadrons plus a neutrino), denoted \thadhad; one leptonically decaying $\tau$-lepton (\taulep) and one \tauhad, denoted \tlhad;\footnote{The \tlhad categories can be split into \tehad and \tmhad when distinguishing the light lepton's flavour is appropriate.} and two \taulep with different flavours, denoted \temu.
The remaining final states, with two same-flavour light leptons (\taue\taue and \taumu\taumu), are not considered due to large uncertainties in \Zee and \Zmm contributions to the expected background.
The dominant background processes after the event selection are \Ztt decays, \ttbar production, and processes with at least one jet misreconstructed as a \tauhad.
Smaller contributions to the background arise from events with \Zll\footnote{In this document, $\ell\,=\,e,\,\mu$.} decays, two weak vector bosons $VV$ (diboson), and \HWW decays.
Templates of the estimated invariant mass of the $\tau\tau$ pairs are built for each process in the signal regions (SR) defined by the event selection and categorisation.
The templates are used as input to a binned maximum-likelihood fit which allows the yields and kinematics of both the signal and the background processes to be measured.
Control regions (CR) enter the fit as event counts and help determine the normalisation of the main backgrounds as well as constrain their uncertainties.
 
This work uses \SI{139}{\ifb} of $pp$ collision data collected at a centre-of-mass energy of \SI{13}{\TeV}, to be compared with \SI{36}{\ifb} for the previous \Htautau cross-section measurements~\cite{HIGG-2017-07}.
It introduces a new reconstructed-event categorisation designed for the improved \emph{stage 1.2} binning~\cite{stxs1.1} of the simplified template cross-section (STXS) framework~\cite{deFlorian:2016spz}.
The treatment of ggF events with Higgs boson produced with a large transverse momentum is refined with three times more categories.
Selected events are categorised with requirements on the transverse momentum of the reconstructed Higgs boson candidate (\pTH) and on the potential additional hadronic jets.
Two new categories targeting production modes where the Higgs boson is created in association with other objects are added, based on requirements on the kinematics and tagged flavour of the jets in the event.
The first targets the production of a Higgs boson in association with a pair of top quarks (\ttH), where both top quarks and both $\tau$-leptons decay hadronically, complementing the explorations in Ref.~\cite{HIGG-2013-26}, and is denoted by tt(0$\ell$)\Hthadthad in the rest of this paper.
The second targets the production of a Higgs boson in association with a vector boson $V$ (\Wboson, \Zboson).
This new category, referred to as V(had)H, focuses on events with a hadronic decay of the $V$ boson while the production of $Z(\to\ell\ell)H$ and $W(\to\ell\nu)H$ events is studied separately~\autocite{HIGG-2014-01}.
Finally, the selection of VBF events was also improved by multivariate techniques.
 
In addition to the new extended categorisation, several improvements to the analysis methodology have been implemented: the object selection has been improved, multivariate discriminants have been optimised to enhance the purity of the SRs in the targeted Higgs boson production modes, the number of simulated background events has been increased significantly and the usage of the \Zll control region has been refined.
The latter relies on a new simplified implementation of the embedding technique~\cite{HIGG-2014-09, CMS-TAU-18-001} which, instead of replacing the reconstructed electrons and muons from \Zll events by equivalent simulated $\tau$-lepton decay products, simply rescales their transverse momentum to that of an equivalent $\tau$-lepton.
 
This document is organised as follows.
\Cref{sec:detector} describes the ATLAS detector.
This is followed in \cref{sec:data_and_sim_samples} by a description of the dataset and Monte Carlo (MC) simulated samples employed in the measurement.
\Cref{sec:object} details the reconstruction of the physics objects.
The event selection and categorisation is described in \cref{sec:event}.
In \cref{sec:bkg}, the estimation of the background processes is discussed with an emphasis on the simplified embedding technique to model \Ztt processes in \cref{sec:ztt} and the data-driven estimates of the processes with at least one jet misidentified as an electron, a muon or a \tauhad in Section~\ref{sec:fake}.
Section~\ref{sec:systematics} presents the systematic uncertainties affecting the measurement and their estimation.
The details of the signal extraction fit are discussed in \cref{sec:stat}, and \cref{sec:results} presents the results of the measurement.
\Cref{sec:conclusion} summarises the conclusions of this work.


\section{The ATLAS detector}
\label{sec:detector}

\newcommand{\AtlasCoordFootnote}{
ATLAS uses a right-handed coordinate system with its origin at the nominal interaction point (IP)
in the centre of the detector and the \(z\)-axis along the beam pipe.
The \(x\)-axis points from the IP to the centre of the LHC ring,
and the \(y\)-axis points upwards.
Cylindrical coordinates \((r,\phi)\) are used in the transverse plane,
\(\phi\) being the azimuthal angle around the \(z\)-axis.
The pseudorapidity is defined in terms of the polar angle \(\theta\) as \(\eta = -\ln \tan(\theta/2)\).
Angular distance is measured in units of \(\Delta R \equiv \sqrt{(\Delta\eta)^{2} + (\Delta\phi)^{2}}\).}
 
The ATLAS detector~\cite{PERF-2007-01} at the LHC covers nearly the entire solid angle around the collision point.\footnote{\AtlasCoordFootnote}
It consists of an inner tracking detector surrounded by a thin superconducting solenoid, electromagnetic and hadron calorimeters,
and a muon spectrometer incorporating three large superconducting air-core toroidal magnets.
 
The inner-detector system (ID) is immersed in a \SI{2}{\tesla} axial magnetic field
and provides charged-particle tracking in the range \(|\eta| < 2.5\).
The high-granularity silicon pixel detector covers the vertex region and typically provides four measurements per track,
the first hit normally being in the insertable B-layer installed before Run~2~\cite{ATLAS-TDR-19,PIX-2018-001}.
It is followed by the silicon microstrip tracker, which usually provides eight measurements per track.
These silicon detectors are complemented by the transition radiation tracker (TRT),
which enables radially extended track reconstruction up to \(|\eta| = 2.0\).
The TRT also provides electron identification information
based on the fraction of hits (typically 30 in total) above a higher energy-deposit threshold corresponding to transition radiation.
 
The calorimeter system covers the pseudorapidity range \(|\eta| <
4.9\).  Within the region \(|\eta|< 3.2\), electromagnetic calorimetry
is provided by barrel and endcap high-granularity lead/liquid-argon
(LAr) calorimeters, with an additional thin LAr presampler covering
\(|\eta| < 1.8\) to correct for energy loss in material upstream of
the calorimeters.  Hadron calorimetry is provided by the
steel/scintillator-tile calorimeter, segmented into three barrel
structures within \(|\eta| < 1.7\), and two copper/LAr hadron endcap
calorimeters.  The solid angle coverage is completed with forward
copper/LAr and tungsten/LAr calorimeter modules optimised for
electromagnetic and hadronic energy measurements respectively.
 
The muon spectrometer (MS) comprises separate trigger and
high-precision tracking chambers measuring the deflection of muons in
a magnetic field generated by the superconducting air-core toroidal
magnets.  The field integral of the toroids ranges between \num{2.0}
and \SI{6.0}{\tesla\metre} across most of the detector.  A set of
precision chambers covers the region \(|\eta| < 2.7\) with three
layers of monitored drift tubes, complemented by cathode-strip
chambers in the forward region, where the background is highest.  The
muon trigger system covers the range \(|\eta| < 2.4\) with
resistive-plate chambers in the barrel, and thin-gap chambers in the
endcap regions.
 
Interesting events are selected by the first-level (L1) trigger system
implemented in custom hardware, followed by selections made by
algorithms implemented in software in the high-level
trigger~\cite{TRIG-2016-01}.  The first-level trigger accepts events
from the \SI{40}{\MHz} bunch crossings at a rate below \SI{100}{\kHz},
which the high-level trigger reduces in order to record events
to disk at about \SI{1}{\kHz}.
 
An extensive software suite~\cite{ATL-SOFT-PUB-2021-001} is used in
the reconstruction and analysis of real and simulated data, in detector
operations, and in the trigger and data acquisition systems of the experiment.


\section{Data and simulated event samples}
\label{sec:data_and_sim_samples}

The data used in this analysis were collected using unprescaled single-lepton, dilepton or $\tau\tau$ triggers~\cite{TRIG-2018-05,TRIG-2018-01,id_trigger,l1topo_trigger} at a centre-of-mass energy of 13~\TeV\ during the 2015--2018 LHC running periods.
Events are selected for analysis only if they are of good quality and if all the relevant detector components are known to have been in good operating condition~\cite{DAPR-2018-01}, which corresponds to a total integrated luminosity of 139.0\,fb$^{-1}$.
 
MC simulated events are used to model most of the backgrounds from SM processes and the \Htautau signal processes.
A summary of all the generators used for the simulation of the signal and background processes is shown in \cref{tab:generators}.
The same event generators as in Ref.~\cite{HIGG-2017-07} were used, but the number of simulated events in each sample was at least quadrupled, which is the factor by which the integrated luminosity grew since the previous publication.
In addition, the total number of simulated \Ztt events was increased by a further factor of approximately four.
This computationally expensive task helps to densely populate the phase space where \Ztt events are produced in association with several jets.
 
All samples of simulated events were processed through the ATLAS detector simulation~\cite{SOFT-2010-01} based on $\GEANT$~\cite{geant}.
The effects of multiple interactions in the same and nearby bunch crossings (pile-up) were modelled by overlaying minimum-bias events, simulated using the soft QCD processes of \PYTHIA[8.186]~\cite{Pythia8} with the A3~\cite{ATL-PHYS-PUB-2016-017} set of tuned parameters and \NNPDF[2.3lo]~\cite{Ball:2012cx} parton distribution functions (PDF).
 
The decays and spin correlations for $\tau$-leptons are handled by \SHERPA for the samples it generated, and by \PYTHIA for the other MC event generators.
The decays and spin correlations have been included in \PYTHIA version 8.150~\cite{ILTEN201477},
and have been thoroughly validated by comparisons with \tauola~\cite{GOLONKA2006818}.
 
\begin{table}[h]
\caption{Overview of the MC generators used for the main signal and background samples. The last column, labelled 'Normalisation', specifies the order of the cross-section calculation used for the normalisation of the simulated samples.}
\label{tab:generators}
\resizebox{\textwidth}{!}{
\begin{tabular}{l c c c c l l}
\toprule
Process               & \multicolumn{2}{c}{Generator}                                   & \multicolumn{2}{c}{PDF set}                                      & Tune  & Normalisation        \\
& ME                                & PS                          & ME                                        & PS                   &       &              \\
\midrule
Higgs boson     &                                   &                             &                                           &                      &                      \\
\midrule
ggF                   & \POWHEGBOX[v2]                           & \PYTHIA[8]                    & \PDFforLHC[15nnlo]                & \CTEQ[6L1]              & AZNLO & N$^{3}$LO QCD + NLO EW    \\
VBF                   & \POWHEGBOX[v2]                           & \PYTHIA[8]                    & \PDFforLHC[15nlo]                                & \CTEQ[6L1]              & AZNLO & NNLO QCD + NLO EW        \\
$VH$                  & \POWHEGBOX[v2]                           & \PYTHIA[8]                    & \PDFforLHC[15nlo]                               & \CTEQ[6L1]              & AZNLO & NNLO QCD + NLO EW          \\
\ttH                  & \POWHEGBOX[v2]                        & \PYTHIA[8]                    & \NNPDF[3.0nnlo] & \NNPDF[2.3lo]              & A14                  & NLO QCD + NLO EW                  \\
\multirow{2}{*}{$tH$} & {\small \textsc{MadGraph5}\_}     & \multirow{2}{*}{\PYTHIA[8]}   & \multirow{2}{*}{\CT[10]} & \multirow{2}{*}{\NNPDF[2.3lo]} & \multirow{2}{*}{A14} & \multirow{2}{*}{NLO} \\
& {\small \textsc{aMC@NLO}}         &                             &                                           &                      &       &              \\
\bbH                 & \POWHEGBOX[v2]                    & \PYTHIA[8] & \NNPDF[3.0nnlo] & \NNPDF[2.3lo]              & A14 & NLO \\
\midrule
Background            &                                   &                             &                                           &                      &                      \\
\midrule
\Vjets (QCD/EW)      & \multicolumn{2}{c}{\SHERPA[2.2.1]} & \multicolumn{2}{c}{\NNPDF[3.0nnlo]} & \SHERPA                                   & NNLO for QCD, LO for EW                                       \\
\ttbar                & \POWHEGBOX[v2]                           & \PYTHIA[8]                    & \NNPDF[3.0nnlo] & \NNPDF[2.3lo]               & A14                  & NNLO + NNLL          \\
Single top            & \POWHEGBOX[v2]                           & \PYTHIA[8]                    & \NNPDF[3.0nnlo] & \NNPDF[2.3lo]               & A14                  & NLO                  \\
Diboson               & \multicolumn{2}{c}{\SHERPA[2.2.1]} & \multicolumn{2}{c}{\NNPDF[3.0nnlo]} & \SHERPA                                   & NLO                                        \\
\bottomrule
\end{tabular}
}
\end{table}
 
\subsection{Higgs boson simulation samples}
The main  Higgs boson production mode at the LHC is ggF with a total expected cross-section of \SI{48.6}{\pico\barn}, followed by VBF (\SI{3.78}{\pico\barn}), associated $VH$ (\SI{2.25}{\pico\barn}),  associated \bbH\ (\SI{0.64}{\pico\barn}) and  \ttH (\SI{0.51}{\pico\barn}) production.
Simulated event samples for these production modes were generated using \POWHEGBOX[v2]~\cite{Frixione:2007nw,Nason:2004rx,Frixione:2007vw,Alioli:2010xd,Hartanto:2015uka}.
The $tH$ process was also considered, but with a cross-section of \SI{0.092}{\pico\barn} its expected contribution was found to be negligible. 
It was simulated with the \MGNLO[2.6.2]~\cite{Alwall:2014hca} generator.
 
For the ggF sample the \PDFforLHC[15nnlo] PDF set~\cite{Butterworth:2015oua} was used, while VBF and $VH$ production samples used the \PDFforLHC[15nlo] PDF set.
The \ttH\ and \bbH\ events were produced with the \NNPDF[3.0nlo] PDF set~\cite{Ball:2014uwa}, and $tH$ events with the \CT[10] PDF set~\cite{Lai:2010vv}.
Parton shower (PS) and non-perturbative effects were modelled with \PYTHIA[8.230]~\cite{Sjostrand:2014zea} with parameter values set according to the \AZNLO tune~\cite{STDM-2012-23}, except for \ttH, \bbH and $tH$ events, which rely on the A14 tune~\cite{ATL-PHYS-PUB-2014-021}.
 
Higgs boson production via gluon--gluon fusion was simulated at next-to-next-to-leading-order (NNLO) accuracy in QCD.
The simulation achieves NNLO accuracy for arbitrary inclusive $gg\to H$ observables by reweighting the Higgs boson rapidity spectrum in \textsc{Hj-MiNLO}~\cite{Hamilton:2012np,Campbell:2012am,Hamilton:2012rf} to that of HNNLO~\cite{Catani:2007vq}.
The gluon--gluon fusion prediction from the MC simulated samples is normalised to the next-to-next-to-next-to-leading-order (N${}^3$LO) cross-section in QCD plus electroweak (EW) corrections at next-to-leading order (NLO)~\cite{deFlorian:2016spz,Anastasiou:2016cez,Anastasiou:2015ema,Dulat:2018rbf,Harlander:2009mq,Harlander:2009bw,Harlander:2009my,Pak:2009dg,Actis:2008ug,Actis:2008ts,Bonetti:2018ukf,Bonetti:2018ukf}.
 
Higgs boson production via vector-boson fusion was simulated at NLO accuracy in QCD.
It is tuned to match calculations with effects due to finite heavy-quark masses and soft-gluon resummations up to next-to-next-to-leading logarithms (NNLL).
The prediction from the MC simulated samples is normalised to an approximate-NNLO QCD cross-section with NLO electroweak corrections~\cite{Ciccolini:2007jr,Ciccolini:2007ec,Bolzoni:2010xr}.
 
Higgs boson production in association with a vector boson was simulated at next-to-leading order accuracy for $VH$ plus one-jet production.
The loop-induced $gg\to ZH$ process was generated separately at leading order in QCD.
The prediction from the MC simulated sample is normalised to cross-sections calculated at NNLO in QCD with NLO electroweak corrections for $p p \to VH$ and at NLO and next-to-leading-logarithm accuracy in QCD for $gg \to ZH$~\cite{Ciccolini:2003jy,Brein:2003wg,Brein:2011vx,Altenkamp:2012sx,Denner:2014cla,Brein:2012ne,Harlander:2014wda}.
 
The production of \ttH events was simulated at NLO accuracy in QCD.
The decays of bottom and charm hadrons were performed by \EVTGEN[1.6.0]~\cite{Lange:2001uf}.
The cross-section used to normalise the \ttH process is calculated at NLO in QCD and electroweak couplings~\cite{deFlorian:2016spz,Beenakker:2002nc,Dawson:2003zu,Yu:2014cka,Frixione:2014qaa}.
The production of \bbH and $tH$ events was simulated at NLO.
The prediction from the MC simulated samples is normalised to cross-sections calculated at NLO in QCD~\cite{Jager:2015hka,demartin2015higgs,Demartin:2016axk}.
 
The normalisation of all Higgs boson samples accounts for the decay branching ratio calculated with HDECAY~\cite{Djouadi:1997yw,Spira:1997dg,Djouadi:2006bz} and \PROPHECY~\cite{Bredenstein:2006ha,Bredenstein:2006rh,Bredenstein:2006nk}.
A Higgs boson mass of \SI{125.09}{\GeV} is assumed in the calculation of the expected cross-sections throughout this measurement.

\subsection{Background processes simulation samples}
The QCD production of \Vjets events was simulated with the \SHERPA[2.2.1]~\cite{Bothmann:2019yzt} generator using NLO matrix elements for up to two partons, and LO matrix elements for up to four partons, calculated with the Comix~\cite{Gleisberg:2008fv} and \OPENLOOPS~\cite{Buccioni:2019sur,Cascioli:2011va,Denner:2016kdg} libraries.
They were matched with the \SHERPA parton shower~\cite{Schumann:2007mg} using the MEPS@NLO prescription~\cite{Hoeche:2011fd,Hoeche:2012yf,Catani:2001cc,Hoeche:2009rj} using the set of tuned parameters developed by the \SHERPA authors.
The \NNPDF[3.0nnlo] set of PDFs~\cite{Ball:2014uwa} was used and the samples are normalised to a NNLO prediction~\cite{Anastasiou:2003ds}.
 
Electroweak production of $\ell\ell jj$, $\ell\nu jj$ and $\nu\nu jj$ final states was generated with \SHERPA[2.2.1], using LO matrix elements with up to two additional parton emissions.
The matrix elements were merged with the \SHERPA parton shower following the \MEPSatLO prescription and using the set of tuned parameters developed by the \SHERPA authors.  Similarly to the QCD \Vjets processes, the \NNPDF[3.0nnlo] set of PDFs was employed.
The samples were produced using the VBF approximation, which avoids an overlap with semileptonic diboson topologies by requiring a t-channel colour-singlet exchange. They are normalised using the \SHERPA cross-section predictions.
 
QCD and electroweak predictions for \Vjets events are grouped in the analysis and collectively referred to as \Vjets in the rest of the paper.
 
The production of \ttbar\ events was modelled by the \POWHEGBOX~v2 generator at NLO with the \NNPDF[3.0nlo] PDF set and the \hdamp parameter\footnote{The \hdamp\ parameter is a resummation damping factor and one of the parameters that controls the matching of \POWHEG matrix elements to the parton shower and thus effectively regulates the high-\pt radiation against which the \ttbar\ system recoils.} set to 1.5\,\mtop~\cite{ATL-PHYS-PUB-2016-020}.
The events were interfaced to \PYTHIA8.230 to model the parton shower, hadronisation, and underlying event, with parameters set according to the A14 tune and using the \NNPDF[2.3lo] set of PDFs.
The decays of bottom and charm hadrons were performed by \EVTGEN as for the \ttH sample.
The \ttbar sample is normalised to the cross-section prediction at NNLO in QCD including the resummation of NNLL soft-gluon terms calculated using \TOPpp[2.0]~\cite{Beneke:2011mq,Cacciari:2011hy,Baernreuther:2012ws,Czakon:2012zr,Czakon:2012pz,Czakon:2013goa,Czakon:2011xx}.
 
Single-top s-channel (t-channel) production was modelled using the \POWHEGBOX[v2]~\cite{Frixione:2007nw,Nason:2004rx,Frixione:2007vw,Alioli:2010xd} generator at NLO in QCD in the five-flavour (four-flavour) scheme with the \NNPDF[3.0nlo] set of PDFs~\cite{Ball:2014uwa}.
The events were interfaced with \PYTHIA[8.230]~\cite{Sjostrand:2014zea} using the A14 tune~\cite{ATL-PHYS-PUB-2014-021} and the \NNPDF[2.3lo] PDF set.
The sample is normalised to the theory prediction calculated at NLO in QCD with \HATHOR~2.1~\cite{Aliev:2010zk,Kant:2014oha}.
 
Diboson production was simulated with the \SHERPA[2.2.1] or 2.2.2 generator depending on the process.
Fully leptonic final states and semileptonic final states, where one boson decays leptonically and the other hadronically, were generated using matrix elements at NLO accuracy in QCD for up to one additional parton and at LO accuracy for up to three additional parton emissions.
Samples for the loop-induced processes $gg \to VV$ were generated using LO-accurate matrix elements for up to one additional parton emission for both the fully leptonic and semileptonic final states.
The matrix element calculations were matched and merged with the \SHERPA parton shower based on Catani--Seymour dipole factorisation~\cite{Gleisberg:2008fv,Schumann:2007mg} using the MEPS@NLO prescription.
The virtual QCD corrections were provided by the \OPENLOOPS library.
The \NNPDF[3.0nnlo] set of PDFs was used~\cite{Ball:2014uwa}, along with the dedicated set of tuned parton-shower parameters developed by the \SHERPA authors.
The samples are normalised to a NLO prediction~\cite{article}.
 
The background originating from \HWW decays was modelled using the same simulation strategy as the \Htautau signal.


\section{Object and event selection}

The topology of \Htautau events requires the reconstruction of electrons, muons, visible products of hadronically decaying $\tau$-leptons (\tauhadvis), jets (along with their $b$-tagging properties) and missing transverse momentum.
The numbers of reconstructed electrons, muons and \tauhadvis in each event are used to define the different channels of the analysis.
Requirements on the number of additional jets in the event are used in the signal region categorisation and to suppress backgrounds.
 
\subsection{Object reconstruction}
\label{sec:object}
 
Tracks measured in the ID are used to reconstruct interaction vertices~\cite{PERF-2015-01}, of which the one with the highest sum of squared transverse momenta of the associated tracks is selected as the primary vertex of the hard interaction.
 
Electrons are reconstructed from topological clusters of energy deposits in the electromagnetic calorimeter which are matched to a track reconstructed in the ID~\cite{EGAM-2018-01}.
They are required to satisfy the \objectsc{Loose} identification criteria, to have $\pT>\SI{15}{\GeV}$, and to be in the fiducial volume of the ID and the high-granularity electromagnetic calorimeters, $|\eta_\text{cluster}|<2.47$.
The transition region between the barrel and endcap calorimeters ($1.37 < |\eta_\text{cluster}| < 1.52$) is excluded except for the \Zll control region where it is kept to facilitate the embedding procedure (see \cref{sec:ztt}).
In the \temu and \tehad channels, the selected electron is further required to satisfy the \objectsc{Medium} identification, which has an associated efficiency of \SI{80}{\percent} to \SI{90}{\percent}, and the \objectsc{Loose} isolation criterion~\cite{EGAM-2018-01} in the signal regions and most control regions, which has an efficiency of \SI{90}{\percent} for \SI{15}{\GeV} candidates, increasing to more than \SI{98}{\percent} for \SI{30}{\GeV} candidates.
In the \tehad channel, the requirement on the electron transverse momentum is further tightened by \SI{1}{\GeV} above the nominal trigger \pT threshold for electrons matched to the single-electron trigger to ensure operation at the trigger's plateau efficiency.
Similarly, in the \temu channel, the requirement is tightened if the event is accepted by the single-electron trigger or the electron--muon trigger.
\Cref{tab:selection:ptcut} summarises the exact requirements used depending on the data-taking period.
 
\begin{table}[tbhp]
\caption{
Transverse momentum thresholds applied to the selected electrons, muons and $\tau_{\text{had-vis}}$  depending on the trigger signature and the data-taking period.
The \pT thresholds of the ATLAS lowest unprescaled triggers during the Run~2 data-taking are reported in Refs.~\cite{ATL-DAQ-PUB-2016-001,ATL-DAQ-PUB-2017-001,ATL-DAQ-PUB-2018-002,ATL-DAQ-PUB-2019-001}.
The electron and muon trigger menu evolution throughout the Run~2 data-taking is discussed in Refs.~\cite{TRIG-2018-05, TRIG-2018-01}.  }
\label{tab:selection:ptcut}
\begin{center}
\resizebox{\textwidth}{!}{
\begin{tabular}{@{}lcc@{}}
\toprule
Trigger signature              & Data-taking period                & \pT threshold [\GeV] used in event selection    \\
\midrule
\multirow{2}{*}{Single electron} & 2015                       & $\pT(e)>25$                     \\
& 2016--2018                  & $\pT(e)>27$                      \\
\midrule
\multirow{2}{*}{Single muon}     & 2015                       & $\pT(\mu)>21$      \\
& 2016--2018                  & ~~~$\pT(\mu)>27.3$                \\
\midrule
One electron, one muon                    & 2015--2018                  & $\pT(e)>18$, $\pT(\mu)>14.7$ \\
\midrule
\multirow{2}{*}{Two \tauhadvis}           & \multirow{2}{*}{2015--2018} & ~~~~~~~$\pT(\textrm{leading}~\tauhadvis)>40$     \\
&                            & $\pT(\textrm{sub-leading}~\tauhadvis)>30$  \\
\bottomrule
\end{tabular}
}
\end{center}
\end{table}

Muons are reconstructed from signals in the MS matched with tracks inside the ID.
They are required to satisfy the \objectsc{Loose} identification criteria~\cite{MUON-2018-03}, corresponding to an efficiency above \SI{97}{\percent} for all muon candidates considered in this analysis, and to have $\pT > \SI{10}{\GeV}$ and $|\eta|<2.5$.
In the \temu and \tmhad channels, the selected muon in the signal regions is further required to satisfy a \objectsc{Tight} isolation criterion~\cite{MUON-2018-03} based on track information.
This requirement has an efficiency increasing from \SI{85}{\percent} to \SI{99}{\percent} for muons with transverse momentum increasing from \SI{10}{\GeV} to \SI{50}{\GeV} and above.
In the \tmhad channel, the requirement on the muon transverse momentum is further tightened to select events in which the single-muon trigger operates with very high efficiency.
Similarly, in the \temu channel, the requirement is further tightened if the event is accepted by the single-muon trigger or the electron--muon trigger.
\Cref{tab:selection:ptcut} summarises the requirements used depending on the data-taking period.
 
Jets are reconstructed using a particle-flow algorithm~\cite{PERF-2015-09} from noise-suppressed positive-energy topological clusters in the calorimeter using the \antikt algorithm with a radius parameter $R=0.4$.
Cleaning criteria are used to identify jets arising from non-collision backgrounds or noise in the calorimeters~\cite{ATLAS-CONF-2015-029}, and events containing such jets are removed.
A jet vertex tagger (JVT)~\cite{PERF-2014-03} is used to remove jets with $\pT<\SI{60}{\GeV}$ and $\abseta<2.5$ that are identified as not being associated with the primary vertex of the hard interaction.
Similarly, pile-up jets in the forward region are suppressed with a \enquote{forward JVT}~\cite{PERF-2016-06} algorithm, exploiting jet shapes and topological jet correlations in pile-up interactions, which is applied to all jets with $\pT<\SI{60}{\GeV}$ and $|\eta|>2.5$.
Only jets with $\pT>\SI{20}{\GeV}$ are considered.
 
Jets with $\pT>\SI{20}{\GeV}$ and $|\eta|<2.5$ containing $b$-hadrons are identified using the DL1r $b$-tagging algorithm~\cite{FTAG-2018-01,ATL-PHYS-PUB-2017-013}.
In the \temu and \tlhad channels, the fixed \SI{85}{\percent} efficiency working point is used, while the \SI{70}{\percent} efficiency working point is used in the \thadhad channel (the target efficiencies being measured in simulated $t\bar{t}$ events).
Since the algorithm is used to veto $b$-tagged jets, the \SI{70}{\percent} efficiency working point offers a looser veto criterion which improves the sensitivity in the \thadhad channel where the backgrounds from $t\bar{t}$ events are less significant.
The rejection factors for $b$-tagged jets initiated by $c$-quarks and light partons are 9.4 (2.6) and 390 (29) respectively for the \SI{70}{\percent} (\SI{85}{\percent}) efficiency working point.
 
Decays of \tauhad are composed of a neutrino and a set of visible decay products, most frequently one or three charged pions and up to two neutral pions and denoted by \tauhadvis.
The reconstruction of the \tauhadvis is seeded by jets reconstructed using the \antikt algorithm~\cite{Cacciari:2008gp}, using calibrated topological clusters~\cite{PERF-2014-07} as inputs, with a radius parameter of $R=0.4$~\cite{ATLAS-CONF-2017-029}.
The jets form \tauhadvis candidates and are additionally required to have $\pT > \SI{10}{\GeV}$ and $|\eta|<2.5$.
Reconstructed tracks are matched to \tauhadvis candidates.
A multivariate discriminant is used to assess whether these tracks are likely to have been produced by the charged \tauhad decay products, and is used to reject tracks originating from other interactions, nearby jets, photon conversions or misreconstructed tracks.
The \tauhadvis objects are required to have one or three associated tracks selected by this discriminant.
Their charge ($q$) is defined as the sum of the measured charges of these associated tracks and must have $|q|=1$.
The \tauhadvis objects must also satisfy the requirements $\pT>\SI{20}{\GeV}$ and $|\eta|< 2.47$, excluding the region $1.37 < |\eta| < 1.52$.
These requirements have an efficiency of about \SI{85}{\percent} (\SI{70}{\percent})  for the majority of hadronic $\tau$ decays with one (three) associated tracks measured in simulated \Ztautau events.
The \tauhadvis energy scale is determined by combining information from the associated tracks, calorimeter clusters and reconstructed neutral pions~\cite{PERF-2014-06} using a multivariate regression technique~\cite{ATLAS-CONF-2017-029} trained in MC samples.
 
To separate the \tauhadvis candidates produced by hadronic $\tau$ decays from those due to jets initiated by quarks or gluons, a recurrent neural network (RNN) identification algorithm~\cite{ATL-PHYS-PUB-2019-033} is constructed employing information from reconstructed charged-particle tracks and calorimeter energy clusters associated with \tauhadvis candidates, as well as high-level discriminating variables.
A separate boosted decision tree discriminant (\objectsc{eBDT}) is also constructed to reject backgrounds arising from electrons misidentified as \tauhadvis (mainly from \Zee events in the \tehad channel in this analysis).
This discriminant is built using information from the calorimeter and the tracking detector, most notably transition radiation information from the TRT system and variables sensitive to the ratio of the energy deposited in the calorimeter and the visible momentum measured from the reconstructed tracks.
In addition, a very loose requirement on the RNN score (corresponding to a percent level efficiency loss for signal \tauhadvis) is applied, as well as a dedicated muon veto criterion, designed to reject muons misreconstructed as \tauhadvis (typically due to large calorimeter energy deposits).
 
In the \thadhad channel, the reconstructed \tauhadvis objects are required to match the two \tauhadvis candidates of the $\tau\tau$ trigger, thus defining the two selected \tauhadvis of the event.
In the \tlhad channel, the \tauhadvis candidate with the highest transverse momentum is the only one kept, and other ones are considered as jets.
This minimum requirement is much looser than the final RNN selection, and leads to a small loss of signal events where a quark- or gluon-initiated jet is taken as the \tauhadvis candidate, quantified to be at the level of 2.5\% (4\%) for the ggF (VBF) production process.
However, this strategy simplifies the treatment of the background processes with jets misidentified as \tauhadvis.
The estimation of this background relies on a control region defined by inverting the final RNN selection.
Picking a minimum requirement aimed at recovering the majority of this signal efficiency loss would sacrifice \SI{30}{\percent} to \SI{40}{\percent} of the statistical power in the control region, and would consequently degrade the estimate of this background (see \cref{sec:fake}).
 
The \tauhadvis objects are further required to fulfil the \objectsc{Medium} identification criteria in the signal regions of the \tlhad and \thadhad channels, which corresponds to an efficiency of 75\% (60\%) for candidates with 1~(3) associated track(s).
In the \tehad channel, for events where the \tauhadvis object has only one associated charged track, the \tauhadvis object is required to pass the \objectsc{Medium} working point of the eBDT algorithm, which corresponds to an 85\% efficiency for candidates which already satisfy the identification requirement.
The transverse momentum requirement for the \tauhadvis objects in the \thadhad final state is tightened to select events recorded with the \tauhadvis trigger operating at its plateau efficiency, as shown in \cref{tab:selection:ptcut}.
In the \tlhad final state, the \tauhadvis transverse momentum requirement is also tightened to $\pT>\SI{30}{\GeV}$ to improve background rejection.
 
The reconstructed objects used in this analysis are not built from disjoint sets of tracks or calorimetric clusters.
It is therefore possible that two different objects share most of their constituents.
An overlap removal procedure is applied to resolve this ambiguity.
This procedure is summarised in \cref{tab:orl}. It uses a definition of angular distance, $\Delta R_y=\sqrt{(\Delta y)^2 + (\Delta\phi)^2}$, that is based on the rapidities $y$ of the objects.
 
\begin{table}[tbhp]
\caption{
Criteria applied to overlapping reconstructed objects. The criteria are listed in the order they are applied.
}
\label{tab:orl}
\begin{center}
\begin{tabularx}{\textwidth}{ccX}
\toprule
Object to remove & Object to keep & Criteria \\
\midrule
electron & electron & If they share the same track, the electron with the highest transverse momentum is kept.\\
\tauhadvis & electron & If $\Delta R_y<0.2$, the electron is kept. \\
\tauhadvis & muon     & If $\Delta R_y<0.2$, the muon is kept. \\
electron & muon & If they share a track, the electron is removed if the muon is associated with a signature in the muon spectrometer, otherwise the muon is removed. \\
jet & electron & Any jet within $\Delta R_y=0.2$  of an electron is removed. \\
jet & muon & Any jet within $\Delta R_y=0.2$ of a muon is removed if it has fewer than three associated tracks. \\
electron & jet & Any electron within $\Delta R_y=0.4$  of a jet is removed. \\
muon & jet & Any muon within $\Delta R_y=0.4$ of a jet is removed. \\
jet & \tauhadvis & Any jet within $\Delta R_y=0.2$ of a \tauhadvis is removed. \\
\bottomrule
\end{tabularx}
\end{center}
\end{table}

The missing transverse momentum vector, $\vecmet$, is reconstructed as the negative vector sum of the transverse momenta of leptons, \tauhadvis and jets, and a \enquote{soft-term}.
The soft-term is calculated as the vectorial sum of the \pT of tracks matched to the primary vertex but not associated with a reconstructed lepton, \tauhadvis or jet~\autocite{PERF-2016-07}.
The magnitude of $\vecmet$ is referred to as the missing transverse momentum, \met.
 
\subsection{Event selection}
\label{sec:event}
 
Events are selected if they contain a \Htautau candidate in one of the final states under study (\temu, \tlhad, \thadhad).
 
The Higgs boson candidate is formed by the vector momentum sum of the visible $\tau$-lepton decay products and \vecmet. Its invariant mass (\mmmc) is calculated using an advanced likelihood-based technique, the Missing Mass Calculator (MMC)~\cite{MMCpaper}, which relies on information about the $\tau$-lepton candidate momenta,
the presence of additional jets, \vecmet and the type of $\tau$-lepton decay.
The addition of information about the number of reconstructed charged and neutral pions~\cite{PERF-2014-06} in hadronic decays of the $\tau$-leptons in new parameterisations for the likelihood function derived using \Ztt MC events are improvements with respect to Ref.~\cite{HIGG-2017-07} and lead to a \SI{1}{\percent} absolute improvement on the width of the reconstructed mass distribution.
 
For each channel a series of selection criteria are applied to enhance the sensitivity to the SM Higgs boson signal and ensure a robust estimate of the invariant mass of the reconstructed \tautau system.
These are summarised in \cref{tab:selection:baseline}.

\begin{table}[tbhp]
\caption{
Summary of the event selection for all sub-channels.
The electron and muon \pT thresholds correspond to the 2016--2018 dataset.
In the
$\tau_{e}\tau_{\mu}$
channel, events recorded with the electron trigger must satisfy $\pT(e)>27$\,\GeV\ and $\pT(\mu)>10$\,\GeV, events recorded with the muon trigger must satisfy  $\pT(e)>15$\,\GeV\ and $\pT(\mu)>27.3$\,\GeV, and events recorded with the electron--muon trigger must satisfy $\pT(e)>18$\,\GeV\ and $\pT(\mu)>14.7$\,\GeV.
Thresholds for the 2015 dataset are given in Table~\ref{tab:selection:ptcut}.
The $b$-veto requirement in the \thadhad channel is not applied in the tt(0$\ell$)$H\to\tau_{\text{had}}\tau_{\text{had}}$ category.
The quantities $x_{1}$ and $x_{2}$ are the momentum fractions carried by the visible decay products of the two $\tau$-leptons in the collinear approximation, as described in the text.  }
\label{tab:selection:baseline}
\begin{center}
\resizebox{\textwidth}{!}{\begin{tabular}{@{}rcccc@{}}
\toprule
Criteria                                 & \temu                  & \multicolumn{2}{c}{\tlhad}                 & \thadhad                  \\
\cmidrule(lr){3-4}
&                        & \tehad                                      & \tmhad &            \\
\cmidrule(r){1-1}\cmidrule(lr){2-2}\cmidrule(lr){3-3}\cmidrule(lr){4-4}\cmidrule(l){5-5}
$N(e)$                                    & 1                      & 1                            & 0            & 0          \\
$N(\mu)$                                  & 1                      & 0                            & 1            & 0          \\
$N(\tauhadvis)$                             & 0                      & 1                            & 1            & 2          \\
$N(b\textrm{-jets})$                               & 0 (85$\%$ WP)   & 0 (85$\%$ WP)         & 0 (85$\%$ WP)  & 0 (70$\%$ WP)           \\
&
& \multicolumn{2}{c}{}
& ($\geq1$ or 2 in ttH categories) \\
$\pT(e)$\,[\GeV]                          & \textgreater\,\numrange{15}{27}
& \textgreater\,\num{27}
&
& \\
$\pT(\mu)$\,[\GeV]                        & \textgreater\,\numrange{10}{27.3}
&
& \textgreater\,\num{27.3}
&\\
\cmidrule(lr){3-4}
$\pT(\tauhadvis)$\,[\GeV]                   &
& \multicolumn{2}{c}{\textgreater\,\num{30}}
& \textgreater\,40, 30                    \\
&
& \multicolumn{2}{c}{ }
& \\
Identification                            & $e$/$\mu$: Medium
& \multicolumn{2}{c}{$e$/$\mu$/\tauhadvis: Medium}
& \tauhadvis: Medium                                                                \\
&
& \multicolumn{2}{c}{ }
& \\
\cmidrule(lr){3-3}
\cmidrule(lr){4-4}
Isolation & $e$: Loose, $\mu$: Tight & $e$: Loose & $\mu$: Tight & \\
&
& \multicolumn{2}{c}{ }
&                                                                                   \\
\cmidrule(lr){2-5}
Charge                                    & \multicolumn{4}{c}{Opposite charge}                    \\
\MET\,[\GeV]                              & \multicolumn{4}{c}{\textgreater\,20}\\
&
& \multicolumn{2}{c}{ }
&                                                                                   \\
\cmidrule(lr){2-2}
\cmidrule(lr){3-4}
\multirow{2}{*}{Kinematics}               & $m_{\tau\tau}^{\mathrm{coll}} > (m_Z - 25)\,\GeV$
& \multicolumn{2}{c}{$\mT < \SI{70}{\GeV}$}
&                                                                                   \\
& $\SI{30}{\GeV} < m_{e\mu} < \SI{100}{\GeV}$
& \multicolumn{2}{c}{ }
&                                                                                   \\
&
& \multicolumn{2}{c}{ }
&                                                                                   \\
\cmidrule(lr){2-4}
\cmidrule(lr){5-5}
Leading jet                               & \multicolumn{3}{c}{$\pT > \SI{40}{\GeV}$}
& $\pT > \SI{70}{\GeV}$, $|\eta| < 3.2$                                             \\
&
& \multicolumn{2}{c}{ }
&                                                                                   \\
\cmidrule(lr){2-2}
\cmidrule(lr){3-4}
\multirow{2}{*}{Angular}                  & $\Delta R_{e\mu} < 2.0$
& \multicolumn{2}{c}{$\Delta R_{\ell\tauhadvis} < 2.5$}
& $0.6 < \Delta R_{\tauhadvis\tauhadvis} < 2.5$                                     \\
& $|\Delta\eta_{e\mu}| < 1.5$
& \multicolumn{2}{c}{  $|\Delta\eta_{\ell\tauhadvis}| < 1.5$ }
& $|\Delta\eta_{\tauhadvis\tauhadvis}| < 1.5$                                       \\
&
& \multicolumn{2}{c}{ }
&                                                                                   \\
\multirow{2}{*}{Coll. app. $x_{1}/x_{2}$} & $0.1 < x_{1} < 1.0$
& \multicolumn{2}{c}{$0.1 < x_{1} < 1.4$}
& $0.1 < x_{1} < 1.4$                                                               \\
& $0.1 < x_{2} < 1.0$
& \multicolumn{2}{c}{ $0.1 < x_{2} < 1.2$}
& $0.1 < x_{2} < 1.4$                                                               \\
&
& \multicolumn{2}{c}{ }
&                                                                                   \\
\bottomrule
\end{tabular}}
\end{center}
\end{table}


In the \temu channel, events must have a single reconstructed electron and a single reconstructed muon satisfying the criteria discussed in \cref{sec:object}.
In order to reject events coming from \Wjets, \Zjets and top processes,\footnote{In the following, `top processes' in the text (`Top' in tables and figures) collectively refer to single and pair production of top quarks.} the charges of the two reconstructed leptons must be of opposite sign, the invariant mass of the $e\mu$ system ($m_{e\mu}$) must be between \SI{30}{\GeV} and \SI{100}{\GeV}, and the collinear mass\footnote{The $\tau\tau$ mass reconstructed in the collinear approximation assumes that the neutrinos from the $\tau$-lepton decays propagate in the same direction as the visible decay products and that the missing transverse momentum is caused solely by those neutrinos~\cite{collinear_approximation}.} ($m_{\tau\tau}^{\mathrm{coll}}$) must be greater than ($m_Z-\SI{25}) \ {\GeV}$.
This last criterion ensures the selected dataset does not include any event considered in the signal regions of the ATLAS measurements of the \HWW process discussed in Ref.~\cite{ATLAS-CONF-2021-014}.
To further reduce backgrounds from top processes, events with a $b$-tagged jet are rejected.
In addition, angular requirements are placed on $\Delta{}R_{e\mu}$ and $|\Delta\eta_{e\mu}|$.  Finally, a $\pT>\SI{40}{\GeV}$ requirement is applied to the leading jet in the event to suppress backgrounds, as the signal final states considered include at least one high-\pT jet.
 
In the \tlhad channel, events must have a single reconstructed light lepton and a single reconstructed \tauhadvis satisfying the criteria discussed in \cref{sec:object}.
In order to reject events coming from \Wjets and top processes, the charges of the reconstructed light lepton and the reconstructed \tauhadvis must be of opposite sign.
The transverse mass of the lepton+\met  system (\mT) is required to be smaller than \SI{70}{\GeV} in order to efficiently suppress \Wjets processes.
To further reduce backgrounds from top processes, an explicit requirement is imposed to reject events with a $b$-tagged jet.
In addition, angular requirements are placed on $\Delta R_{\ell\tauhadvis}$ and $|\Delta\eta_{\ell\tauhadvis}|$.
The requirement on the leading jet transverse momentum in the event is the same as for the \temu channel.
 
In the \thadhad channel, events must have exactly two reconstructed \tauhadvis objects satisfying the criteria discussed in \cref{sec:object}.
In order to maintain low thresholds for the \pT of the \tauhadvis, additional criteria for the angular separation of the two \tauhadvis and the presence of an additional jet in the event were added to the lowest unprescaled $\tau\tau$ trigger during the Run~2 data-taking.
The additional criteria were imposed on the regions-of-interest (ROI) defining the \tauhadvis candidates at the L1 trigger.
In order to ensure that the ROIs of the two reconstructed \tauhadvis do not have overlapping cores, the criterion $\Delta{}R_{\tauhadvis\tauhadvis}>0.6$ is applied.
The extra-jet trigger criterion mentioned above translates into a requirement on the presence of at least one jet with $|\eta|<3.2$ and \pT greater than \SI{70}{\GeV}.
Similarly to the \temu and \tlhad channels, the charges of the two reconstructed \tauhadvis must be of opposite sign in order to reject events coming from \Wjets and top processes.
Events with $b$-tagged jets are rejected, except for the tt(0$\ell$)\Hthadthad signal region (see next \cref{sec:selection:categorisation}).
 
Finally, criteria concerning \met and the fraction of the $\tau$-lepton's momentum carried by its visible decay products, computed with the \vecmet components decomposed into the collinear approximation (defined as $x_{1}$ and $x_{2}$ for leading and sub-leading reconstructed visible $\tau$-lepton candidates respectively) are applied to improve the invariant mass estimation in the three channels.
 
Assuming SM predictions, about 2920 \Htautau events (330, 1410, and 1180 events in the \temu, \tlhad, and \thadhad channels respectively) are expected to be reconstructed and satisfy the event selection from the $\approx 440 \cdot 10^{3}$ \Htautau events that were produced with $|y_{H}| < 2.5$ during the LHC Run 2.
In data, \num{204442} events are selected.
 
\subsection{Event categorisation}
\label{sec:selection:categorisation}
 
The categorisation of selected events targets the four dominant Higgs boson production modes (see \cref{sec:intro}), uses their unique and characteristic signatures and is designed to closely match the production bins within the \emph{stage 1.2} of the STXS framework.
Bins of the full \emph{stage 1.2} scheme are merged to match the available sensitivity of the selected \Htautau events. Both the STXS bins and the event categories are illustrated in \Cref{fig:stxs_sketch}.
 
Requirements on the reconstructed Higgs boson transverse momentum, \pTH, and on properties of additional jets are described in the following.
Events in the VBF, V(had)H and tt(0$\ell$)\Hthadthad categories are further split with BDT taggers into two subcategories, the first (suffixed \_1) with enhanced signal fractions and the second (suffixed \_0) containing the remaining events.
All taggers are designed inclusively for all $\tau\tau$ decay modes and the variables are chosen to avoid any potential bias in the \mmmc distribution.
For each tagger, this is verified by comparing templates of the \mmmc distribution for signal and background processes between the relevant subcategories.
The taggers are described in the following and their input variables are listed in \cref{tab:mva:variables}.

\begin{table}
\caption{ Variables used in the four multivariate taggers employed in
the analysis.  For each tagger, the presence or absence of a $\bullet$
indicates whether the variable is used or not.
The symbol $\tau$ stands for any reconstructed $\tau$-lepton
candidate (electron, muon or $\tau_{\text{had-vis}}$) as appropriate in each
channel. The symbols $\tau\tau$ and $jj$ indicate the vectorial sums of the momenta of
two visible $\tau$-lepton candidates and of the two leading jets, respectively.  The
Higgs boson candidate \PH is formed by the vector sum of the two $\tau$-lepton
candidates' momenta and \vecmet.  The \PW candidate is built as the
pair of non-$b$-tagged jets in the event with invariant mass closest
to $m_W$. The top-quark candidate is built as the system of the $W$
candidate and a $b$-tagged jet in the event with invariant mass
closest to \mtop.}
\label{tab:mva:variables}
\centering
\begin{tabular}{@{}clcccc@{}}
\toprule
& Variable                                           & VBF       & V(had)H   & ttH vs \ttbar & ttH vs \Ztt \\
\midrule
\multirow{10}{*}{\rotatebox{90}{Jet properties}}
& Invariant mass of the two leading jets                   & $\bullet$ & $\bullet$ &               &             \\
& \pTjj                                              & $\bullet$ & $\bullet$ &               &             \\
& Product of $\eta$ of the two leading jets                & $\bullet$ &           &               &             \\
& Sub-leading jet \pT                                 & $\bullet$ &           &               &             \\
& Leading jet $\eta$                                 &           &           &               & $\bullet$   \\
& Sub-leading jet $\eta$                              &           &           &               & $\bullet$   \\
& Scalar sum of all jets \pT                         &           &           & $\bullet$     & $\bullet$   \\
& Scalar sum of all $b$-tagged jets \pT              &           &           &               & $\bullet$   \\
& Best \PW-candidate dijet invariant mass            &           &           & $\bullet$     & $\bullet$   \\
& Best \Pqt-quark-candidate three-jet invariant mass &           &           & $\bullet$     & $\bullet$   \\
\midrule
\multirow{7}{*}{\rotatebox{90}{Angular distances}}
& \Dphi between the two leading jets                        & $\bullet$ &           &               &             \\
& \Deta between the two leading jets   & $\bullet$ & $\bullet$ &               &             \\
& \DR between the two leading jets                         &           & $\bullet$ &               &             \\
& \dRtautaujj                                        &           & $\bullet$ &               &             \\
& \dRtautau                                          &           & $\bullet$ & $\bullet$     &             \\
& Smallest \DR(any two jets)                           &           &           & $\bullet$     &             \\
& $|\Deta(\tau,\tau)|$                               &           &           & $\bullet$     & $\bullet$   \\
\midrule
\multirow{3}{*}{\rotatebox{90}{$\tau$ prop.}}
& \pTtautau                                          &           &           & $\bullet$     &             \\
& Sub-leading $\tau$ \pT                              &           &           &               & $\bullet$   \\
& Sub-leading $\tau$ $\eta$                           &           &           &               & $\bullet$   \\
\midrule
\multirow{2}{*}{\rotatebox{90}{\parbox{1cm}{\centering \PH\\ cand.}}}
& \pTHjj                                             & $\bullet$ & $\bullet$ &               &             \\
& \pTHoverpTjj                                       &           & $\bullet$ &               &             \\
\midrule
\multirow{2}{*}{\rotatebox{90}{\vecmet}}
& Missing transverse momentum \MET                     &           & $\bullet$ & $\bullet$     & $\bullet$   \\
& Smallest \Dphi($\tau, \vecmet$)                    &           &           &               & $\bullet$   \\
\bottomrule
\end{tabular}
\end{table}


\subsubsection*{tt(0$\ell$)\Hthadthad categorisation}
The final state targeted in the tt(0$\ell$)\Hthadthad category includes six jets and two of these jets are initiated by the hadronisation of a $b$-quark.
However, to enhance the signal acceptance, the selection allows exactly one of these two numbers to be off by one unit.
Therefore, the event selection in the tt(0$\ell$)\Hthadthad category requires the presence of either six jets with \pT greater than \SI{20}{\GeV} including at least one $b$-tagged jet or five jets including at least two $b$-tagged jets.
The events satisfying these criteria are not considered by the analysis reported in Ref.~\cite{HIGG-2013-26}.
 
The signal-enhancing separation in this category uses two BDTs: one BDT is optimised to enhance \ttH signal events over \Ztt background events, while the second BDT is optimised to enhance \ttH signal events over $t\bar{t}$ background events.
A variety of two-dimensional combinations of requirements on the two BDT scores were studied, using the expected counting-experiment statistical significance \footnote{The \enquote{Poisson-Binomial model}in Ref.~\cite{ATL-PHYS-PUB-2020-025}.}, including an estimate of the systematic uncertainties in the background normalisations, as an estimator for their performance; none was found to outperform a simple rectangular requirement in the plane formed by the two BDT scores, and this was the requirement ultimately selected.
Of all Higgs boson events selected in the ttH\_0 (ttH\_1) categories \SI{74}{\percent} (\SI{92}{\percent}) are due to the \ttH process.
 
All other event categories in the \thadhad channel require that no $b$-tagged jets with $\pT>\SI{20}{\GeV}$ and $|\eta|<2.5$ are present.

\subsubsection*{VBF categorisation}
 
The VBF categories are designed to select Higgs bosons produced from the fusion of two vector bosons emitted by two quarks of the colliding protons.
The scattered quarks give rise to two high-\pT jets with a large rapidity gap and therefore large invariant mass \mjj.
This signature allows VBF events to be experimentally distinguished from the other Higgs production modes and \Ztautau events.
 
To match the STXS $qq\to H$ particle-level \pTjet requirement and \mjj binning, events selected in the VBF categories must have $\mjj>\SI{350}{\GeV}$ and \pT of the sub-leading jet greater than $\SI{30}{\GeV}$.
Additional selection criteria are applied to enhance the VBF Higgs production mode relative to the \Ztt background.
The product of the pseudorapidities of the two leading jets ($\eta(j_0)\times\eta(j_1)$) is required to be negative (i.e.\ jets must be in opposite hemispheres of the detector).
The absolute difference in pseudorapidity ($|\detajj|$) is required to be greater than 3.
Finally, the visible decay products of the $\tau$-leptons are required to be reconstructed in the rapidity gap between the VBF jets.
 
The VBF tagger is optimised by treating both the ggF \Htautau and \Ztt events as backgrounds and relies solely on observables based on the kinematics of the two leading jets (see \cref{tab:mva:variables}).
While the expected contribution from ggF \Htautau events is small, the considerably larger theoretical uncertainty associated with its cross-section prediction in this kinematic phase space can significantly enlarge the systematic uncertainty of the VBF production cross-section measurement.
 
The BDT score requirement used to define the categories was optimised to give the smallest uncertainty in the VBF cross-section, and provides a selection where the fraction of VBF events among all Higgs boson events is about \SI{94}{\percent} (\SI{63}{\percent}) in the VBF\_1 (VBF\_0) region.
 
\subsubsection*{V(had)H categorisation}
 
To match the STXS $qq\to V(\to qq)H$ particle-level \pTjet requirement and \mjj binning, events selected in the V(had) categories must satisfy $\SI{60}{\GeV}<\mjj<\SI{120}{\GeV}$ and \pT of the sub-leading jet greater than \SI{30}{\GeV}.
 
The V(had)H tagger was trained by treating all Higgs events produced by processes other than $VH$ as background.
The BDT score requirement used to define the two categories was optimised to give the smallest uncertainty for the V(had)H cross-section, and provides a selection where the expected fraction of V(had)H among all Higgs boson events is \SI{66}{\percent} (\SI{24}{\percent}) in the VH\_1 (VH\_0) category.

\subsubsection*{Boost categorisation}
 
Events failing to meet the criteria of the VBF, V(had)H and ttH categories but having high-\pT Higgs candidates are considered for the `boost' categories targeting ggF events with large Higgs boson transverse momentum.
The reconstructed Higgs boson transverse momentum, \pTH, is determined from the Higgs boson candidate defined by the vectorial sum of the momenta of the visible decay products of the $\tau$-leptons and \vecmet.
Events in the boost category must satisfy $\pTH>\SI{100}{\GeV}$.
To match the STXS $gg\to H$ particle-level requirements, events are further categorised by \pTH value and by the total number of jets with \pT greater than \SI{30}{\GeV} (\njetsthirty).
Events with $\pTH<\SI{200}{\GeV}$ are separated into 1-jet and $\ge$2-jet categories, while for $\pTH>\SI{200}{\GeV}$ events with at least one jet are considered without further jet-multiplicity separation of the events.
\Cref{tab:sel:boost:cat} describes the boost phase-space  categorisation.
 
\begin{table}[h!]
\caption{Definition of the six categories in the boosted phase space.}
\label{tab:sel:boost:cat}
\begin{center}
\begin{tabular}{
@{}lllll@{}
}
\toprule
\multirow{2}{*}{\njetsthirty} & \multicolumn{4}{c}{\pTH bins in \gev}                                                   \\
& [100, 120]       & [120, 200]       & [200, 300]                  & $\left[300, \infty\right[$             \\
\midrule
Exactly 1                     & boost\_0\_1J   & boost\_1\_1J   & \multirow{2}{*}{boost\_2} & \multirow{2}{*}{boost\_3} \\
\cmidrule(r){1-3}
At least 2                    & boost\_0\_ge2J & boost\_1\_ge2J &                           &                           \\
\bottomrule
\end{tabular}
\end{center}
\end{table}
 
The three analysis channels are therefore split into six kinematic
categories in the boost phase space for a total of eighteen categories
in the fit performed for the cross-section measurement.
 
\subsubsection*{Summary}
 
\begin{figure}[htbp]
\centering
\includegraphics[width=\textwidth]{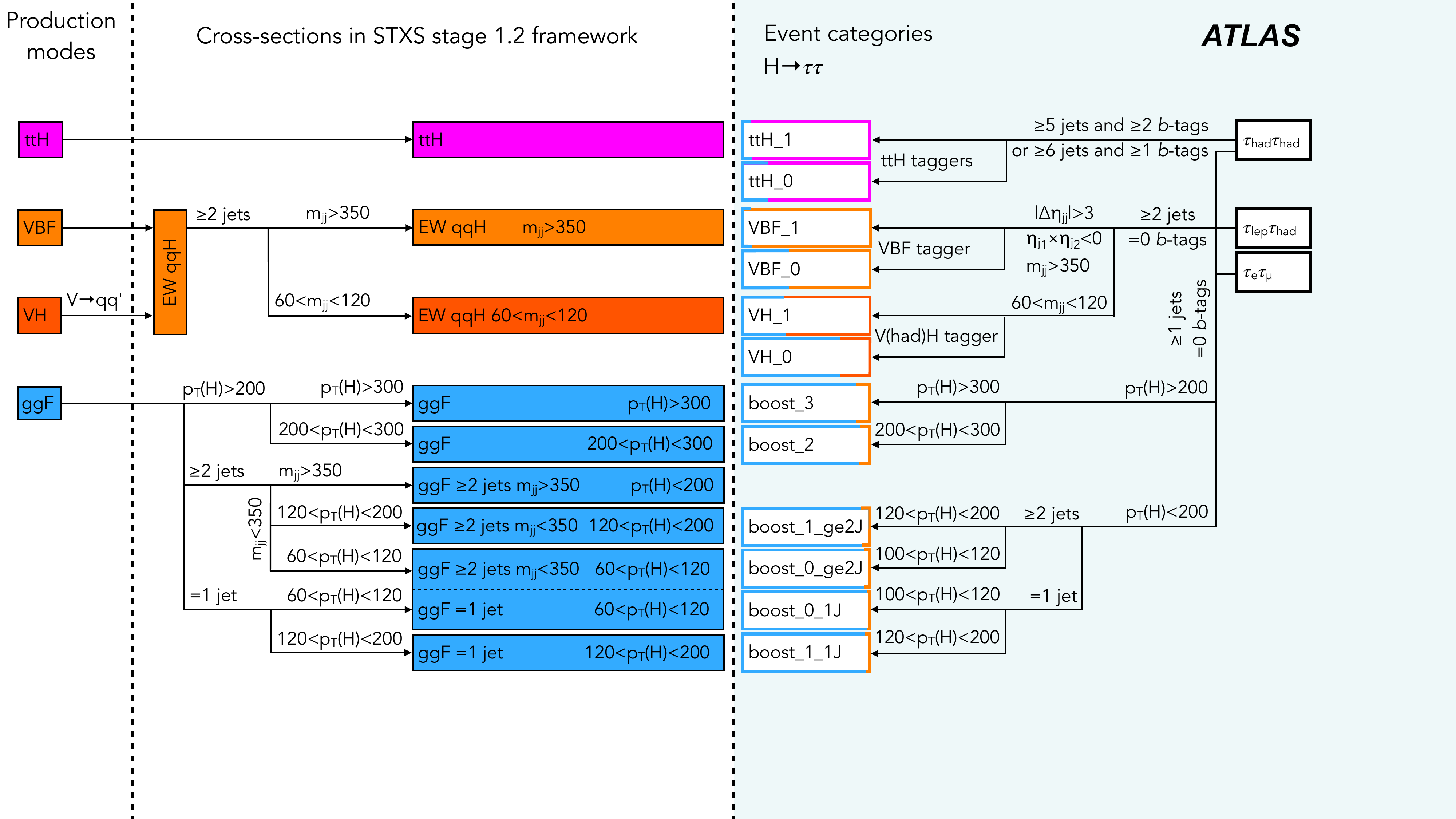}
\caption[STXS sketch]{
Sketch of the event categorisation and the targeted cross-sections in the STXS \emph{stage 1.2} framework (bins).
The relative contributions to each event category from the two most dominant STXS bins are indicated by the two colours used along the width of the category box.
The requirements on \pTH and \mjj are given in units of $\GeV$.}
\label{fig:stxs_sketch}
\end{figure}

\begin{figure}
\begin{center}
\begin{tabular}{c}
\includegraphics[width=\linewidth]{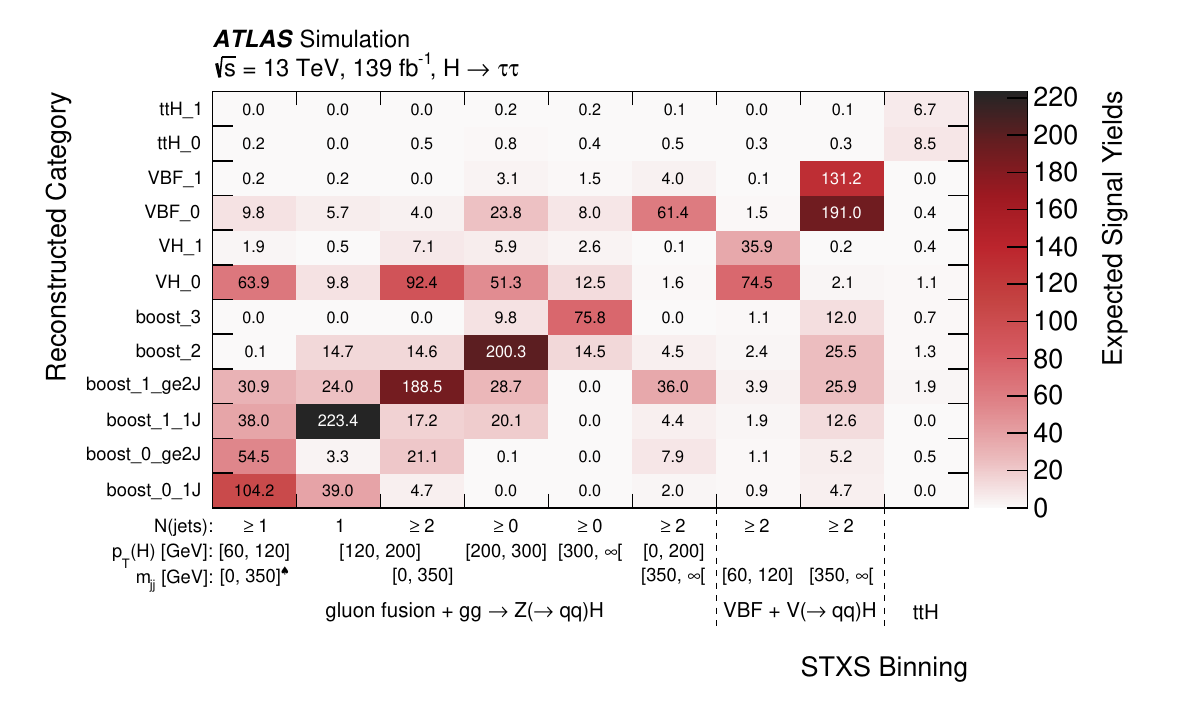} \\
(a) \\
\includegraphics[width=\linewidth]{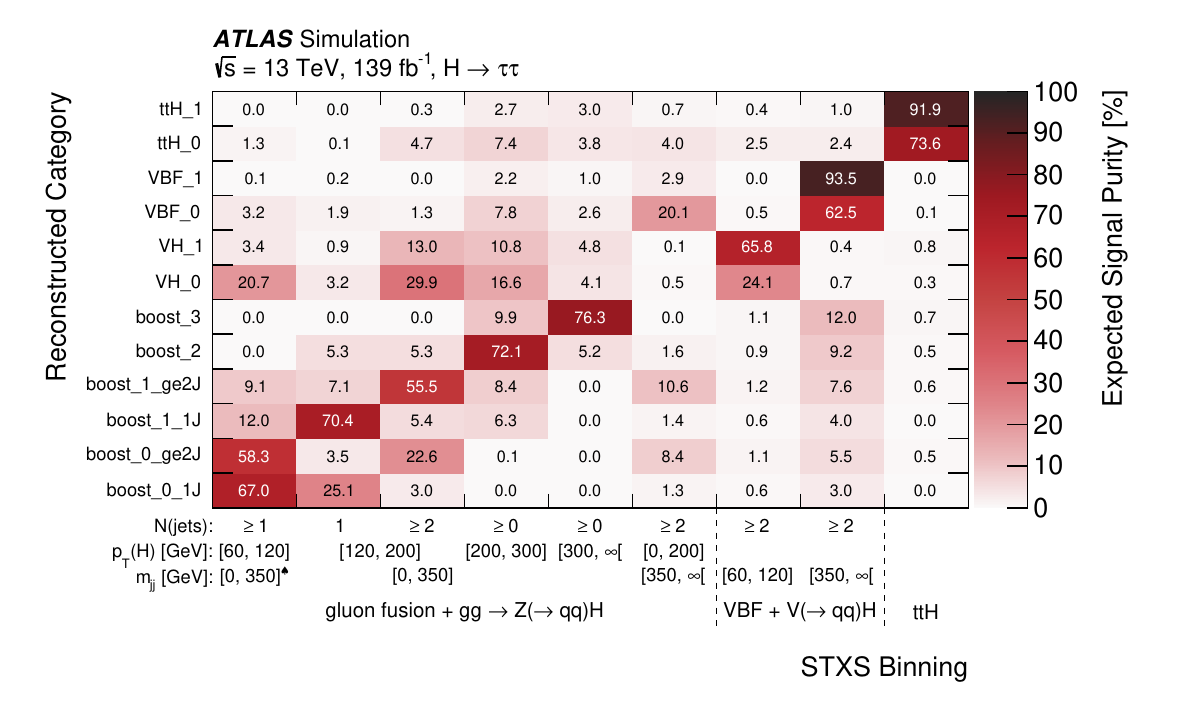} \\
(b) \\
\end{tabular}
\caption{
(a)~Expected $H\to\tau\tau$ signal yield in each of the reconstructed-event
categories of the analysis ($y$-axis) for each of the nine measured STXS
bins ($x$-axis).  (b)~Relative contribution of each of the nine
measured STXS bins to the total $H\to\tau\tau$ signal expectation in each
reconstructed-event category.  The spades symbol ($\spadesuit$)
indicates that the criteria for \mjj only apply to events with at least
two reconstructed jets.  Yields are summed over the three $\tau\tau$
decay channels ($\tau_{e}\tau_{\mu}$, $\tau_{\text{lep}}\tau_{\text{had}}$, $\tau_{\text{had}}\tau_{\text{had}}$).
\label{fig:signal:yields}}
\end{center}
\end{figure}

Nine bins of the STXS framework are targeted in the measurement presented in this paper and are illustrated in \Cref{fig:stxs_sketch}.
The expected signal yields for each of these bins is presented in \Cref{fig:signal:yields}(a), while \Cref{fig:signal:yields}(b) illustrates the relative population of these nine bins in each reconstruction category described in this section.
Events selected in each reconstruction category are used to build templates of the \mmmc variable for each of the nine bins.
As illustrated in \Cref{fig:signal:yields}, ggF events produced with \pTH\,<\,\SI{200}{\GeV} and two additional jets forming a system with \mjj\,>\,\SI{350}{\GeV} are mainly reconstructed in the VBF\_0 category (\SI{61}{\percent}) and the boost\_1\_ge2J category (\SI{36}{\percent}).
It is difficult to select these events in only a single category but through the simultaneous usage of all the categories, their production rate can be measured.
In contrast, the reconstructed ggF event candidates satisfying \SI{60}{\GeV}\,<\,\pTH\,<\,\SI{120}{\GeV} are further separated into those produced with a single jet (boost\_0\_1J) and those produced with two jets forming a system with \mjj~<\,\SI{350}{\GeV} (boost\_0\_ge2J).
However, the categorisation does not provide enough sensitivity to measure these two contributions individually and they are therefore combined.

\clearpage


\section{Background modelling}
\label{sec:bkg}

The expectations from SM processes other than the \Htautau signal in the phase space of the analysis are evaluated using a mixture of simulations and data-driven techniques.
Processes with \tauhadvis, prompt light leptons or light leptons from $\tau$-lepton decays are estimated through simulations.
Among these, \Zttjets and top processes are dominant, and dedicated control regions are employed to validate the simulations of both processes and to constrain their normalisation in the signal regions.
For the \Zttjets background, a control region enriched in \Zlljets events is defined as described in \cref{sec:ztt}.
In the \temu and \tlhad channels, control regions enriched in top-induced processes are defined by replacing the $b$-jet veto from the event selection (see \cref{tab:selection:baseline}) with a requirement of at least one $b$-tagged jet.
 
Using these control regions, the templates of the \mmmc observable from the simulations are checked in each event category (see \cref{sec:selection:categorisation}).
Very good agreement with the data is observed.
 
Smaller background contributions are due to diboson, \Zlljets and \HWW processes.
They are normalised to their theoretical expectations.
Contributions from light- and heavy-flavour jets misidentified as electrons, muons or \tauhadvis, as well as non-prompt electrons or muons, collectively referred to as misidentified $\tau$ background, are estimated using data-driven techniques.
Their estimation is detailed in \cref{sec:fake}.
 
\begin{figure}[h!]
\begin{center}
\begin{tabular}{cc}
\includegraphics[width=0.5\linewidth]{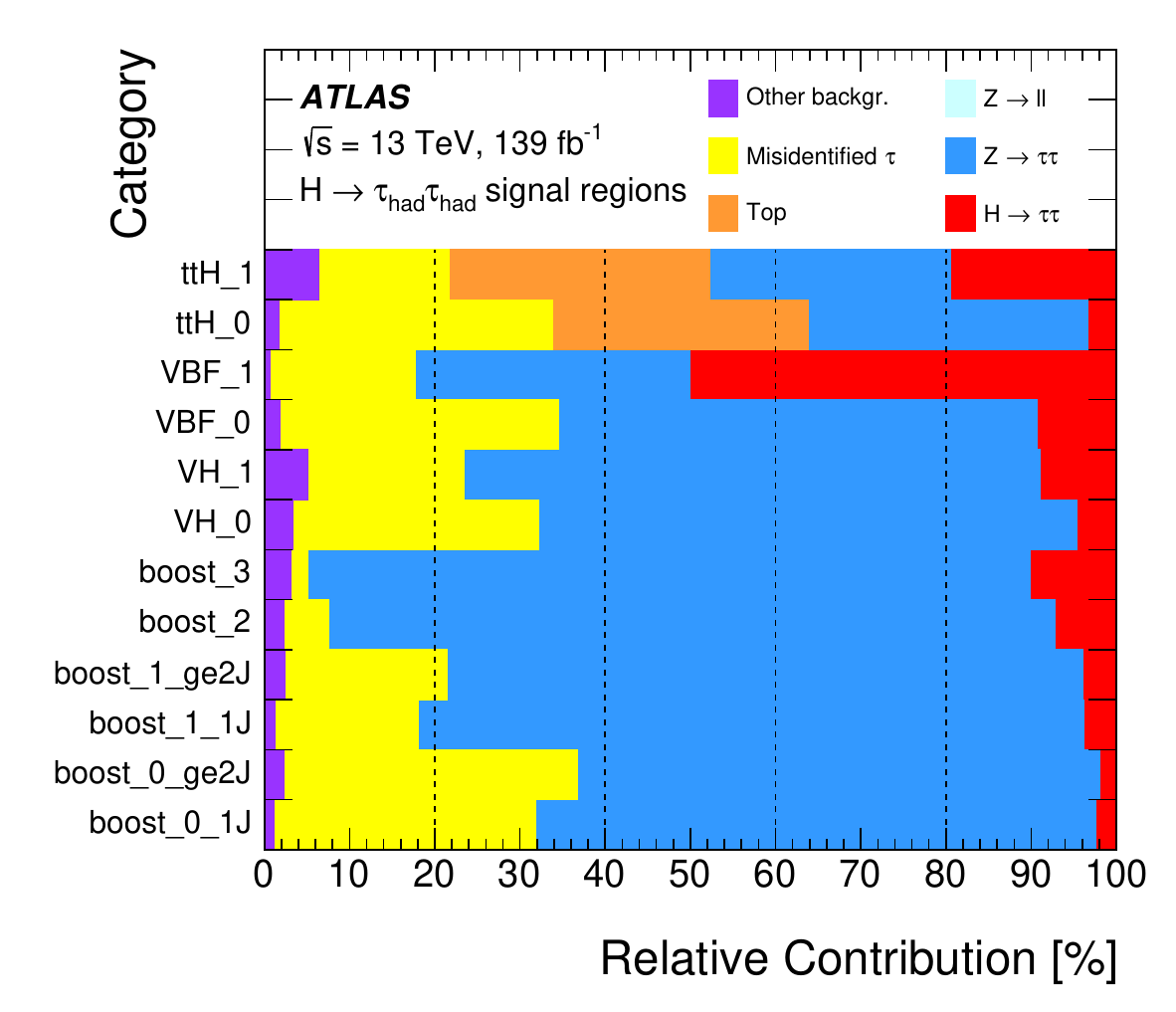} &
\includegraphics[width=0.5\linewidth]{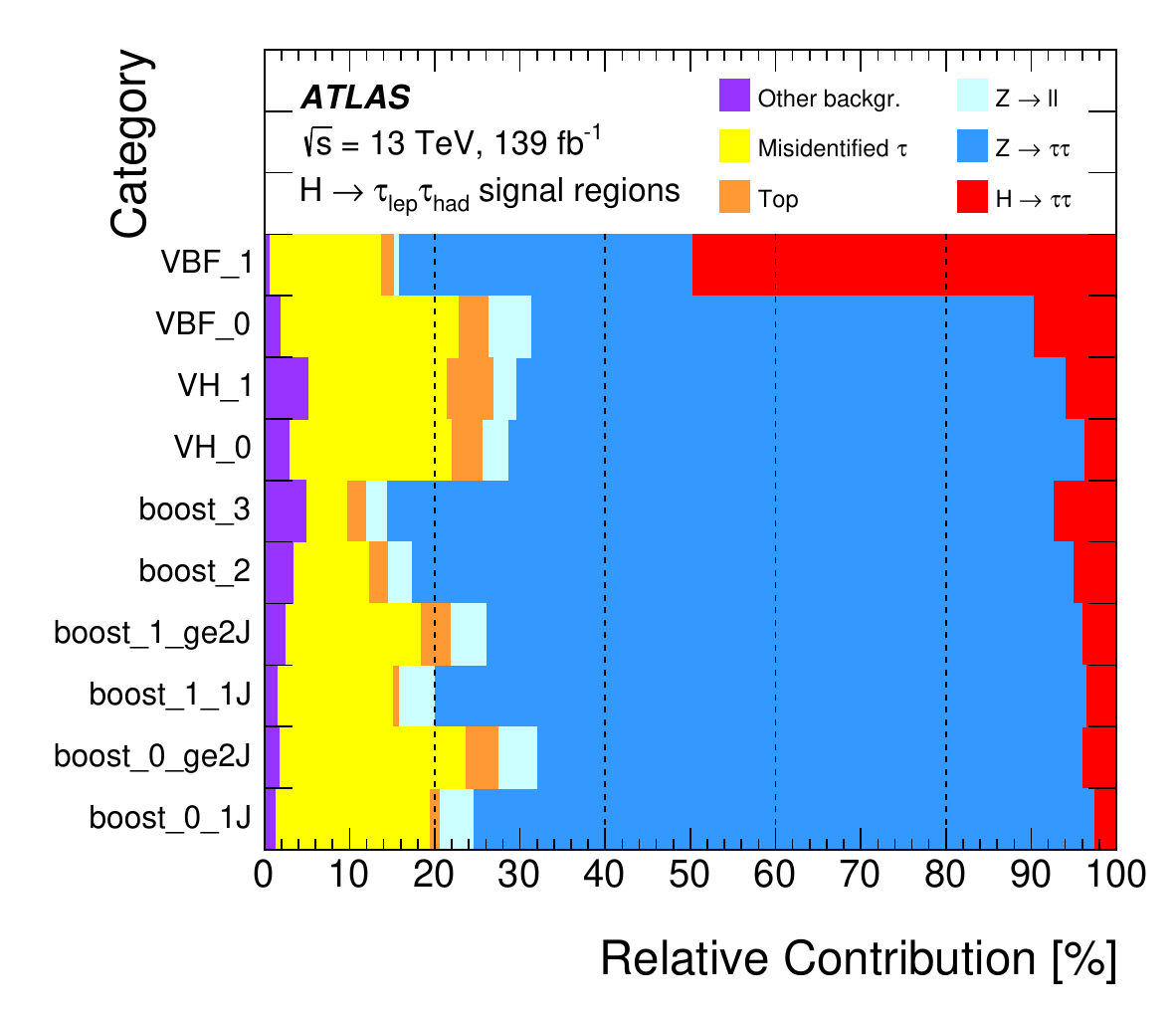} \\
(a) \thadhad & (b) \tlhad \\
\includegraphics[width=0.5\linewidth]{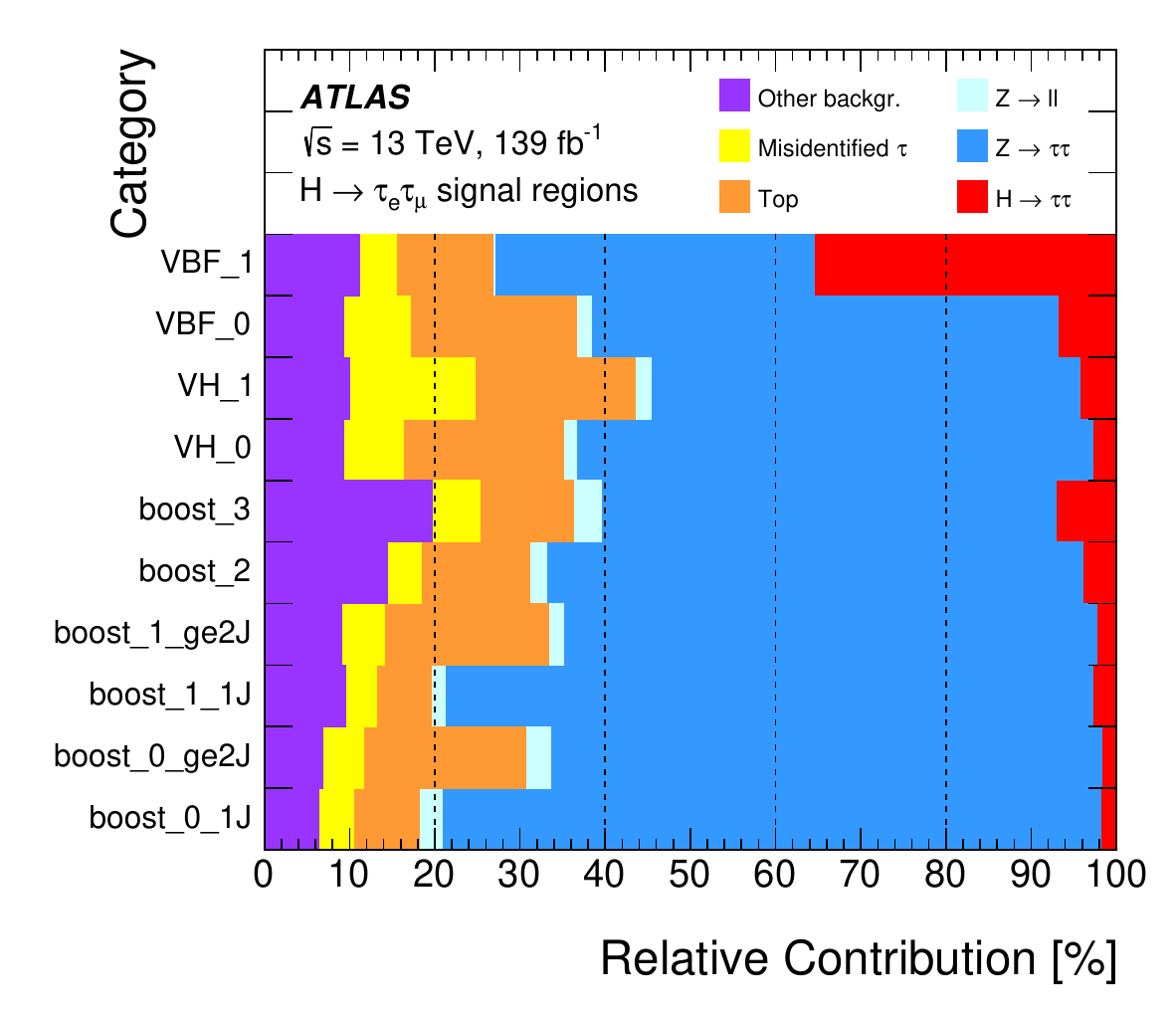} & \\
(c) \temu & \\
\end{tabular}
\caption{
Relative contribution of each process to the total measured yields in each category of the analysis for the (a)~$\tau_{\text{had}}\tau_{\text{had}}$, (b)~$\tau_{\text{lep}}\tau_{\text{had}}$ and (c)~$\tau_{e}\tau_{\mu}$ channels, within 100\,\GeV < $m_{\tau\tau}^{\text{MMC}}$ < 150\,\GeV.
`Other backgr.' includes diboson and \HWW processes.}
\label{fig:cat:composition}
\end{center}
\end{figure}

\Cref{fig:cat:composition} illustrates the measured composition of the selected events in each category of the analysis.
 
\clearpage 
\subsection{\Ztt background modelling using \Zll events}
\label{sec:ztt}

Events from the \Zttjets process form the dominant source of background in this measurement.
They account for \SI{79}{\percent} of the background across all signal regions, and up to \SI{90}{\percent} of the background in the most boosted regime investigated in the analysis.
They are estimated using MC simulations validated with data.
The predictions from these MC simulations are corrected using dedicated control regions based on the \Zlljets process with kinematic properties of the events similar to those of the corresponding signal regions as explained in the following.
 
In order to mimic as well as possible the \PZ boson kinematics and the associated production of jets in \Zttjets events selected in the signal regions, the selected \Zlljets events are modified through a simplified implementation of the embedding procedure.
The kinematic properties of the \PZ boson are reconstructed with a much better resolution in the \Zll decay channel than in the \Ztt one due to the absence of neutrinos and the excellent momentum resolution of the ATLAS detector for electrons and muons.
While the original method presented in Refs.~\cite{HIGG-2014-09, CMS-TAU-18-001} relied on substituting the detector signatures of the objects before re-reconstructing the event, the simplified embedding consists of a rescaling of the transverse momentum of each reconstructed lepton through parameterisations, followed by a recomputation of all the relevant kinematic quantities in the analysis.
The method used entails a significant reduction of complexity.
 
Embedding techniques are of particular interest in this analysis, where no statistically significant study of the \Zttjets background can be performed in data without looking at the signal regions.
In this context, the simplified embedding can be applied to data events passing the \Zlljets selection, thus obtaining a \Ztautau control region that is orthogonal to the signal region.
This control region can also be used to measure the \Ztautau normalisation in a phase space relevant to this measurement.
 
The \Zlljets events are selected using the single-lepton triggers and are required to have exactly two electrons or two muons with opposite charge.
The selected electrons and muons must satisfy the identification and isolation criteria defined in \cref{tab:selection:baseline}.
Additionally, the invariant mass of the dilepton system must be above \SI{80}{\GeV}.
The selected sample contains about $9.3\cdot10^6$ data events and \SI{99}{\percent} of them are expected to come from \Zlljets processes.
A small contribution from diboson and top processes with two electrons or two muons in the final state is also expected and the embedding procedure is also applied to them.
Contributions from processes with jets misidentified as leptons were found to be negligible.
Selected events in data and simulation are then randomly separated into three subsets to provide a statistically independent control region for each of the \temu, \tlhad and \thadhad signal regions.
 
Weights derived in simulations are applied to each event to remove the kinematic biases and normalisation effects introduced by the electron and muon trigger, reconstruction, identification, and isolation algorithms.
The four-vectors of the reconstructed electrons and muons are used to pair each lepton in the \Zlljets event with a scaling term, which parameterises the effects of $\tau$-lepton decay kinematics and of the energy calibration algorithms for $\tau$-leptons with similar four-vectors.
The scaling term is derived as a function of the transverse momentum and the pseudorapidity of the $\tau$-lepton before it decays.
The original four-vectors of the electrons and muons are scaled using this term so that they match those of the visible reconstructed decay products of either leptonically or hadronically decaying $\tau$-leptons.
The \Zlljets event yields are then reweighted to account for the expected efficiencies of the reconstruction, identification and calibration steps for the visible $\tau$ decay products.
 
\begin{figure}[h!]
\begin{center}
\begin{tabular}{cc}
\includegraphics[width=0.5\linewidth]{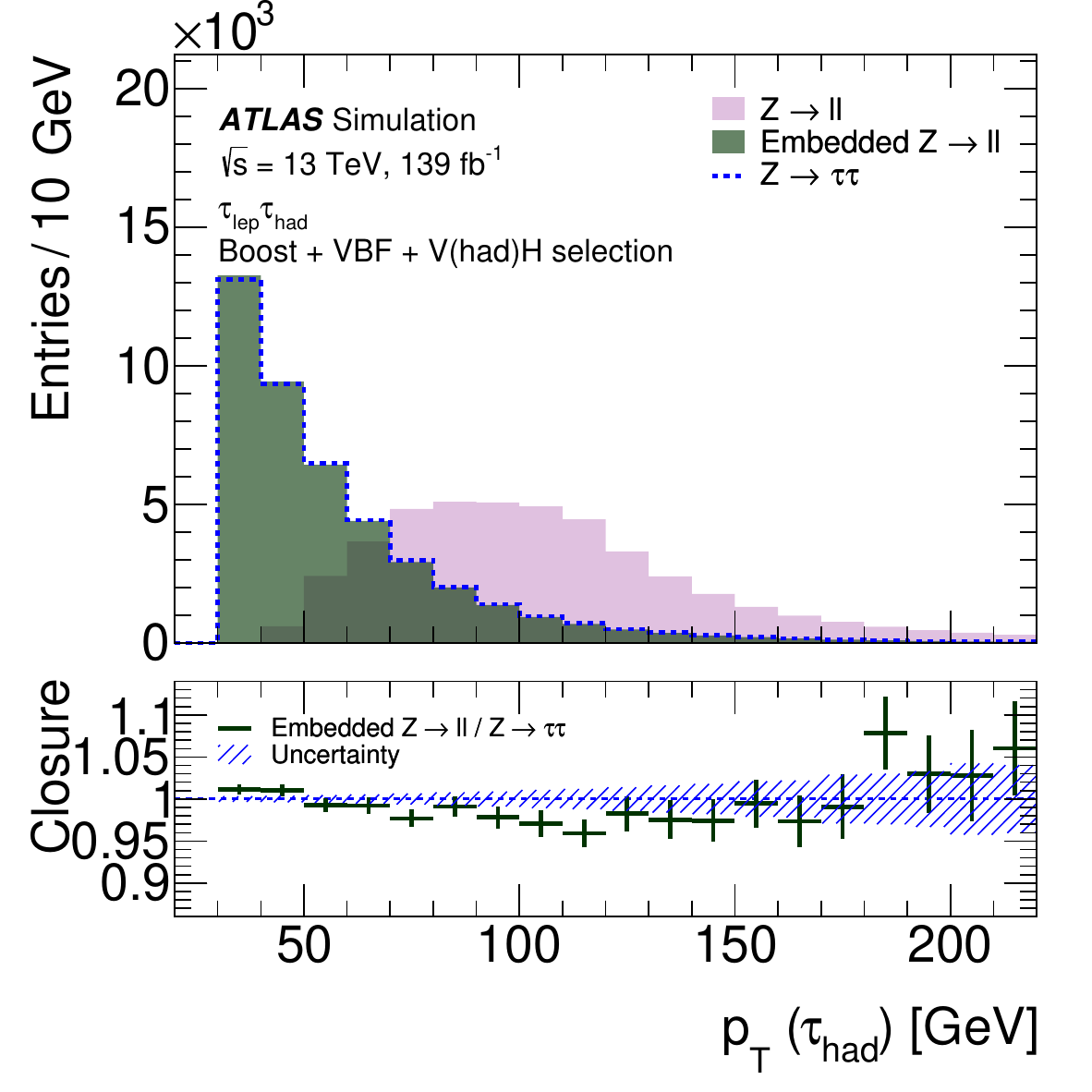} &
\includegraphics[width=0.5\linewidth]{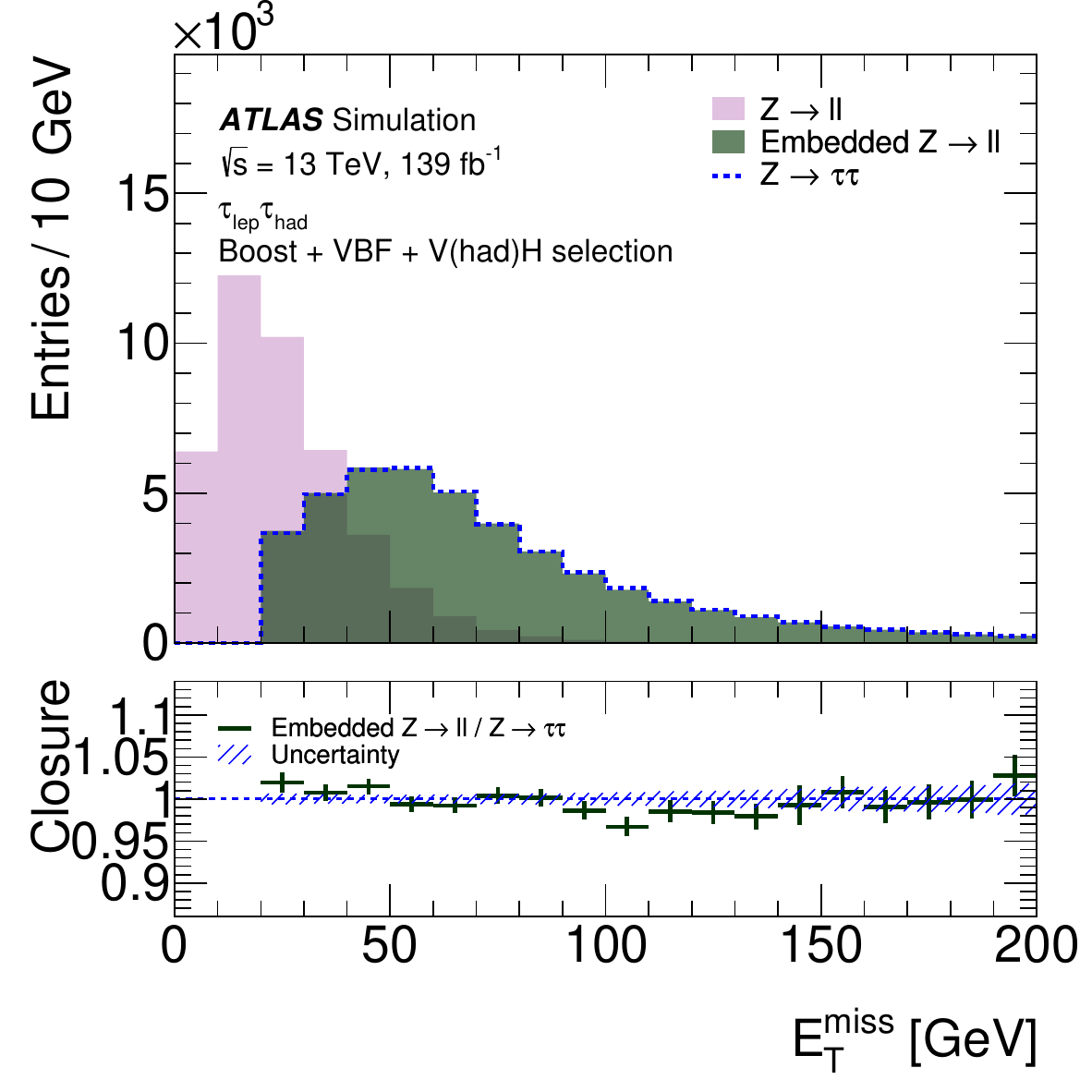} \\
(a) \pT(\tauhad) & (b) \met \\
\end{tabular}
\caption{
Comparison of kinematic quantities for $Z\to\ell\ell$  simulated events in the $\tau_{\text{lep}}\tau_{\text{had}}$ channel before (light purple histogram) and after (dark green histogram) the embedding procedure, in the boost, VBF and V(had)H phase spaces combined.
The distribution for $Z\to\tau\tau$  simulated events (dashed blue line) is also shown.
(a)~\pT distribution of the simulated $\tau_{\text{had}}$ in the event. For the $Z\to\ell\ell$  events, the reconstructed lepton with the highest \pT in the event is shown. For the $Z\to\ell\ell$  events after the embedding procedure, a scaling term is applied to the \pT of the lepton chosen to mimic the $\tau_{\text{had}}$ as described in the text.
(b)~\met distribution.
The bottom panels display the ratio of embedded $Z\to\ell\ell$ events to $Z\to\tau\tau$  events.
The error bars display the statistical uncertainties in the ratio and the dashed blue band illustrates the statistical uncertainty in the $Z\to\tau\tau$ simulation.
\label{fig:validation:embedding}
}
\end{center}
\end{figure}

The per-lepton weights assume collinearity of the $\tau$-lepton and its visible decay products and cannot take into account any correlation between the \PZ boson decay products.
All event variables used in the signal region definitions are recalculated using the kinematics of the new final-state physics objects, and a weight is applied to each event to account for the expected trigger efficiency associated with these objects.
The implementation of the new embedding procedure is validated by comparing \Zll simulated events, after applying this procedure, with \Ztt simulations, where both the kinematic and spin-correlation effects are modelled correctly.
\Cref{fig:validation:embedding} shows good agreement between the distributions of the two samples for two illustrative cases and indicates that the assumptions made in calculating the weights have negligible impact on the relevant observables.
 
All uncertainties affecting the reconstructed physics objects used in embedding are propagated through the full procedure, including those associated with the parameterisations.
Dedicated uncertainties affecting each control region are assigned to account for the differences in modelling observed between the \Ztt and embedded \Zll MC predictions, which are expected to come from approximations associated with the simplified embedding procedure.
These uncertainties are derived by studying the change in the data-to-simulation normalisation factors as events are moved between different control regions to cover the observed acceptance mismodeling.
They are found to be at the \SI{1}{\percent} level and cover for the residual non-closure observed in \cref{fig:validation:embedding}.
 
\begin{figure}[h!]
\begin{center}
\begin{tabular}{@{}cc@{}}
\includegraphics[width=0.5\textwidth]{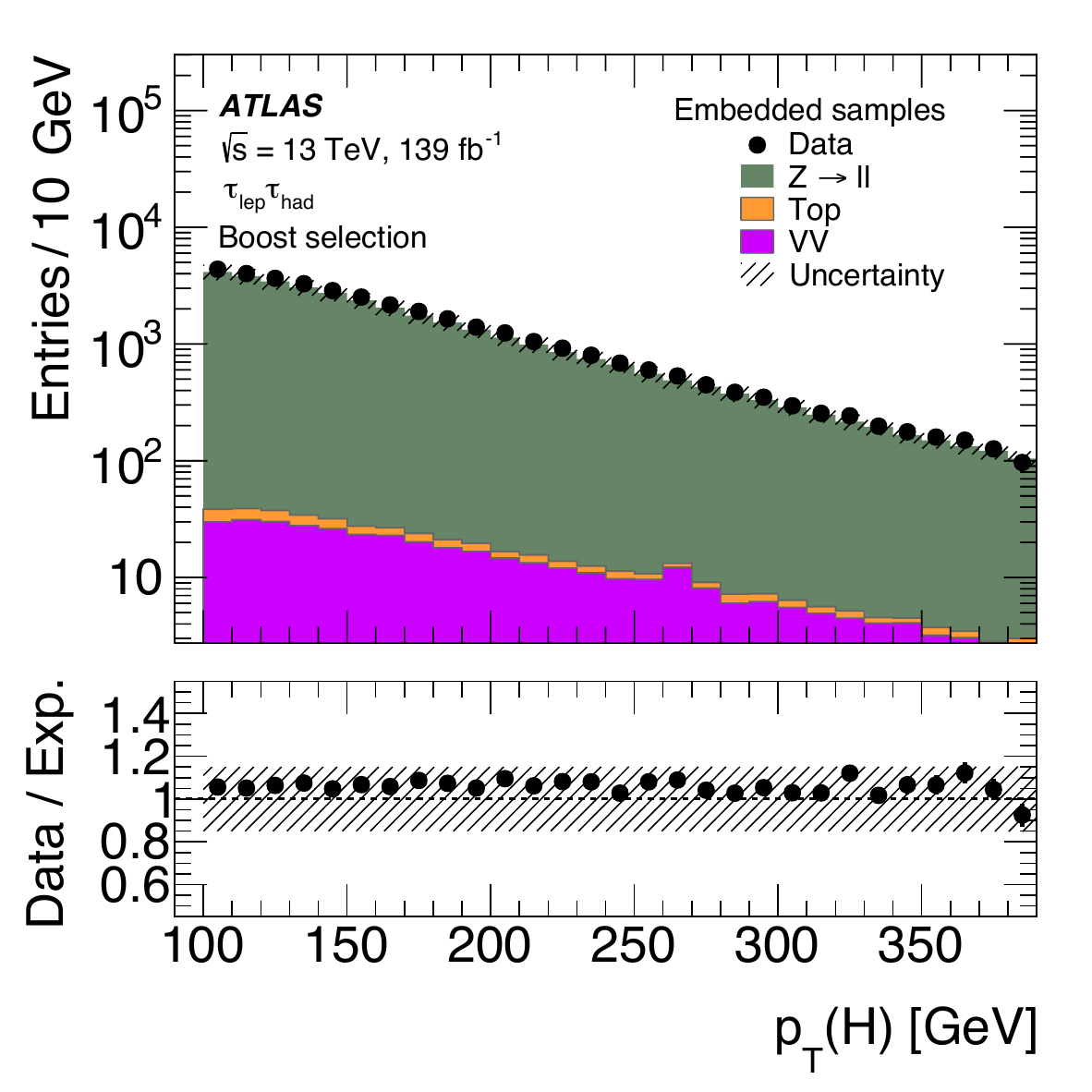} &
\includegraphics[width=0.5\textwidth]{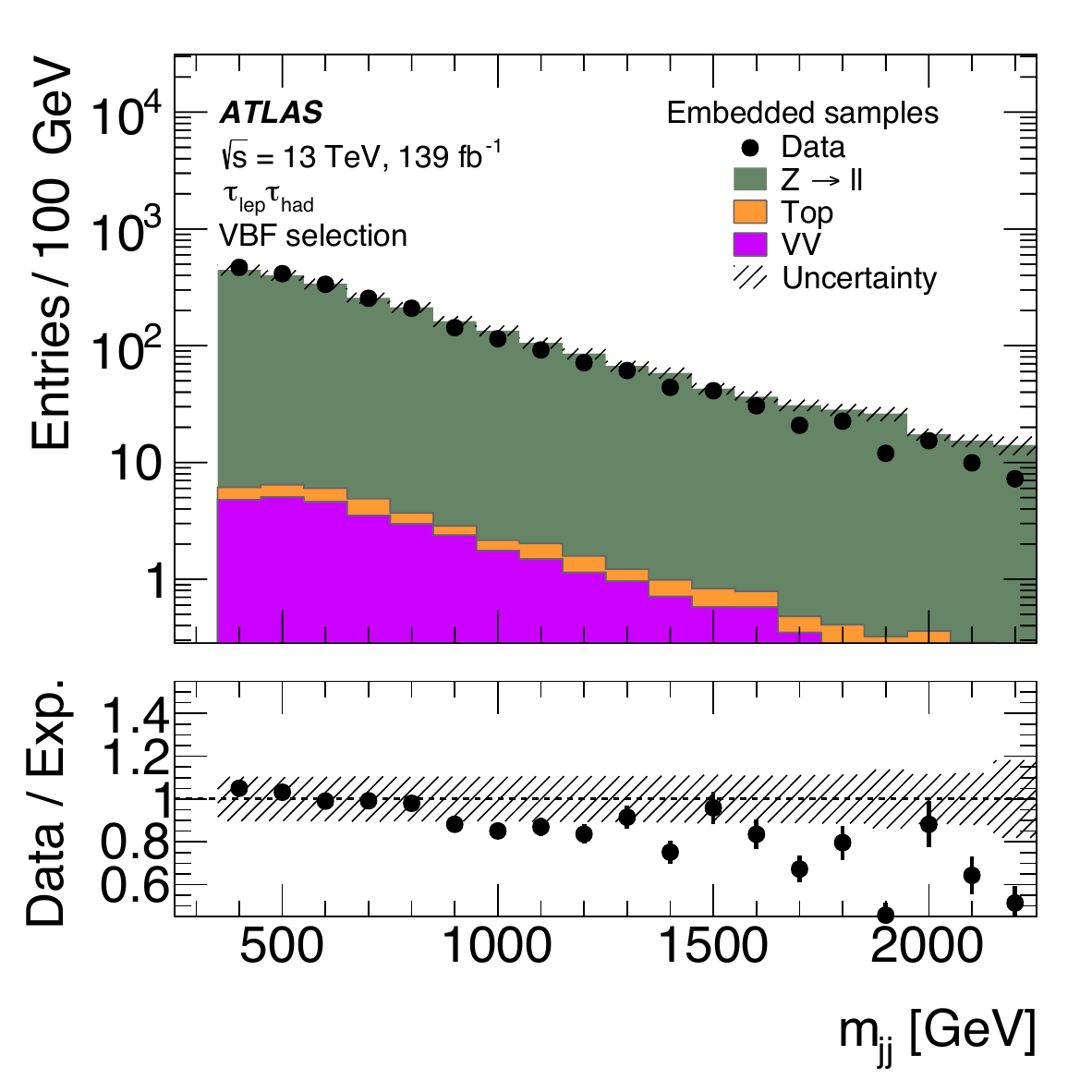} \\
(a) \pTH in boost categories & (b) \mjj in VBF categories \\
\includegraphics[width=0.5\textwidth]{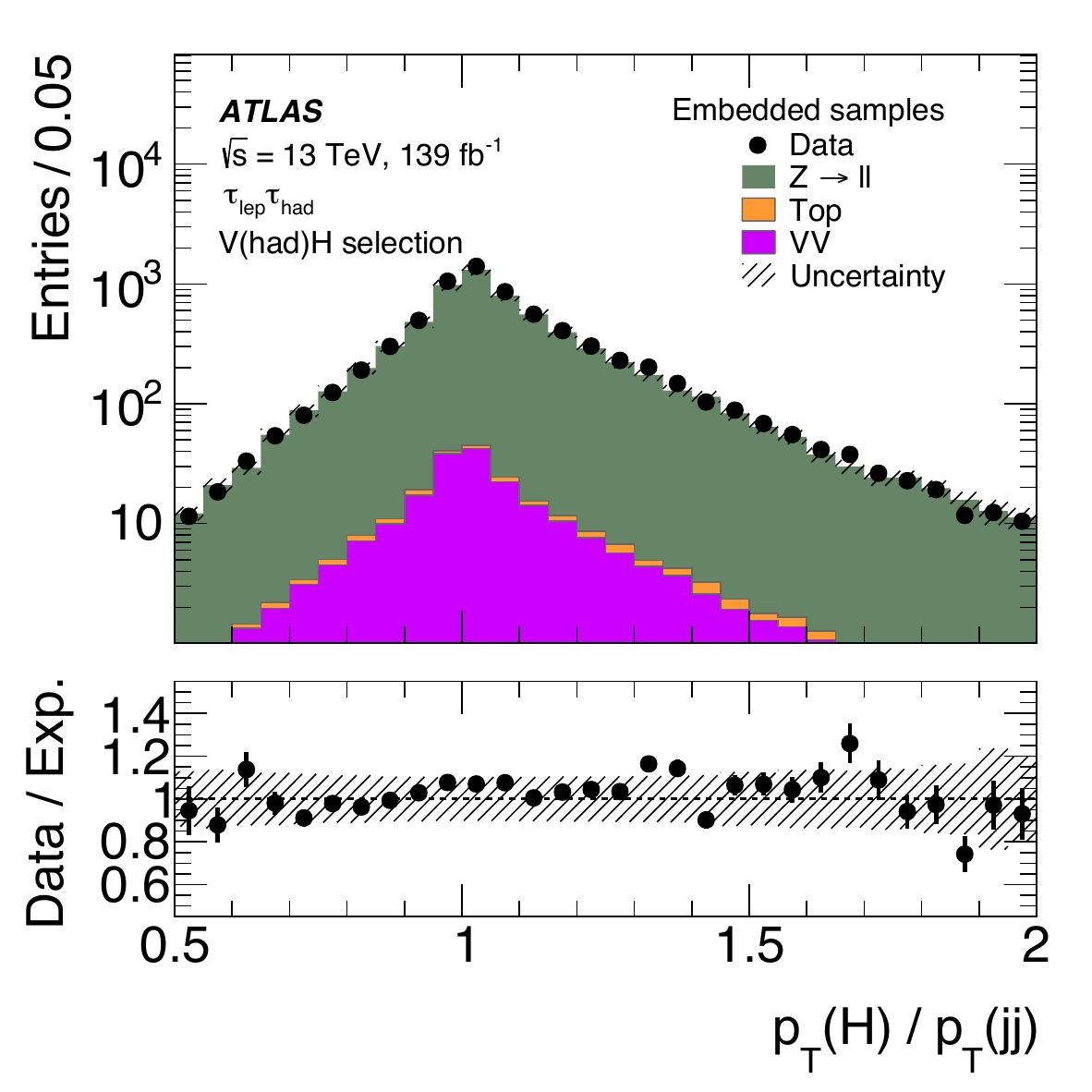} & \\
(c) \pTHoverpTjj in V(had)H categories & \\
\end{tabular}
\caption{
Comparison between MC simulation prediction and data in the \Zlljets-enriched control regions.
The embedding procedure is applied to the data and simulation samples to mimic the $\tau_{\text{lep}}\tau_{\text{had}}$  event selection: (a)~\pTH in the boost categories, (b)~\mjj in the VBF categories and (c)~\pTHoverpTjj in the V(had)H categories.
The bottom panels show the level of agreement between the embedded data and the embedded simulation samples.
The uncertainty is the sum in quadrature of the statistical uncertainty of the simulated events and the systematic uncertainties of the simulation.
Only the acceptance uncertainties in each category are considered.
The shape variations, translating to potential bin-by-bin changes, were estimated to be minor and are not displayed.
\label{fig:embedding:results}}
\end{center}
\end{figure}

Distributions for this control region, and a comparison with the embedding of all the simulated background processes, are shown in \Cref{fig:embedding:results}.
The observed discrepancies are consistent with the results reported in dedicated measurements of the \Zjets processes~\cite{STDM-2016-01,STDM-2016-09}.
The impact of this mismodelling on the analysis is alleviated by the use of control regions mimicking the event selection criteria after the embedding procedure is applied to data and simulated events.
 
\clearpage 
\subsection{Data-driven estimate of misidentified $\tau$ processes}
\label{sec:fake}

Processes with at least one jet misidentified as an electron, muon or \tauhad are collectively referred to as misidentified $\tau$ background.
They account for a fraction of the total background ranging from \SI{5}{\percent} to \SI{25}{\percent}, with less importance in the more boosted categories.
They are evaluated in a similar fashion in the three channels of the analysis.
First, data events are selected using the same criteria as for the SRs with the exception of the criteria for electron or muon identification and isolation and the criteria for \tauhadvis identification.
These criteria are loosened or inverted depending on the specific methodology used in each channel.
Then, transfer factors are computed in dedicated control regions.
These factors are used to correct for the kinematic and normalisation differences between the events with altered isolation or identification criteria and the SRs.
 
In the \temu channel, the misidentified $\tau$ background is estimated using the matrix-method technique~\cite{ATLAS-CONF-2014-058}.
Data events are selected by removing the lepton isolation criteria from the nominal selection, and loosening the identification criteria for electrons.
The expected number of fake leptons in the SR is computed from a system of equations relating the efficiencies for real ($\epsilon_{\textsc{r}}$) and fake leptons ($\epsilon_{\textsc{f}}$) to the observed event yields.
The efficiencies are estimated separately for electrons and muons and are parameterised as a function of the \pT and $\eta$ of the leptons.
The real-lepton efficiencies $\epsilon_{\textsc{r}}$ are estimated using simulations, while the fake-lepton efficiencies $\epsilon_{\textsc{f}}$ are measured using data events selected to have two leptons of the same charge.
For the latter, the contribution from events with real leptons is subtracted using MC simulations; they account for approximately \SI{35}{\percent} of the 1333 selected events.
 
Dedicated uncertainties estimated for these predictions account for: statistical uncertainties in the derived efficiencies (${\sim}10\%$), dependencies of $\epsilon_{\textsc{f}}$ on the numbers of jets and $b$-tagged jets in the final state (${\sim}35\%$), the dependency of $\epsilon_{\textsc{r}}$ on whether they are measured in \ttbar, \Zlljets or \Zttjets events (${\sim}15\%$), and the uncertainty associated with the normalisation of the contribution from real leptons during the measurement of $\epsilon_{\textsc{f}}$ (${\sim}15\%$).
 
In the \tlhad channel, the misidentified $\tau$ background refers to events with a jet misidentified as a \tauhadvis.
Contributions with a real \tauhad and a jet misidentified as an electron or a muon are estimated from simulations to be negligible.
The misidentified $\tau$ background is evaluated using the fake-factor technique~\cite{HIGG-2013-21}.
Data events are selected if they satisfy a very loose requirement on the \tauhadvis identification score but do not satisfy the \objectsc{Medium} working point criteria (such events are `reverse-identified').
All other criteria of the nominal selection of the \tlhad channel are applied.
Residual contributions from processes with real \tauhadvis satisfying this requirement are evaluated using simulations and subtracted accordingly.
They account for approximately \SI{18}{\percent} of the \num{136500} selected events.
 
The distribution of the misidentified $\tau$ background component in the SR is obtained by multiplying the contribution of the data events selected by the reverse-identified criterion with a fake factor defined as the ratio of misidentified \tauhadvis that respectively pass or fail the \objectsc{Medium} working point of the \tauhadvis identification algorithm.
These fake factors are parameterised as a function of the \pT and track multiplicity of the \tauhadvis.
Two sets of fake factors are derived in separate regions and then combined for the final estimate.
The first set is derived in a region enriched in \Wjets processes obtained by inverting the SR criterion for $\mT$ (see \cref{tab:selection:baseline}).
The second set is derived in a control region enriched in QCD multijet processes obtained by inverting the isolation criteria for the selected electron or muon.
An estimate of the fraction of events expected to originate from QCD multijets is used to determine the relative weighting of both sets of fake factors; it is parameterised as a function of the \pT and $\eta$ of the \tauhadvis candidate.
This estimate is obtained by scaling the number of events in the second control region by the ratio of events where the light lepton respectively fails or passes the isolation requirements, measured in another QCD-multijet-enriched region where the \taulep and \tauhad have the same charge.
 
Uncertainties in the fake factors are estimated, and account for statistical uncertainties in the fake factors and their relative weighting (${\sim}15\%$), for uncertainties associated with the subtraction of the residual contributions from processes with real \tauhad (${\sim}10\%$), and for uncertainties in the flavour composition (${\sim}10\%$), taken from comparisons between the predicted and observed backgrounds in a dedicated validation region.
 
In the \thadhad channel, the misidentified $\tau$ background is also determined using a fake-factor approach.
The method differs slightly from the one used in the \tlhad channel: the fake factors are parameterised to simultaneously account for processes with one or two jets misidentified as \tauhadvis.
Additionally, the reconstructed \tauhadvis candidates are matched to their high-level-trigger counterparts.
The fake factors are estimated in the \Wjets-enriched region defined for the \tlhad channel, but with the addition of the trigger-matching requirement in the \tauhadvis definition.
 
Two alternative sets of fake factors are computed in control regions defined with two \tauhadvis.
The first alternative set is derived by inverting the requirement on the  $\Deta(\tauhadvis, \tauhadvis)$ variable with respect to the signal region.
The second is derived by requiring the charges of the two \tauhadvis to have the same sign.
The difference between these two alternative sets and the nominal fake factors derived in the \Wjets-enriched control region is used to estimate the uncertainty in the composition of the misidentified $\tau$ background (${\sim}15\%$).
Two additional uncertainties in the misidentified $\tau$ background estimate in the \thadhad channel are considered: the statistical uncertainty of the fake-factor calculation (${\sim}15\%$), and uncertainties related to the parameterisation choice for the fake factors (${\sim}5\%$).
 
In the \temu and \tlhad channels, the analysis employs control regions enriched in top processes.
In these control regions, heavy-flavour jets misidentified as electrons or muons represent \SI{70}{\percent} to \SI{80}{\percent} of the expected contributions for the \temu channel, while  for the \tlhad channel about \SI{25}{\percent} of misidentified \tauhadvis originate from heavy-flavour jets.
To estimate these contributions, the data-driven estimate described above is repeated with the $b$-jet veto replaced by a $b$-tagged jet requirement to mimic the control region selection.
 
\begin{figure}[h!]
\centering
\begin{tabular}{cc}
\includegraphics[width=0.5\textwidth]{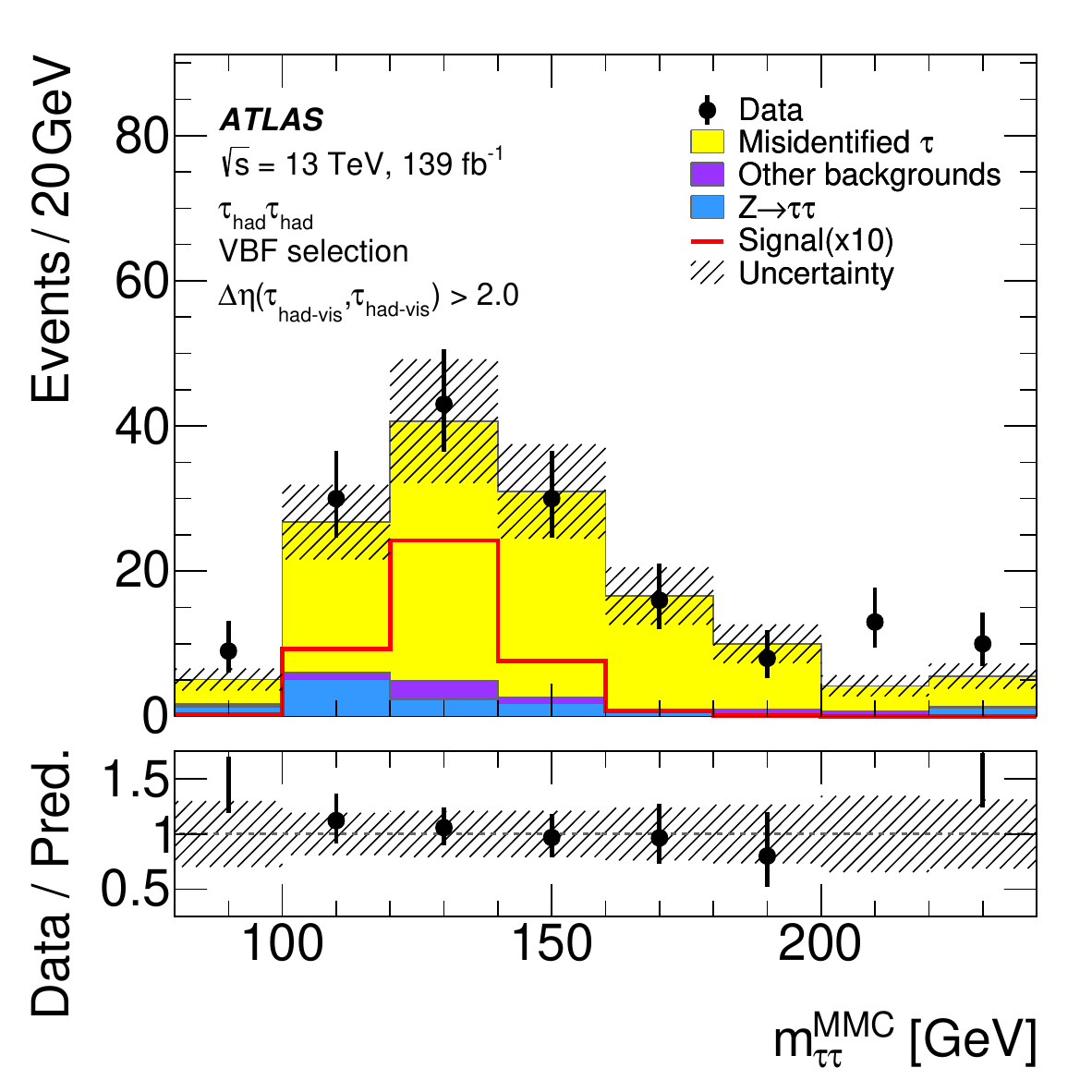} &
\includegraphics[width=0.5\textwidth]{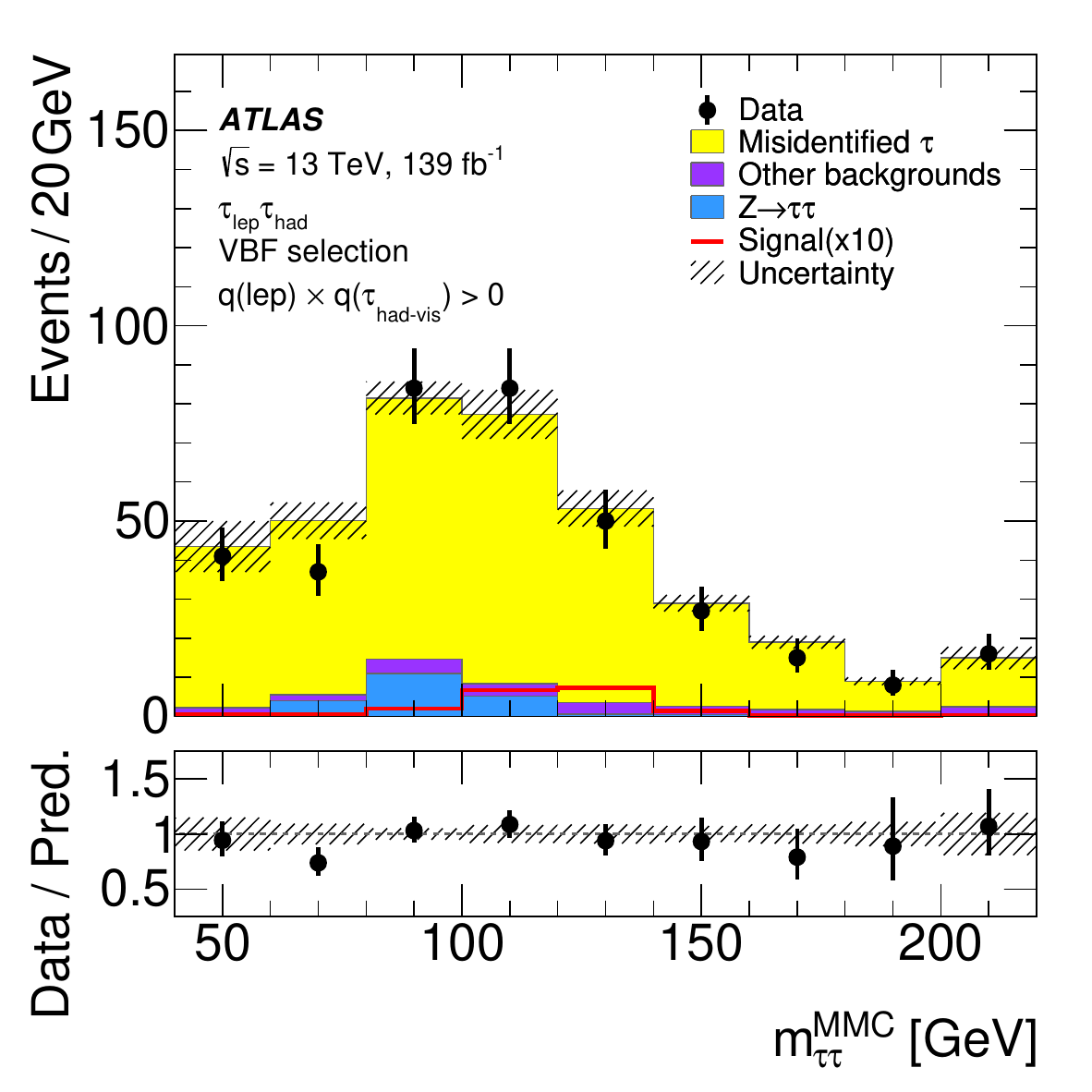} \\
(a) \thadhad & (b) \tlhad \\
\includegraphics[width=0.5\textwidth]{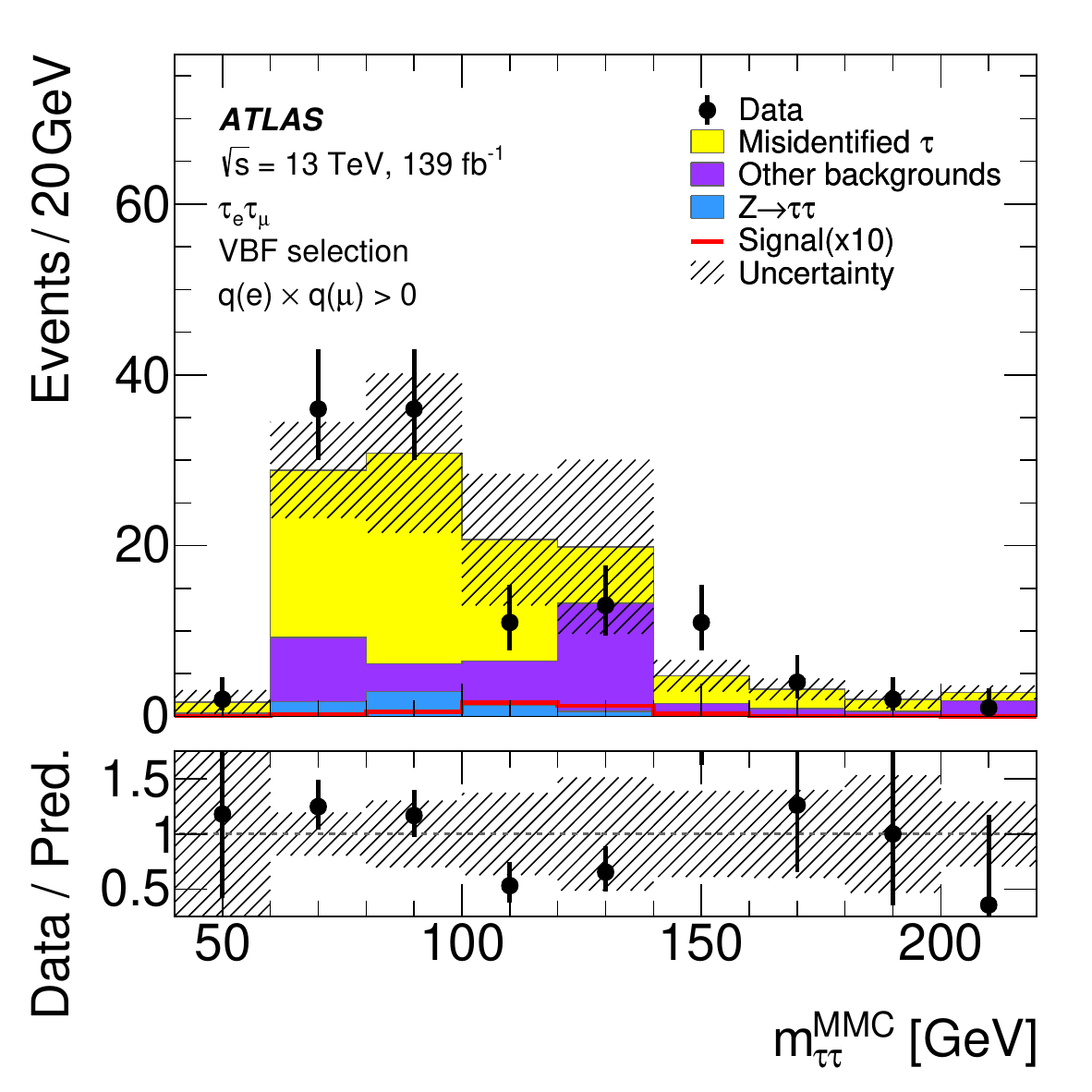} & \\
(c) \temu & \\
\end{tabular}
\caption{
Validation of the data-driven estimate of the processes with jets misidentified as \tauhadvis in the VBF categories: (a)~events with $\Delta \eta (\tau_{\text{had-vis}}, \tau_{\text{had-vis}})>2.0$ in the $\tau_{\text{had}}\tau_{\text{had}}$ final state, (b)~events with a light lepton and $\tau_{\text{had-vis}}$
of the same charge in the $\tau_{\text{lep}}\tau_{\text{had}}$  channel, and (c)~events with same-charge leptons in the $\tau_{e}\tau_{\mu}$ final state.
The hashed band represents the statistical uncertainty due to the limited size of the simulated samples and the systematic uncertainty of the data-driven estimate.
\label{fig:fakes}}
\end{figure}

The modelling of the misidentified $\tau$ background was validated in dedicated regions for each channel.
In the \thadhad channel, the validation region selects events with $\Delta \eta (\tau_{\text{had-vis}}, \tau_{\text{had-vis}})>2.0$.
In the \tlhad channel, the validation region contains events with a light lepton and $\tau_{\text{had-vis}}$ of the same charge.
Finally, events with same-charge leptons are used as the validation region for the \temu channel.
\Cref{fig:fakes} illustrates the modelling of the misidentified $\tau$ background in the validation region for each channel.
Good agreement between the observed data and the prediction is seen in all cases.
 
\clearpage 
\section{Systematic uncertainties}

\label{sec:systematics}
Systematic uncertainties affect the yields in the various signal and control regions as well as the distribution shape of the main fit observable (\mmmc).
They can be assigned to three main groups: the experimental uncertainties, the theoretical uncertainties for the backgrounds and the theoretical uncertainties for the signal.
They are detailed in the following sections.
Their impact on the measured $pp\to\Htautau$ cross-section is summarised in \cref{tab:sig::1poi-breakdown}.
Systematic uncertainty sources are parameterised in the statistical analysis using nuisance parameters with Gaussian priors (see \cref{sec:stat}).
 
\begin{table}[h!]
\caption{
Summary of the different sources of uncertainty in decreasing order of their impact on $\sigma(pp\to\Htautau)$.
Their observed and expected fractional impacts, both computed by the fit, are given, relative to the $\sigma(pp\to\Htautau)$ value.
Experimental uncertainties for reconstructed objects combine efficiency and energy/momentum scale and resolution uncertainties.
Background sample size  includes the bin-by-bin statistical uncertainties in the simulated backgrounds as well as statistical uncertainties in misidentified $\tau$ backgrounds, which are estimated using data.}
\label{tab:sig::1poi-breakdown}
\begin{center}
\begin{tabular}{
@{}l
S
S
@{}
}
\toprule
\multirow{2}{*}{Source of uncertainty}      & \multicolumn{2}{c}{Impact on $\Delta\sigma$\,/\,$\sigma(pp\to\Htautau)$} $[\%]$ \\
& \multicolumn{1}{c}{Observed} & \multicolumn{1}{c}{Expected}             \\
\midrule
Theoretical uncertainty in signal           & 8.7                          & 8.5                                      \\
Jet and \etmiss                             & 4.5                          & 4.2                                      \\
Background sample size                      & 4.0                          & 3.7                                      \\
Hadronic $\tau$ decays                      & 2.1                          & 2.1                                      \\
Misidentified $\tau$                        & 2.0                          & 2.0                                      \\
Luminosity                                  & 1.8                          & 1.8                                      \\
Theoretical uncertainty in \Zjets\ processes & 1.7                          & 1.2                                      \\
Theoretical uncertainty in top processes    & 1.1                          & 1.1                                      \\
Flavour tagging                              & 0.4                          & 0.5                                      \\
Electrons and muons                         & 0.4                          & 0.4                                      \\
\midrule
Total systematic uncertainty                & 12.0                         & 11.4                                     \\
Data sample size                            & 7.2                          & 6.7                                      \\
Total                                       & 13.9                         & 13.2                                     \\
\bottomrule
\end{tabular}
\end{center}
\end{table}
 
\subsection{Experimental uncertainties}
 
In addition to the object misidentification rate already discussed in \cref{sec:fake}, experimental systematic uncertainties include those on the trigger, reconstruction, identification and isolation efficiencies for the final-state particle candidates, and their energy scale and resolution.
These uncertainties affect the shape of the \mmmc distribution, the background yields and the signal cross-section through their effects on the acceptance and the migration between different event categories.
 
The dominant experimental uncertainties in the measurement of the $pp\to\Htautau$ cross-section are related to the jet energy scale and resolution, to the \tauhadvis candidate identification and energy scale, and to the object misidentification rates, as shown in \cref{tab:sig::1poi-breakdown}.
The uncertainties related to the reconstruction and identification of electrons and muons and the jet $b$-tagging efficiency have only a minor impact on the measurement.
 
The jet energy scale uncertainty consists of components related to the in situ calibration of jets as well as pile-up, the extrapolation to higher transverse momentum, and uncertainties related to the different responses to quark- and gluon-initiated jets.
The latter is of particular importance and covers both the uncertainties in the response of the detector to particular jet flavours and the uncertainty in the response due to the unknown fractions of quark- and gluon-initiated jets within the sample.
The jet energy scale uncertainty for central jets ($|\eta|<1.2$) varies from \SI{1}{\percent} for a wide range of jet \pt ($\SI{250}{\GeV}<\pt<\SI{2000}{\GeV}$), to \SI{5}{\percent} for very low \pt jets (\SI{20}{\GeV}) and \SI{3.5}{\percent} for very high \pt jets ($>\SI{2.5}{\TeV}$).
The relative jet energy resolution is measured in a dedicated analysis~\cite{JETM-2018-05} and ranges from ($24\pm5$)\% at \SI{20}{\GeV} to ($6\pm0.5$)\% at \SI{300}{\GeV}.
 
The uncertainties in the \tauhadvis identification efficiency are in the range of \SI{2}{\percent} to \SI{6}{\percent}, while the trigger efficiency and the eBDT efficiency uncertainties are of the order of \SI{1}{\percent} to \SI{1.5}{\percent} and \SI{1}{\percent} to \SI{2}{\percent}, respectively.
All these uncertainties are parameterised as a function of the \tauhadvis \pt\ and number of associated tracks (identification and trigger efficiency) or $\tau$ decay mode (eBDT efficiency).
As this analysis is highly sensitive to the \tauhadvis reconstruction efficiency uncertainty due to the introduction of the \Zll control regions, this efficiency is left as a free parameter in the fit and measured in situ; the associated uncertainty is found to be at the \SI{2}{\percent} level.
For the \tauhadvis energy scale, the total uncertainty is in the range of \SI{1}{\percent} to \SI{4}{\percent}, arising from a combination of measurements: a direct measurement with \Ztaumutauhad events, measurements of the calorimeter response to single particles, and comparisons between simulations using different detector geometries or $\GEANT$ physics lists.
This uncertainty is also parameterised as a function of the \tauhadvis \pt\ and number of associated tracks~\cite{ATLAS-CONF-2017-029}.
 
All of the above uncertainties affecting the different hard objects are propagated through the \etmiss calculation.  Additional uncertainties associated with the scale and resolution of the soft-term of the \etmiss~\cite{PERF-2016-07} are also considered.
 
The luminosity uncertainty is considered only for the background samples whose normalisations are not determined in data (diboson, \Zll, non-(\Htautau) Higgs) and to derive the signal cross-sections from the measured yields.
The uncertainty in the combined 2015--2018 integrated luminosity is 1.7$\%$ \cite{ATLAS-CONF-2019-021}, obtained using the LUCID-2 detector \cite{Avoni:2018iuv} for the primary luminosity measurements.
 
\subsection{Background theoretical uncertainties}
Theoretical uncertainties are considered for the two main background contributions in this analysis: \Zjets\ and \ttbar.
The normalisation of these backgrounds is determined in the fit to the data in the signal and control regions (see \cref{sec:stat}).
The theoretical uncertainties of \Zjets and \ttbar are therefore parameterised to account for the migration across the analysis regions and to account for their impact on the \mmmc templates in each region.
 
For \Zjets, uncertainties were considered for renormalisation (\muR), factorisation (\muF) and resummation scale (\muQ) variations, for the jet-to-parton matching scheme (\ckkw), for variations in the choice of $\alphas$ value, and for the choice of PDFs.
Uncertainties from missing higher orders were evaluated~\cite{Bothmann:2016nao} using six variations of the QCD \muR and \muF scales in the matrix elements by factors of $0.5$ and $2$, avoiding the extreme variations in opposite directions.
Uncertainties in the nominal PDF set were evaluated using 100 replica variations; an uncertainty is derived in each bin of the \mmmc templates by evaluating the $\pm 1\sigma$ spread of the 100 replica variations.
The effect of the uncertainty in the strong coupling constant $\alphas$ was assessed by variations of $\pm 0.001$.
The resummation scale uncertainties were estimated using generator-level parameterisations derived from samples with \muQ varied by factors of 2 and 0.5 from its nominal value.
Similarly, the jet-to-parton matching uncertainties were estimated using generator-level parameterisations derived from samples with the \ckkw parameter set to \SI{15}{\GeV} and \SI{30}{\GeV}, compared to the nominal value of \SI{20}{\GeV}.
 
For \ttbar, uncertainties were considered for the choice of matrix element and parton shower generators, the choice of model for initial- and final-state radiation (ISR and FSR respectively), and the choice of PDFs.
The uncertainty due to ISR was estimated by simultaneously varying the \hdamp parameter and the \muR and \muF scales, and
propagating the \alphas uncertainties through the Var3c parameter of the A14 tune as described in Ref.~\cite{ATL-PHYS-PUB-2017-007}.
The impact of FSR was evaluated by varying the \muR scale for emissions from the parton shower by factors of 2 and 0.5.
The impact of using a different matrix element was evaluated by comparing the nominal \ttbar sample with an event sample produced using \MGNLO[2.6.0] instead of \POWHEGBOX[v2] but keeping the same parton shower model.
The impact of using a different parton shower and hadronisation model was evaluated by comparing the nominal \ttbar sample with an event sample which was interfaced with \HERWIG[7.04]~\cite{Bahr:2008pv,Bellm:2015jjp} instead of \PYTHIA[8] and used the H7UE set of tuned parameters~\cite{Bellm:2015jjp} and the \MMHT[lo] PDF set~\cite{Harland-Lang:2014zoa}.
 
The \NNPDF[3.0lo] replicas were used to evaluate the PDF uncertainties for the nominal PDF.
For both \Zjets and \ttbar, the central value of the PDF was additionally compared with the central values of the \CT[14nnlo]~\cite{Dulat:2015mca} and \MMHT[nnlo]~\cite{Harland-Lang:2014zoa} PDF sets.
 
Theory uncertainties in \Zjets and \ttbar predictions represent a sub-leading contribution, compared to signal theoretical uncertainties and experimental uncertainties (see \cref{tab:sig::1poi-breakdown}).
 
For renormalisation and factorisation scale variations and PDF uncertainties, their impact on the extrapolation factor between each SR and its corresponding \Zll control region, and on the shape of the \mmmc distribution, is treated as uncorrelated across the different categories.
This choice is driven by the structure of the statistical analysis, which employs a dedicated control region to constrain the \Zjets prediction in each signal region.
 
\subsection{Signal theoretical uncertainties}
Signal theoretical uncertainties are the dominant source of uncertainty for this analysis.
For each signal process, several sources of uncertainty are considered, including the uncertainty in the total inclusive cross-section (evaluated only for the $pp\to\Htautau$ cross-section measurement), the parton-shower and hadronisation model effect and the migration uncertainties among the STXS bins.
The migration uncertainties stem from the determination of the kinematic quantities used in the STXS framework as well the expected relative contribution of each process in the signal regions.
These uncertainties can affect signal acceptance in the various SRs as well as the \mmmc shape.
For all production modes, uncertainties are estimated for the PDF and \alphas, the parton shower and hadronisation model, and missing higher orders in the matrix element calculation.
PDF and \alphas uncertainties were estimated using the \PDFforLHC[15nlo] set of eigenvectors.
The impact of using a different parton shower and hadronisation model is evaluated by comparing the nominal sample with an event sample which was interfaced with \HERWIG[7] instead of \PYTHIA[8].
The effects on the signal expectations are treated as uncorrelated between the production modes, and the comparison leads to the largest uncertainty in the $pp\to\Htautau$ cross-section measurement.
Uncertainties from missing higher orders are calculated following the methodology outlined in Refs.~\cite{deFlorian:2016spz,bendavid2018les} and are determined as follows.
 
For the ggF process, 15 main sources of uncertainty were considered.
Four of these are jet-multiplicity-related uncertainties due to missing high-order corrections, and are estimated using the approach described in Refs.~\cite{deFlorian:2016spz,Stewart:2011cf}.
Three uncertainties parameterise the uncertainties in modelling the Higgs boson \pt and the 0-jet bin, one of which encapsulates the treatment of the top-quark mass in the loop corrections.
Three uncertainties take into account dijet mass migrations across the STXS bin boundaries.
Finally, three uncertainties are considered for the modelling of the ggF process in the VBF phase space.
Two of them are derived using the method described in Ref.~\cite{PhysRevD.87.093008}, from the study of the selection of exactly two or at least three jets.
The third one is derived from the comparison of the \POWHEG prediction with \MGNLO samples using the FxFx prescriptions~\cite{Frederix:2012ps} to merge the jet multiplicities and it also applies to the $VH$ phase space.
As the ggF process in the VBF phase space is difficult to model, the impact of increasing its contribution in the VBF\_1 category was estimated.
Doubling its contribution induced a \SI{7}{\percent} shift in the apparent VBF production cross-section.

For the VBF and $VH$ processes, ten uncertainties related to the STXS categorisation were considered: one related to the inclusive cross-section of the process, one related to the two-jet requirement, one related to the Higgs boson \pt selection at \SI{200}{\GeV}, one related to the \pt balance between the Higgs boson and the dijet system in events with two or three jets, and six uncertainties taking into account dijet mass migrations across the STXS bin boundaries.
 
For the \ttH process, six other uncertainties are included: one related to the inclusive cross-section of the process, and five migration uncertainties related to Higgs boson \pt boundaries in the STXS scheme.


\section{Statistical analysis}
\label{sec:stat}

\begin{figure}[h!]
\begin{center}
\includegraphics[width=0.9\linewidth]{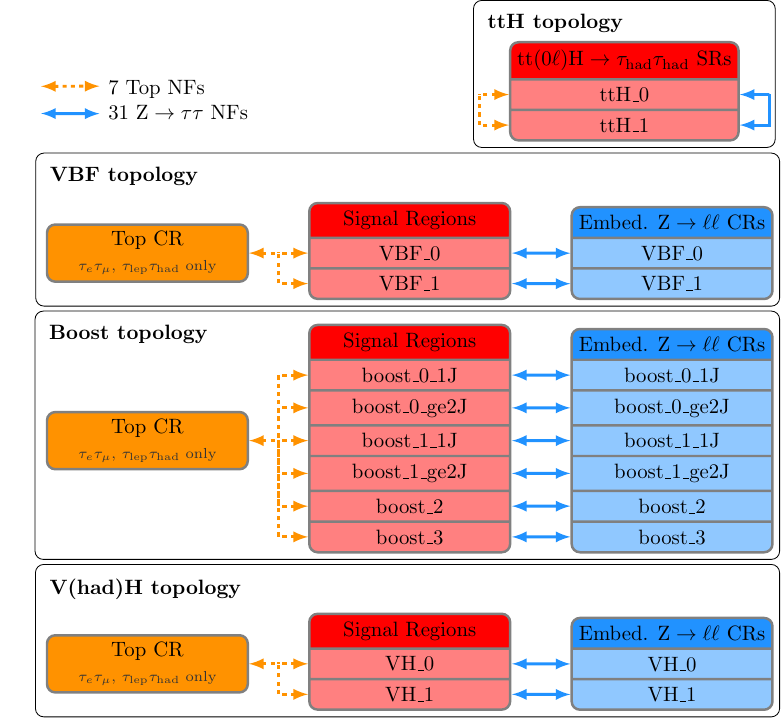}
\caption{
Graphical representation of the regions considered in the likelihood function and the normalisation factors (NFs) defined in the analysis.
The four unfilled black boxes represent the four main topologies targeted in this measurement.
Within each unfilled black box, the dark filled coloured boxes represent from left to right, the Top control regions, the signal regions and the $Z\to\ell\ell$  control regions.
When applicable the subcategories are represented by a light filled colour.
Each blue solid arrowed-line represents a normalisation factor that applies to the \Zttjets process in the signal regions and to the \Zlljets process in the $Z\to\ell\ell$  control regions.
Each orange dashed arrowed-line represents a normalisation factor that applies to the top processes in the signal regions and to the top processes in the Top control regions.
The arrow ends of each line indicate which regions are connected by each normalisation factor.
In the likelihood function, there are signal regions and $Z\to\ell\ell$  control region for each final state in the VBF, boost and V(had)H topologies.
Therefore, the ten signal regions and $Z\to\ell\ell$  control regions are repeated three times.
The Top control regions are only used in the $\tau_{e}\tau_{\mu}$  and $\tau_{\text{lep}}\tau_{\text{had}}$  final states.
Additionally, only one Top control region is considered by each topology.
}
\label{fig:fit_model}
\end{center}
\end{figure}

A statistical analysis of the collected data is performed to measure the $pp\to\Htautau$ cross-sections.
The procedure relies on a likelihood function constructed as the product of Poisson probability terms over the bins of the input distributions.
The uncertainties affecting the model (see \cref{sec:systematics})  are included in the likelihood function through nuisance parameters that are constrained by Gaussian probability terms that multiply the Poisson probability terms.
The parameters of interest (POIs) of the model are estimated by maximising the likelihood.
The likelihood function comprises 32 signal regions and 36 control regions.
In each signal region, Poisson terms describe the expected event counts in each bin of the \mmmc distribution, while in each control region a single Poisson term describes the total expected event yield in that region.
\Cref{fig:fit_model} illustrates the usage of the signal and control regions in the construction of the likelihood function.
The test statistic is constructed from the profile likelihood ratio and the confidence intervals on the parameters of interest are derived unsing the asymptotic approximation~\cite{Cowan:2010js}.
 
The normalisation of the \Ztt background is left as a freely floating parameter in the fit in several regions.
Each signal region in the boost, VBF and V(had)H categories is paired with an associated embedded \Zll control region and both share a common \Ztt normalisation factor.
Additionally, a common \Ztt normalisation factor is shared between the ttH\_0 and ttH\_1 signal regions.
In total, 31 floating normalisation factors are defined in order to constrain the yields of the \Ztt background in the signal regions.
The normalisation of the top processes is also allowed to float freely with six normalisation factors defined for boost, VBF, and V(had)H signal regions in the \temu and \tlhad channels separately and one for the ttH categories in the \thadhad channel.
The other backgrounds are normalised to their expected cross-section and the integrated luminosity of the recorded data.
 
In the signal regions, a smoothing procedure is applied to remove potentially large local fluctuations in the \mmmc templates caused by the limited size of the MC samples used to build the templates. The \mmmc template of uncertainties that are subject to large statistical fluctuations is smoothed, and uncertainties that have a negligible impact on the final results are pruned away sample-by-sample and region-by-region.
 
The \mmmc  discriminant distributions in each SR are binned in a way that maximises the significance of each targeted signal production mode, taking into account the full uncertainties.
Effectively, this leads to a fine binning near the resonant \Ztt peak with coarser binning further away from it.
 
Three different measurements are performed.
They include the branching ratio of $H\to\tau\tau$ and are performed with true Higgs boson rapidity $|y_H|<2.5$.
They differ in the definition of the POIs (see also \Cref{fig:stxs_sketch}):
\begin{enumerate}
\item \emph{$pp\to\Htautau$ cross-section}:
a single POI, corresponding to the $pp\to\Htautau$ cross-section, is estimated by the fit. In the likelihood function, the signal yields in each category are parameterised as the product of the $pp\to\Htautau$ cross-section, the integrated luminosity and the efficiency (including the acceptance of the ATLAS detector) of the selection for a SM Higgs boson with a mass of \SI{125.09}{\GeV}.
In this measurement, the relative contributions to the $pp\to\Htautau$ cross-section from the various production modes are fixed to the SM predictions.
 
\item \emph{Cross-sections per production mode}:
four POIs, corresponding to the cross-sections of the four dominant production modes (ggF, VBF, $VH$, \ttH) of the Higgs boson, are estimated by the fit. In this configuration, the event yields in the likelihood function are the sum of those from each individual production mode, parameterised as a function of the POI similarly to the way for the first measurement.
 
\item \emph{Reduced Simplified Template Cross-Sections}:
nine POIs, corresponding to the cross-sections of merged bins of the STXS \emph{stage 1.2} framework shown in \Cref{fig:signal:yields}, to which this analysis is sensitive, are determined by the fit.
The cross-sections for \ttH production and for $\text{VBF}+{qq}\to{V}(\to{qq})H$ production are measured.
The latter is measured for events with particle-level dijet mass between \SI{60}{\GeV} and \SI{120}{\GeV} or above \SI{350}{\GeV}.
In addition, the cross-section of ggF production is measured in six bins of the phase space.
One of them is a combination of two bins in the \emph{stage 1.2} prescription: events with one jet and intermediate \pTH (\SIrange{60}{120}{\GeV}) are measured together with events with two or more jets, low \mjj ($<\SI{350}{\GeV}$) and the same intermediate \pTH.
\end{enumerate}

\clearpage
\section{Results}
\label{sec:results}

\begin{figure}[h!]
\begin{center}
\begin{tabular}{ccc}
\includegraphics[width=0.32\linewidth]{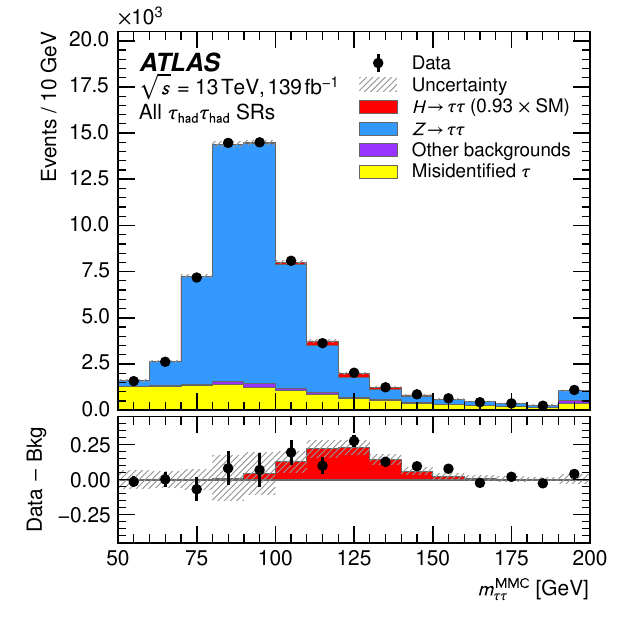} &
\includegraphics[width=0.32\linewidth]{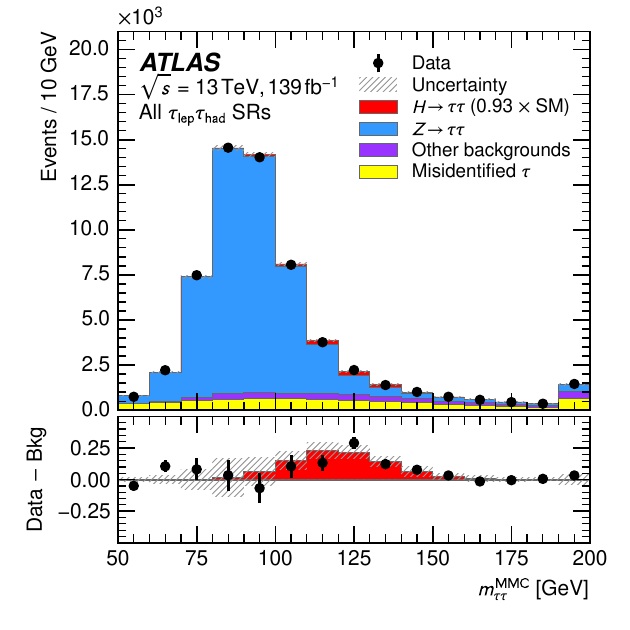} &
\includegraphics[width=0.32\linewidth]{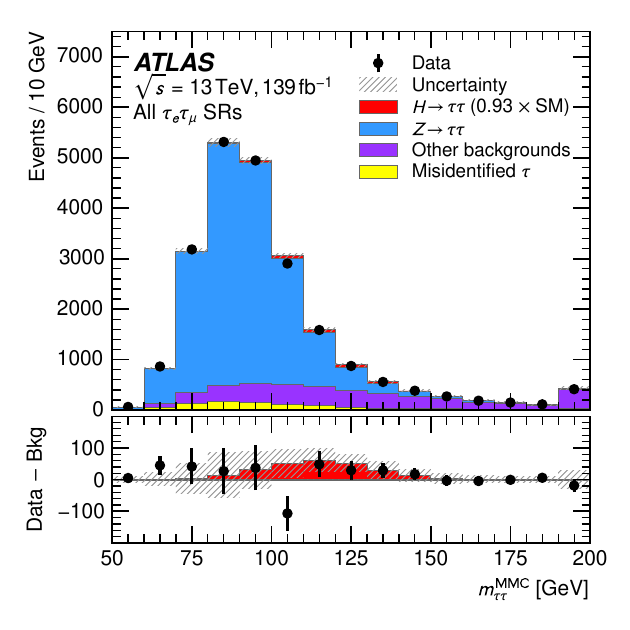} \\
(a) \thadhad & (b) \tlhad & (c) \temu \\
\end{tabular}
\caption{
Distribution of the reconstructed $\tau\tau$ invariant mass (\mmmc) for all events in the (a) $\tau_{\text{had}}\tau_{\text{had}}$, (b) $\tau_{\text{lep}}\tau_{\text{had}}$  and (c) $\tau_e\tau_\mu$ signal regions.
The bottom panel shows the differences between the numbers of observed data events and expected background events (black points).
The observed Higgs boson signal, corresponding to $(\sigma\times B)/(\sigma\times B)_{\text{SM}}\,=\,0.93$, is shown with a filled red histogram.
Entries with values above the $x$-axis range are shown in the last bin of each distributions.
The dashed band indicates the total uncertainty on the total predicted yields.
The prediction for each sample is determined from the likelihood fit performed to measure the $pp\to\Htautau$ cross-section.
\label{fig:result:all_categories}}
\end{center}
\end{figure}

\begin{figure}[h!]
\begin{center}
\begin{tabular}{ccc}
\includegraphics[width=0.32\linewidth]{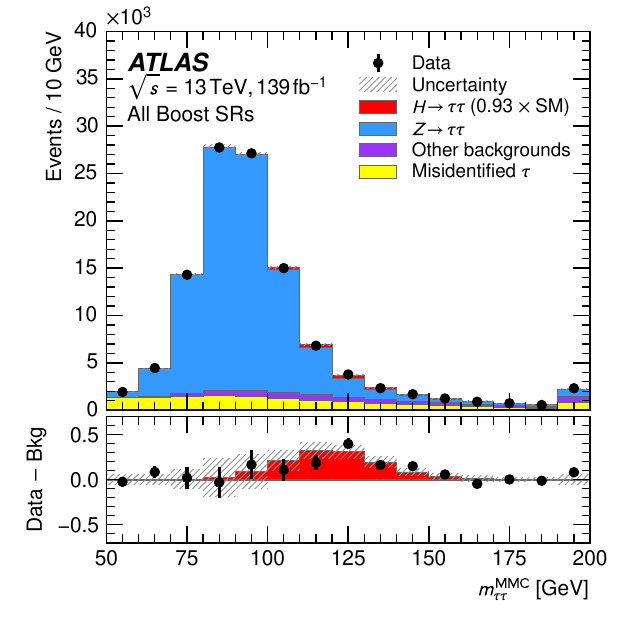} &
\includegraphics[width=0.32\linewidth]{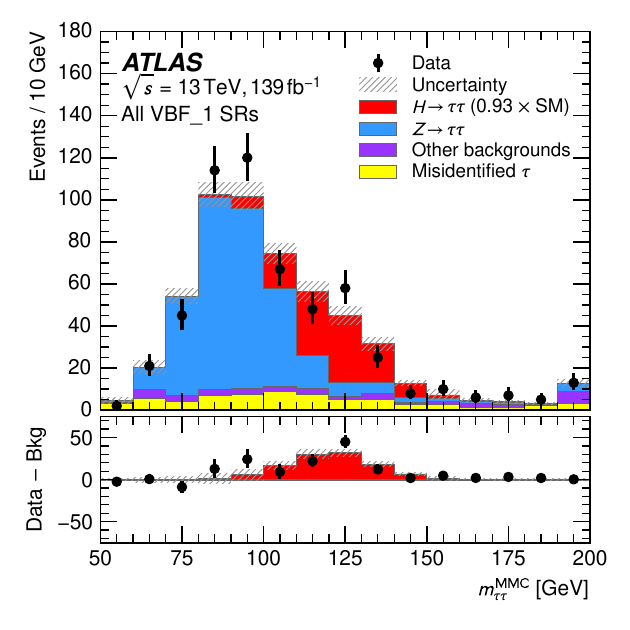} &
\includegraphics[width=0.32\linewidth]{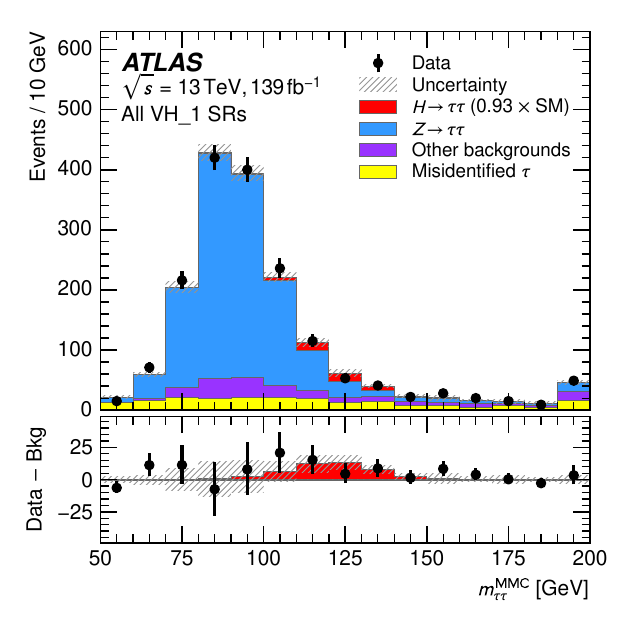} \\
(a) boost & (b) VBF\_1 & (c) VH\_1 \\
\end{tabular}
\caption{
Distribution of the reconstructed $\tau\tau$ invariant mass (\mmmc) for all events in the (a) boost, (b) VBF\_1 and (c) VH\_1 signal regions.
The bottom panel shows the differences between the numbers of observed data events and expected background events (black points).
The observed Higgs boson signal, corresponding to $(\sigma\times B)/(\sigma\times B)_{\text{SM}}\,=\,0.93$, is shown with a filled red histogram.
Entries with values above the $x$-axis range are shown in the last bin of each distributions.
The dashed band indicates the total uncertainty on the total predicted yields.
The prediction for each sample is determined from the likelihood fit performed to measure the $pp\to\Htautau$ cross-section.
\label{fig:result:all_channels_boost_vbf_vh}}
\end{center}
\end{figure}

\begin{figure}[h!]
\begin{center}
\begin{tabular}{ccc}
\includegraphics[width=0.32\linewidth]{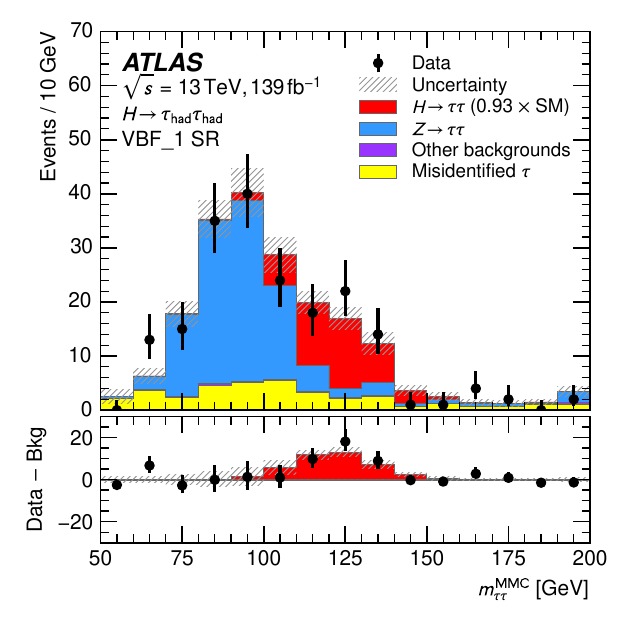} &
\includegraphics[width=0.32\linewidth]{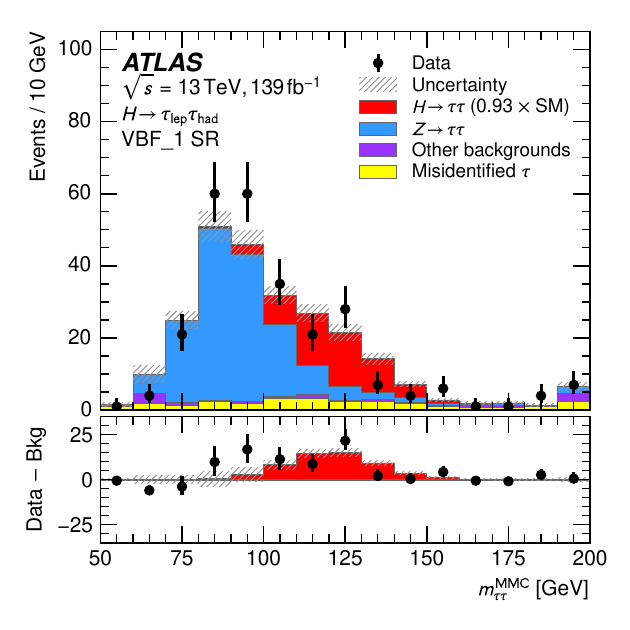} &
\includegraphics[width=0.32\linewidth]{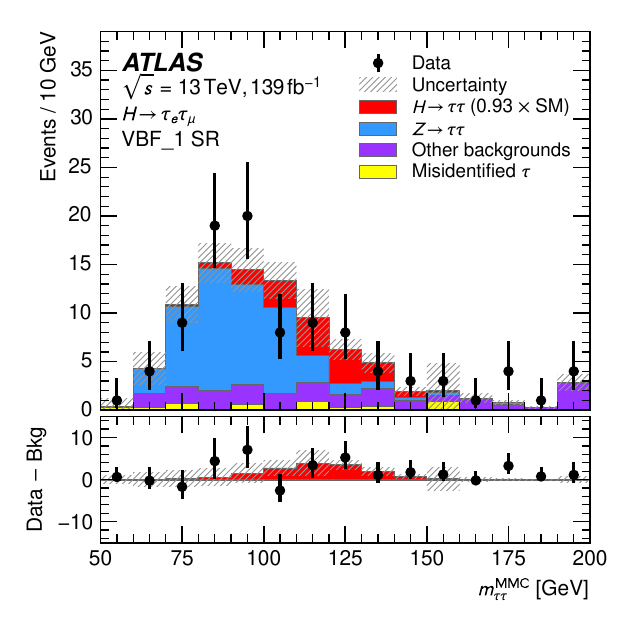} \\
(a) \thadhad & (b) \tlhad & (c) \temu \\
\end{tabular}
\caption{
Distribution of the reconstructed $\tau\tau$ invariant mass (\mmmc) for all events in the VBF\_1 categories of (a) \thadhad, (b) \tlhad and (c) \temu signal regions.
The bottom panel shows the differences between the numbers of observed data events and expected background events (black points).
The observed Higgs boson signal, corresponding to $(\sigma\times B)/(\sigma\times B)_{\text{SM}}\,=\,0.93$, is shown with a filled red histogram.
Entries with values above the $x$-axis range are shown in the last bin of each distributions.
The dashed band indicates the total uncertainty on the total predicted yields.
The prediction for each sample is determined from the likelihood fit performed to measure the $pp\to\Htautau$ cross-section.
\label{fig:result:vbf_1}}
\end{center}
\end{figure}

\begin{figure}[h!]
\begin{center}
\begin{tabular}{cc}
\includegraphics[width=0.47\linewidth]{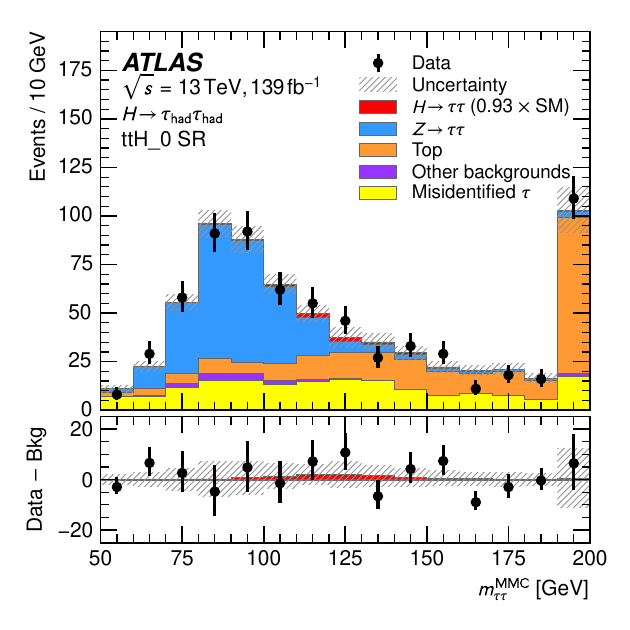} &
\includegraphics[width=0.47\linewidth]{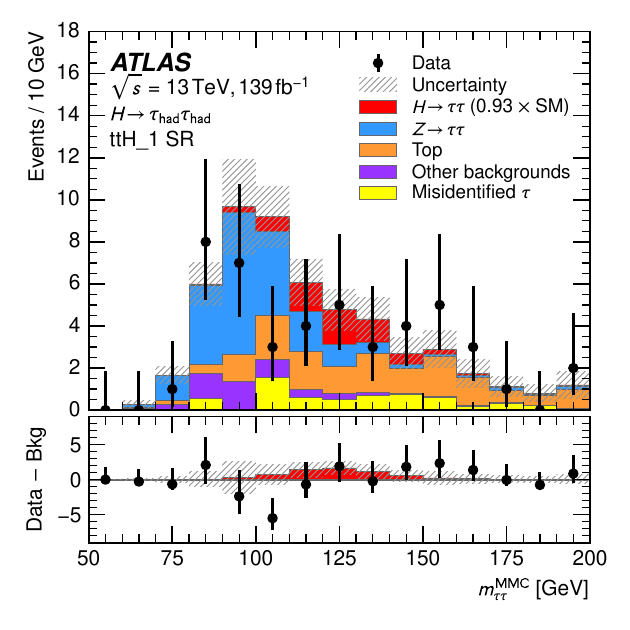} \\
(a) ttH\_0 & (b) ttH\_1 \\
\end{tabular}
\caption{
Distribution of the reconstructed $\tau\tau$ invariant mass (\mmmc) for all events in the (a) ttH\_0 and (b) ttH\_1 categories of the $\tau_{\text{had}}\tau_{\text{had}}$  channel.
The bottom panel shows the differences between the numbers of observed data events and expected background events (black points).
The observed Higgs boson signal, corresponding to $(\sigma\times B)/(\sigma\times B)_{\text{SM}}\,=\,0.93$, is shown with a filled red histogram.
Entries with values above the $x$-axis range are shown in the last bin of each distributions.
The dashed band indicates the total uncertainty on the total predicted yields.
The prediction for each sample is determined from the likelihood fit performed to measure the $pp\to\Htautau$ cross-section.
\label{fig:result:tth}
}
\end{center}
\end{figure}

The results of the statistical analysis (see \cref{sec:stat}) performed for the $pp\to\Htautau$ cross-section measurement are presented in \Cref{fig:result:all_categories,fig:result:all_channels_boost_vbf_vh,fig:result:vbf_1,fig:result:tth}.
Additional figures displaying the results of the total cross-section measurement with the binning used in the statistical analysis are available in the appendix.
The observed event yields and predictions as computed by the fit in the signal regions of the analysis are reported in \cref{tab:yield:vbfvh:hh,tab:yield:vbfvh:lh,tab:yield:vbfvh:emu,tab:yield:boost:hh,tab:yield:boost:lh,tab:yield:boost:emu}.
Excellent agreement is observed between the data and the expectations.
All measurements include the branching ratio of $H\to\tau\tau$ and are performed with true Higgs boson rapidity $|y_H|\,<\,2.5$.
 
The $pp\,\to\,\Htautau$ cross-section is measured to be \mbox{$2.94\pm0.21 \text{(stat)} ^{+\,0.37}_{-\,0.32} \text{(syst)}$\,pb}, in agreement with the SM predictions (\mbox{$3.17\pm0.09$\,pb}) with a $p$-value of $0.58$.
 
\begin{figure}[h!]
\begin{center}
\begin{tabular}{cc}
\includegraphics[width=0.49\linewidth]{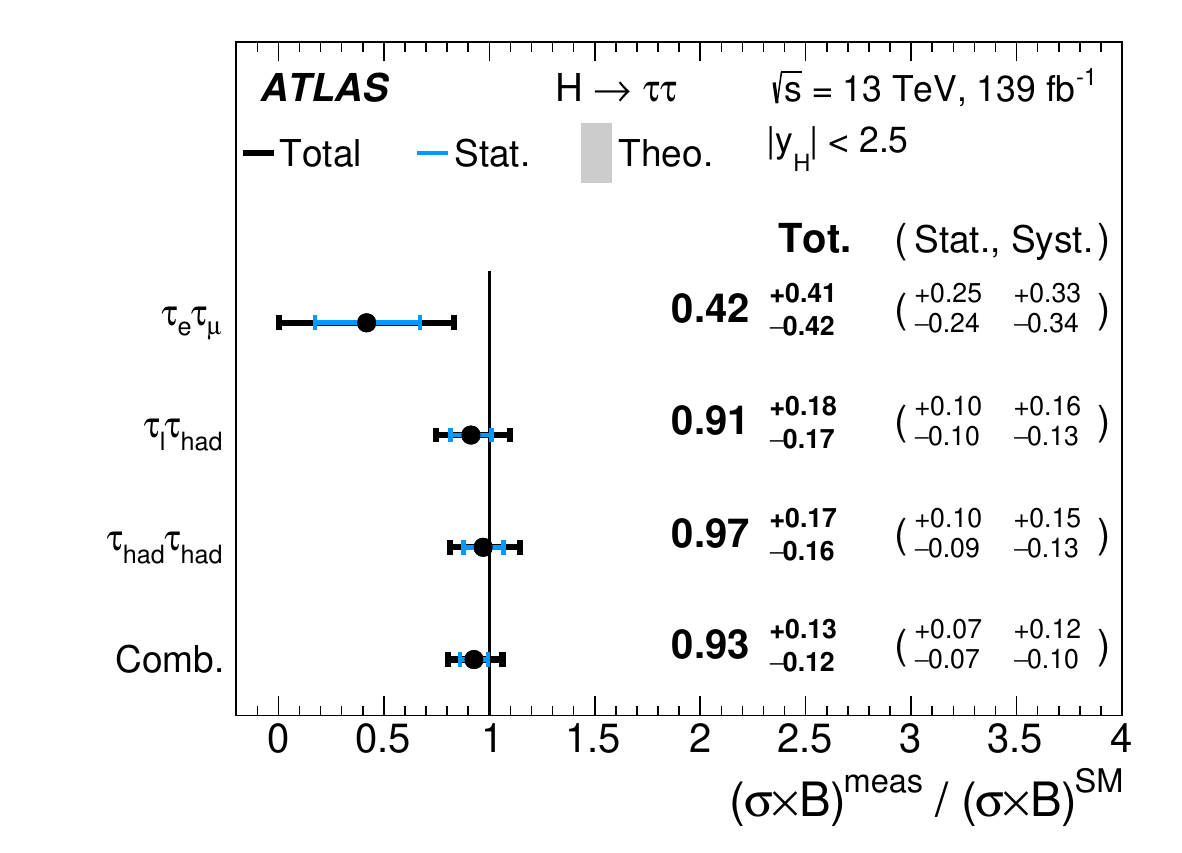} &
\includegraphics[width=0.49\linewidth]{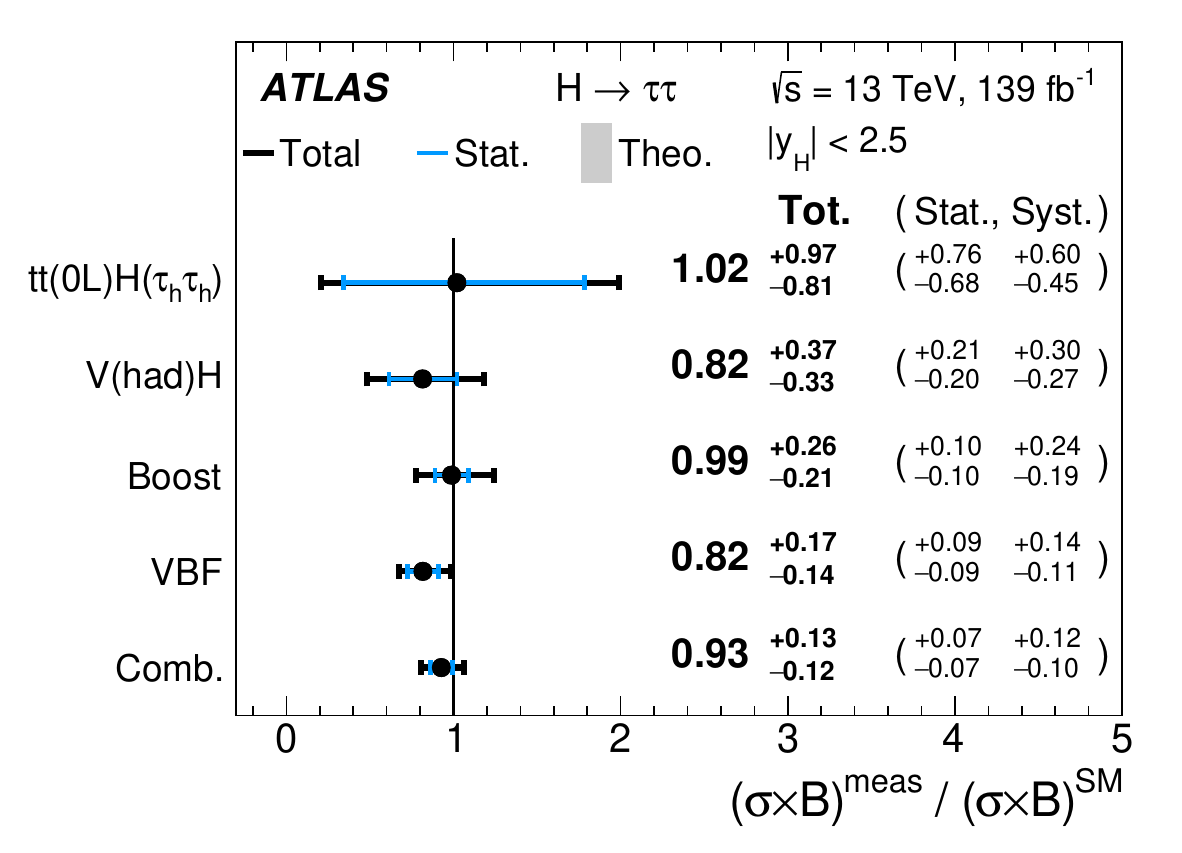} \\
(a) & (b) \\
\end{tabular}
\caption{
The measured values for $\sigma_{H}\times B(\Htautau)$ relative to the SM  expectations when only the data of (a)~individual channels or (b)~individual categories are used.
The total $\pm1\sigma$ uncertainty in the measurement is indicated by the black error bars, with the individual contribution from the statistical uncertainty in blue.
The results have been extracted performing a fit for the inclusive cross-section measurement.}
\label{fig:res:split}
\end{center}
\end{figure}

The measurement is also performed in the \thadhad, \tlhad and \temu final states separately and in the boost, VBF, V(had)H and tt(0$\ell$)\Hthadthad categories.
The results are illustrated in \Cref{fig:res:split}.
The $p$-values for the compatibility of the measurements are $0.30$ across $\tau$-lepton decay modes and $0.72$ across kinematic categories.
 
\begin{figure}[h!]
\begin{center}
\begin{tabular}{cc}
\includegraphics[width=0.49\linewidth]{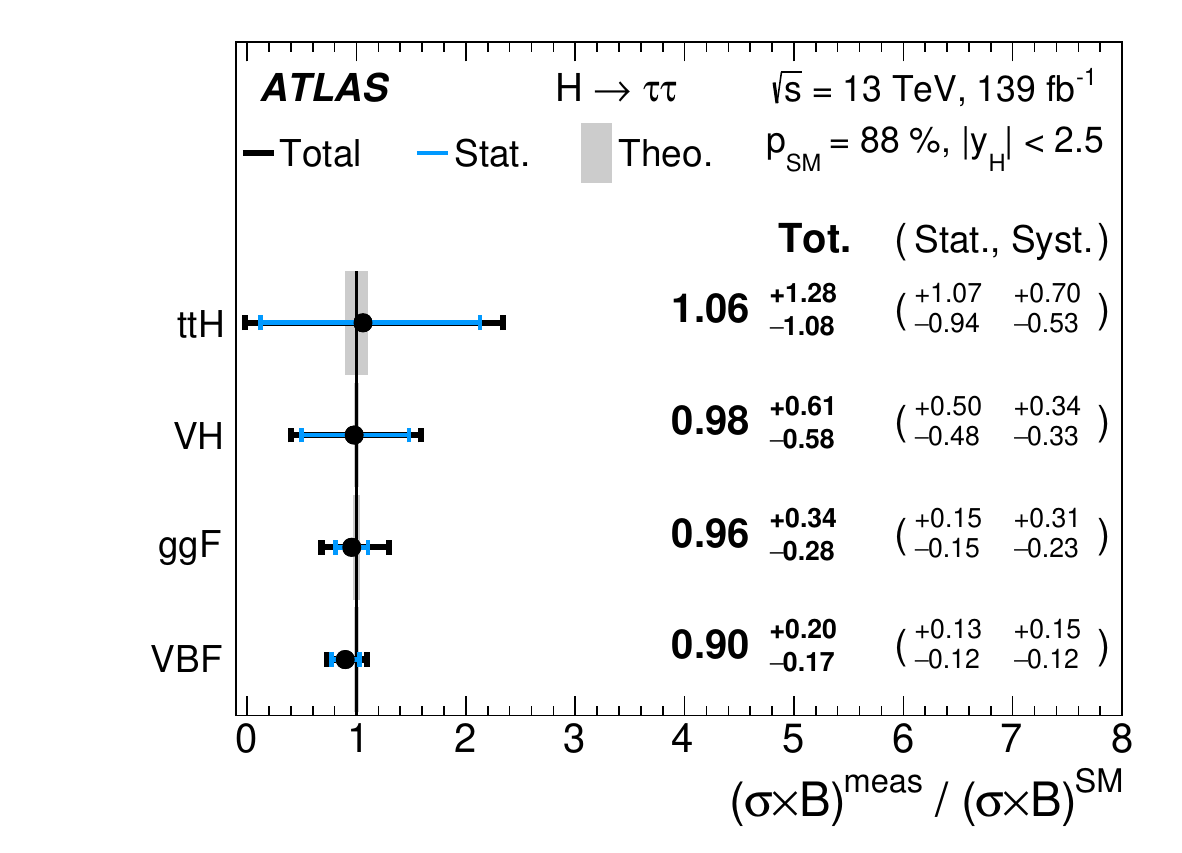} &
\includegraphics[width=0.49\linewidth]{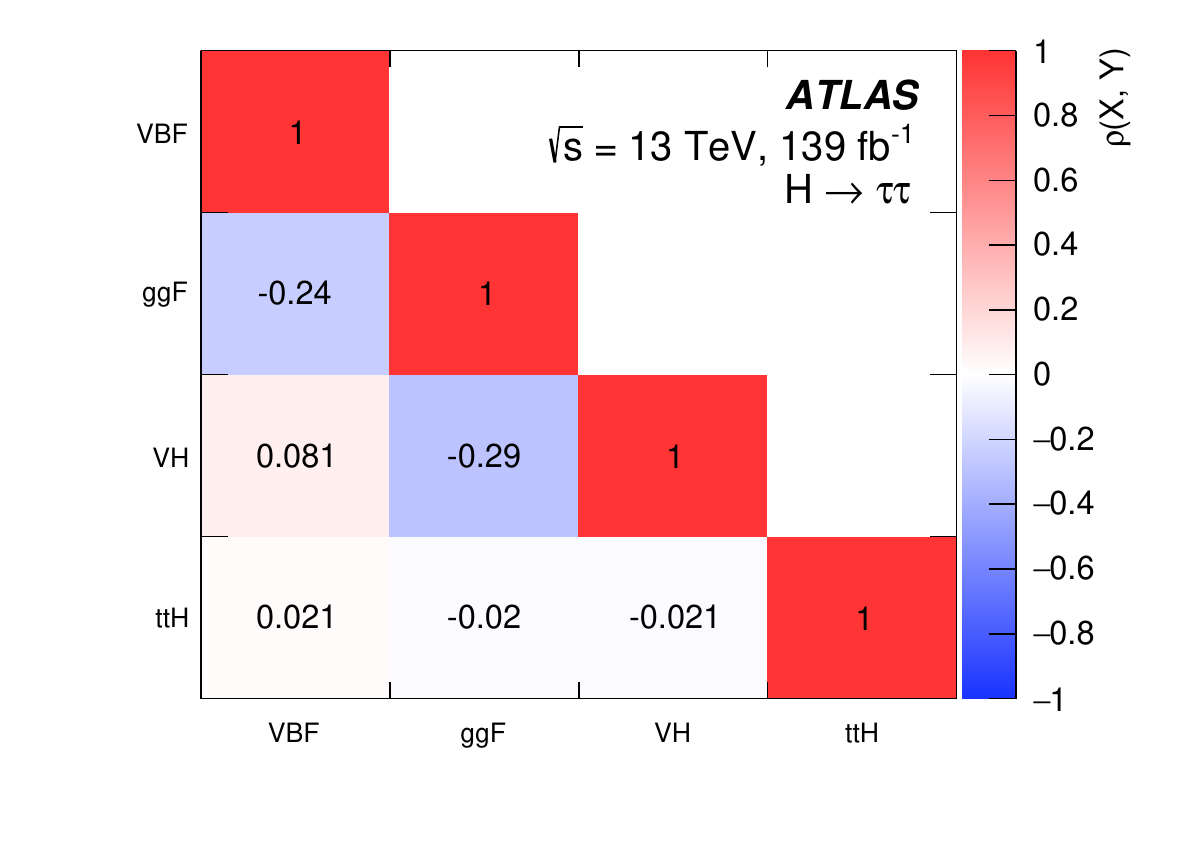} \\
(a) & (b) \\
\end{tabular}
\caption{
(a)~The measured values for $\sigma_{H}\times B(\Htautau)$ relative to the SM  expectations in the four dominant production modes.
The total $\pm1\sigma$ uncertainty in the measurement is indicated by the black error bars, with the individual contribution from the statistical uncertainty in blue.
(b)~The measured correlations between each parameter of interest in the measurement of the cross-sections per production mode.
The results have been extracted performing a fit for the production mode cross-section measurements. The measured values for $\sigma_{H}\times B(\Htautau)$ along with the corresponding correlation matrix are available in the HEPData repository~\cite{hepdata}. }
\label{fig:res:4poi}
\end{center}
\end{figure}

The same dataset is subsequently used to measure the production cross-section for the Higgs boson in the four dominant production mechanisms.
The results are illustrated in \Cref{fig:res:4poi}(a) and reported in \cref{tab:res:4poi} with a breakdown of the uncertainties.
They are all consistent with the SM predictions, with a $p$-value of $0.98$.
The measurement establishes the observation of the VBF production of the Higgs boson in the $\tau\tau$ decay channel with an observed (expected) significance of $5.3\sigma$~($6.2\sigma$).
 
The VBF production cross-section measurement is the most precise of the four dominant production mechanisms.
The theoretical uncertainties in VBF production are smaller than in the other channels, and the VBF\_1 categories represent the best combination of high signal yields and purity in this measurement.
The measured VBF cross-section is \mbox{$0.197\pm0.028 \text{(stat)} ^{+\,0.032}_{-\,0.026} \text{(syst)}$\,pb}.
The second most precisely measured cross-section is that of ggF, \mbox{$2.7\pm0.4 \text{(stat)} ^{+\,0.9}_{-\,0.6} \text{(syst)}$\,pb}, corresponding to an observed (expected) significance of $3.9\sigma$~($4.6\sigma$).
The $VH$ and \ttH production modes are determined with lower precision.
The measured $VH$ cross-section is \mbox{$0.12\pm0.06 \text{(stat)}\pm0.04 \text{(syst)}$\,pb}, while the \ttH cross-section is \mbox{$0.033 ^{+\,0.033}_{-\,0.029} \text{(stat)} ^{+\,0.022}_{-\,0.017} \text{(syst)}$\,pb}.
\Cref{fig:res:4poi}(b) illustrates the observed correlation between the measured cross-section parameters in the fit.
The ggF cross-section exhibits an anti-correlation of \SI{24}{\percent} and \SI{29}{\percent} with the VBF and $VH$ cross-sections respectively.
This is caused by a significant contribution of ggF events to the VBF\_0, VH\_0 and VH\_1 categories as illustrated by \Cref{fig:signal:yields}.
The simultaneous measurement of the cross-sections of the four dominant production modes is compatible with the SM expectations, with a $p$-value of $0.88$.
 
\begin{figure}[h!]
\begin{center}
\begin{tabular}{c}
\includegraphics[width=0.7\linewidth]{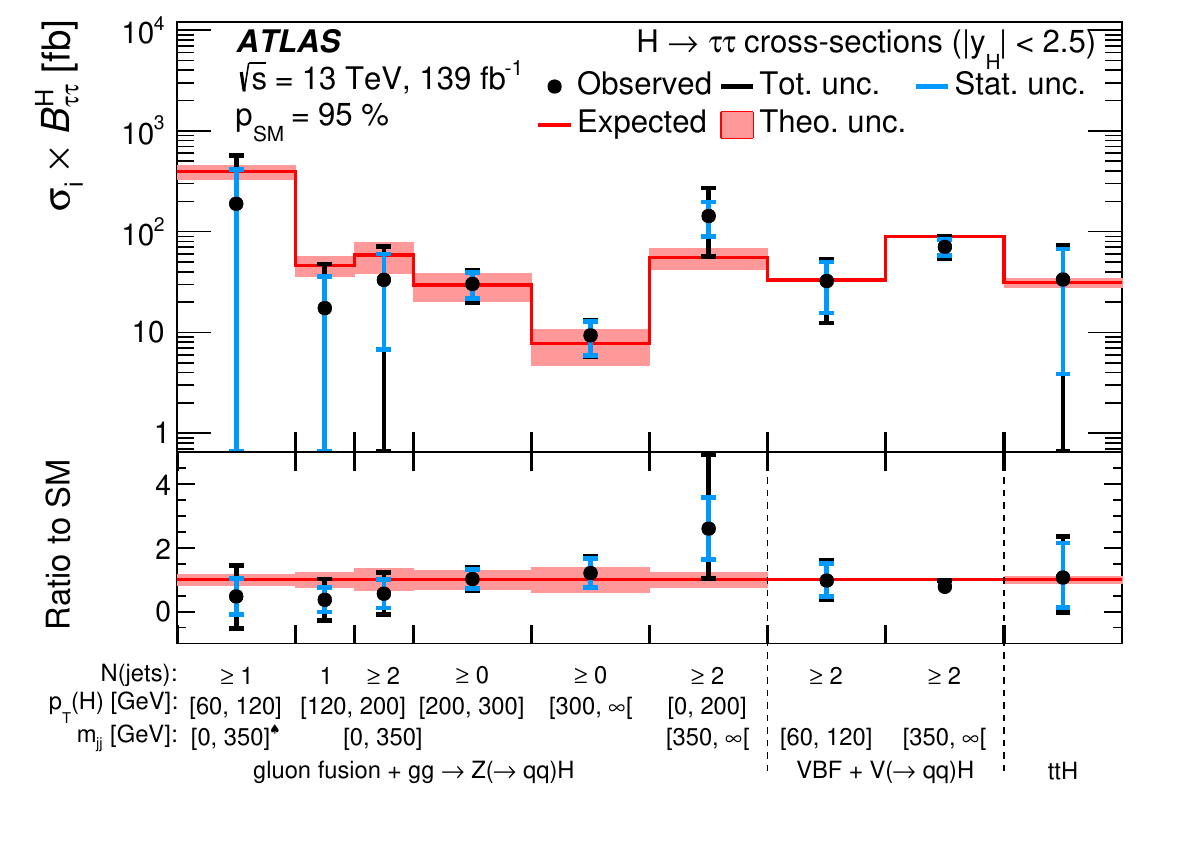} \\
(a) \\
\includegraphics[width=0.7\linewidth]{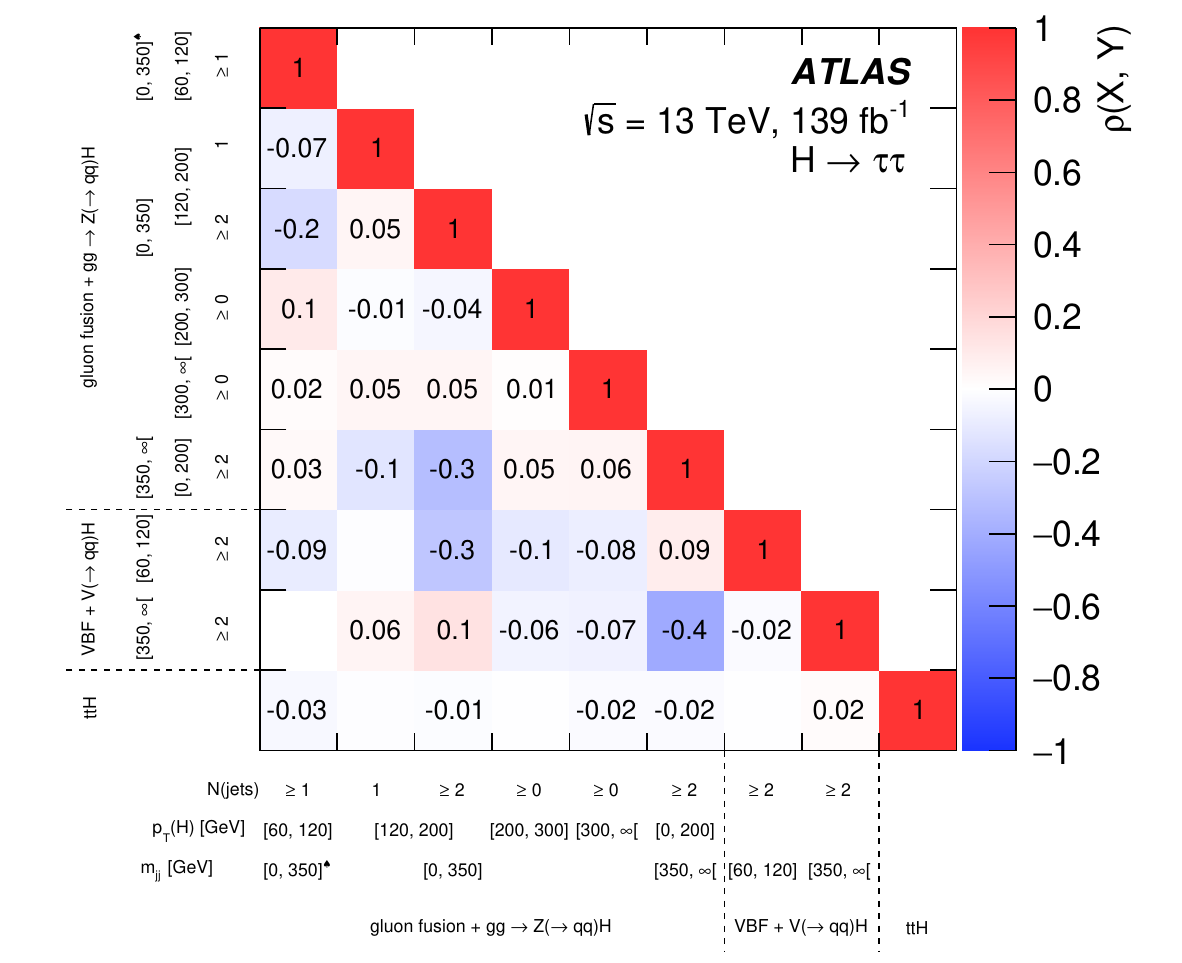} \\
(b) \\
\end{tabular}
\caption{
(a)~The measured values for $\sigma_{H}\times B(\Htautau)$ relative to the SM  expectations in the nine fiducial volumes defined in the STXS  measurement.
Also shown is the result from the combined fit.
The total $\pm1\sigma$ uncertainty in the measurement is indicated by the black error bars, with the individual contribution from the statistical uncertainty in blue.
(b)~The measured correlations between each pair of parameters of interest in the STXS measurement.
The spades symbol ($\spadesuit$) indicates that the criteria for \mjj only apply to events with at least two reconstructed jets.
The measured values for $\sigma_{H}\times B(\Htautau)$ along with the corresponding correlation matrix are available in the HEPData repository~\cite{hepdata}. }
\label{fig:res:9poi}
\end{center}
\end{figure}

Finally, the $pp\to\Htautau$ cross-sections are measured as a function of \pTH, \njetsthirty and \mjj in a reduced set of the bins of the \emph{stage 1.2} of the STXS framework.
The results, illustrated in \Cref{fig:res:9poi}(a), are reported in \cref{tab:res:9pois}.
They are in very good agreement with the SM expectations.
The gluon--gluon fusion + $gg\to{Z}(\to{qq})H$ production mode is measured in four \pTH intervals starting at \SI{60}{\GeV}.
For \pTH values between \SI{120}{\GeV} and \SI{200}{\GeV}, the measurements are further separated depending on the number of jets in the event.
The best precision is obtained in the \pTH interval between \SI{200}{\GeV} and \SI{300}{\GeV} and in the \pTH regime above \SI{300}{\GeV}.
The cross-sections are determined with an uncertainty of \SI{37}{\percent} and \SI{42}{\percent} respectively.
 
The EW production mode includes the VBF and $qq\to{V}(\to{qq})H$ processes and is measured in \mjj intervals.
In the interval with \mjj between \SI{60}{\GeV} and \SI{120}{\GeV}, the measurement has an uncertainty of \SI{63}{\percent}.
The EW production mode for events with \mjj greater than \SI{120}{\GeV} is measured with an uncertainty of \SI{26}{\percent} and is the most precise cross-section determined within the simplified template cross-section framework in this paper.
It exhibits an anti-correlation of approximately 40$\%$ with the cross-section for gluon--gluon fusion events produced in the same interval ($\mjj>\SI{350}{\GeV}$) as illustrated on \Cref{fig:res:9poi}(b).

\clearpage
\begin{table}[htbp]
\begin{center}
\caption{
Observed event yields and predictions as computed by the fit in the VBF, V(had)H and tt(0$\ell$)$H\to\tau_{\text{had}}\tau_{\text{had}}$ signal regions of the $\tau_{\text{had}}\tau_{\text{had}}$  channel.
In the VBF and V(had)H categories, the top processes are estimated with the other backgrounds (diboson, \HWW) by the fit.
Uncertainties include statistical and systematic components.
The prediction for each sample is determined from the likelihood fit performed to measure the $pp\to\Htautau$ cross-section.
\label{tab:yield:vbfvh:hh}
}
\resizebox{\textwidth}{!}{
\begin{tabular}{
@{}
l
S[table-format=4.3]@{$\,\pm\,$}S[table-format=2.3]
S[table-format=3.2]@{$\,\pm\,$}S[table-format=2.2]
S[table-format=5.2]@{$\,\pm\,$}S[table-format=3.2]
S[table-format=3.3]@{$\,\pm\,$}S[table-format=2.3]
S[table-format=3.3]@{$\,\pm\,$}S[table-format=2.3]
S[table-format=2.3]@{$\,\pm\,$}S[table-format=1.3]
@{}
}
 
\toprule
& \multicolumn{4}{S}{{VBF \thadhad}} &  \multicolumn{4}{S}{{V(had)H \thadhad}} &  \multicolumn{4}{S}{{tt(0$\ell$)\Hthadthad}} \\
\cmidrule(lr){2-5}
\cmidrule(lr){6-9}
\cmidrule(l){10-13}
& \multicolumn{2}{c}{VBF\_0} & \multicolumn{2}{c}{VBF\_1} & \multicolumn{2}{c}{VH\_0} & \multicolumn{2}{c}{VH\_1} & \multicolumn{2}{c}{ttH\_0} & \multicolumn{2}{c}{ttH\_1}  \\
\midrule
\Ztautau                   & 2051 & 50 & 115  & 11 & 4636 & 84  & 539 & 24 & 265 & 24 & 20 & 4 \\
Fake                       & 1027 & 68 & 39.6 & 5.3 & 1627 & 110 & 112 & 10 & 182 & 17 & 6.5 & 1.7 \\
Top                        & \multicolumn{2}{c}{}& \multicolumn{2}{c}{} & \multicolumn{2}{c}{} &\multicolumn{2}{c}{} & 239 & 26 & 15.2 & 3.4 \\
Other backgrounds          & 57.1 & 5.2 & 1.7 & 0.6   & 209  & 12  & 43.1 & 2.9 & 15.7 & 2.1 & 5.0 & 0.8 \\
\midrule
ggF, \Htautau              & 38.5 & 9.5 & 3.2 & 1.8   & 72   & 14  & 6.8 & 1.7 & 1.96 & 0.41 & 0.42 & 0.08 \\
VBF, \Htautau              & 72   & 10  & 40  & 5     & 8.1  & 1.3 & 0.5 & 0.1 & 0.23 & 0.03 & \multicolumn{2}{c}{$<0.01$}      \\
$WH$, \Htautau             & 1.00 & 0.14 & \multicolumn{2}{c}{$<0.01$} & 15.2 & 2.5  & 9.8  & 1.5 & 0.24 & 0.03 & 0.034 & 0.005      \\
$ZH$, \Htautau             & 0.8  & 0.1  & \multicolumn{2}{c}{$<0.01$} & 12.4 & 2.3  & 5.4  & 1.1 & 0.69 & 0.15 & 0.15 & 0.02       \\
$t\overline{t}H$, \Htautau & 0.19 & 0.03 & \multicolumn{2}{c}{$<0.01$} & 0.52 & 0.07 & 0.22 & 0.03 & 7.5 & 1.6 & 5.4 & 1.3       \\
$tH$, \Htautau             & 0.41 & 0.06 & \multicolumn{2}{c}{$<0.01$} & 0.25 & 0.03 & 0.07 & 0.01 & 0.92 & 0.13 & 0.41 & 0.06 \\
$bbH$, \Htautau            & 0.10 & 0.02 &   \multicolumn{2}{c}{$<0.01$} & 0.14 & 0.02 & 0.015 & 0.002 & 0.27 & 0.04 & 0.09 & 0.02 \\
\midrule
Total background           & 3135 & 84 & 156  & 12  & 6472 & 136 & 694 & 26 & 703 & 33 & 46.6 & 5.3 \\
Total signal               & 113 & 15  & 43.6 & 5.2 & 109 & 16 & 23.0 & 3.2 & 12 & 2 & 6.6 & 1.4 \\
\midrule
Total                      & 3248 & 84 & 200 & 12 & 6581 & 135 & 717 & 26 & 715 & 33 & 53.3 & 5.5     \\

\midrule
Data                            & \multicolumn{1}{l}{$3318\phantom{.000}$} & &  \multicolumn{1}{l}{$197\phantom{.00}$} & & \multicolumn{1}{l}{$\quad 6532\phantom{.000}$} & & \multicolumn{1}{l}{$720\phantom{.000}$} &  & \multicolumn{1}{l}{$727\phantom{.000}$} & &  \multicolumn{1}{l}{$49\phantom{.000}$} &                  \\
\end{tabular}}
\end{center}
\end{table}

\begin{table}[htbp]
\begin{center}
\caption{
Observed event yields and predictions as computed by the fit in the VBF and V(had)H signal regions of the $\tau_{\text{lep}}\tau_{\text{had}}$  channel.
Uncertainties include statistical and systematic components.
The prediction for each sample is determined from the likelihood fit performed to measure the $pp\to\Htautau$ cross-section.
\label{tab:yield:vbfvh:lh}
}
\resizebox{0.85\textwidth}{!}{
\begin{tabular}{
@{}
l
S[table-format=4.3]@{$\,\pm\,$}S[table-format=2.3]
S[table-format=3.4]@{$\,\pm\,$}S[table-format=2.4]
S[table-format=4.2]@{$\,\pm\,$}S[table-format=3.2]
S[table-format=3.3]@{$\,\pm\,$}S[table-format=2.3]
@{}
}
 
\toprule
& \multicolumn{4}{S}{{VBF \tlhad}} &  \multicolumn{4}{S}{{V(had)H \tlhad}}\\
\cmidrule(lr){2-5}
\cmidrule(l){6-9}
& \multicolumn{2}{c}{VBF\_0} & \multicolumn{2}{c}{VBF\_1} & \multicolumn{2}{c}{VH\_0} & \multicolumn{2}{c}{VH\_1} \\ \midrule
\Ztautau                   & 2362 & 59 & 162 & 12 & 6724 & 112 & 535 & 23   \\
Fake                       & 611 & 49 & 30 & 3 & 1315 & 126 & 80 & 8   \\
Top                        & 107 & 12 & 5.3 & 1.5 & 243 & 25 & 27 & 5   \\
Other backgrounds          & 139 & 17 & 5.8 & 2.4 & 396 & 39 & 50 & 4 \\
\midrule
ggF, \Htautau              & 71 & 28 & 3.5 & 1.2 & 87.3 & 20.3 & 5.1 & 2.2  \\
VBF, \Htautau              & 84.4 & 11.2 & 52 & 6 & 9.3 & 1.6 & 0.5 & 0.2 \\
$WH$, \Htautau             & 0.83 & 0.11 & 0.011 & 0.002 & 17.4 & 2.6 & 8.1 & 1.2   \\
$ZH$, \Htautau             & 0.86 & 0.12 & \multicolumn{2}{c}{$<0.01$} & 13.3 & 2.5 & 5.0 & 0.9   \\
$t\overline{t}H$, \Htautau & 0.10 & 0.01 & \multicolumn{2}{c}{$<0.01$} & 0.35 & 0.05 & 0.13 & 0.02 \\
$tH$, \Htautau             & 0.26 & 0.04 & 0.023 & 0.003 & 0.17 & 0.03 & 0.030 & 0.004 \\
$bbH$, \Htautau            & 0.09 & 0.01 & \multicolumn{2}{c}{} & 0.16 & 0.02 & \multicolumn{2}{c}{} \\
\midrule
Total background           & 3219 & 75 & 203 & 13 & 8678 & 143 & 692 & 24    \\
Total signal               & 158 & 30 & 56 & 7 & 128 & 23 & 19 & 3   \\
\midrule

Total                      & 3377 & 76 & 259 & 13 & 8806 & 143 & 711 & 24     \\
\midrule
Data                        & \multicolumn{1}{l}{$3402\phantom{.000}$} && \multicolumn{1}{l}{$267\phantom{.0000}$} && \multicolumn{1}{l}{$8780\phantom{.00}$} && \multicolumn{1}{l}{$743\phantom{.000}$}         &      \\
\bottomrule
\end{tabular}
}
\end{center}
\end{table}
 
\begin{table}[htbp]
\begin{center}
\caption{
Observed event yields and predictions as computed by the fit in the VBF and V(had)H signal regions of the $\tau_{e}\tau_{\mu}$ channel.
Uncertainties include statistical and systematic components.
The prediction for each sample is determined from the likelihood fit performed to measure the $pp\to\Htautau$ cross-section.
\label{tab:yield:vbfvh:emu}
}
\resizebox{0.85\textwidth}{!}{
\begin{tabular}{
@{}
l
S[table-format=4.3]@{$\,\pm\,$}S[table-format=2.3]
S[table-format=3.4]@{$\,\pm\,$}S[table-format=2.4]
S[table-format=4.3]@{$\,\pm\,$}S[table-format=2.3]
S[table-format=3.4]@{$\,\pm\,$}S[table-format=2.4]
@{}
}
 
\toprule
& \multicolumn{4}{S}{{VBF \temu}} &  \multicolumn{4}{S}{{V(had)H \temu}}\\
\cmidrule(lr){2-5}
\cmidrule(l){6-9}
& \multicolumn{2}{c}{VBF\_0} & \multicolumn{2}{c}{VBF\_1} & \multicolumn{2}{c}{VH\_0} & \multicolumn{2}{c}{VH\_1} \\
\midrule
\Ztautau                   & 820   & 29 & 49.3 & 6.5 & 2424 & 58 & 186 & 13    \\
Fake                       & 90    & 21 & 3.3  & 5.3 & 214 & 42 & 33 & 14  \\
Top                        & 165   & 15 & 9    & 2   & 346 & 29 & 33 & 5  \\
Other backgrounds          & 96.1  & 8.5 & 11.9 & 1.6 & 259 & 23 & 28 & 2   \\
\midrule
ggF, \Htautau              & 12.7  & 3.2 & 1.05 & 0.31 & 26 & 5 & 1.8 & 0.6   \\
VBF, \Htautau              & 22    & 3   & 14.3 & 1.8 & 2.41 & 0.43 & 0.13 & 0.02 \\
$WH$, \Htautau             & 0.21  & 0.03 & \multicolumn{2}{c}{} & 4.3 & 0.7 & 2.5 & 0.4  \\
$ZH$, \Htautau             & 0.14  & 0.02 & \multicolumn{2}{c}{} & 3.6 & 0.8 & 1.5 & 0.3  \\
$t\overline{t}H$, \Htautau & \multicolumn{2}{c}{$<0.01$} & \multicolumn{2}{c}{$<0.01$} & 0.08 & 0.01 & 0.021 & 0.003  \\
$tH$, \Htautau             & 0.13  & 0.02 & 0.014 & 0.002 & 0.038 & 0.005 & 0.018 & 0.002 \\
$bbH$, \Htautau            & 0.026 & 0.004 & \multicolumn{2}{c}{} & 0.046 & 0.006 & \multicolumn{2}{c}{} \\
\midrule
Total background           & 1171  & 35  & 73.5 & 9.4 & 3243 & 63 & 281 & 18    \\
Total signal               & 35.4  & 3.7 & 15.3 & 1.2 & 36.5 & 5.2 & 6 & 1  \\
\midrule
Total                      & 1206  & 35 & 88.8 & 8.6 & 3280 & 63 & 287 & 18    \\

\midrule
Data                       & \multicolumn{1}{l}{$1215\phantom{.000}$} && \multicolumn{1}{l}{$\quad 98\phantom{.00}$} && \multicolumn{1}{l}{$3277\phantom{.000}$} && \multicolumn{1}{l}{$286\phantom{.0000}$} &       \\
\bottomrule
\end{tabular}
} 
\end{center}
\end{table}

\begin{table}[htbp]
\begin{center}
\caption{
Observed event yields and predictions as computed by the fit in the boost signal regions of the $\tau_{\text{had}}\tau_{\text{had}}$  channel.
Uncertainties include statistical and systematic components.
The prediction for each sample is determined from the likelihood fit performed to measure the $pp\to\Htautau$ cross-section.
\label{tab:yield:boost:hh}
}
\resizebox{\textwidth}{!}{
\begin{tabular}{
@{}
l
S[table-format=6.3]@{$\,\pm\,$}S[table-format=4.3]
S[table-format=5.2]@{$\,\pm\,$}S[table-format=3.2]
S[table-format=6.3]@{$\,\pm\,$}S[table-format=4.3]
S[table-format=5.2]@{$\,\pm\,$}S[table-format=3.2]
S[table-format=5.2]@{$\,\pm\,$}S[table-format=3.2]
S[table-format=5.2]@{$\,\pm\,$}S[table-format=3.2]
@{}
}
 
\toprule
&  \multicolumn{12}{S}{{Boost \thadhad}}\\
\cmidrule(l){2-13}
\pTH[\gev]              & \multicolumn{4}{S}{{$[100, 120]$}} & \multicolumn{4}{S}{{$[120, 200]$}} & \multicolumn{2}{S}{{$[200, 300]$}} & \multicolumn{2}{S}{{$\left[300, \infty\right[$}} \\
\cmidrule(lr){2-5}\cmidrule(lr){6-9}
\njetsthirty       & \multicolumn{2}{S}{{$=1$}} & \multicolumn{2}{S}{{$\geq\,2$}} & \multicolumn{2}{S}{{$=1$}} & \multicolumn{2}{S}{{$\geq\,2$}} & \multicolumn{2}{S}{{$\geq\,1$}} & \multicolumn{2}{S}{{$\geq\,1$}} \\
\midrule
\Ztautau           &  5635 & 115 & 2640 & 67 & 11863 & 134 & 10076 & 125 & 7252 & 93 & 973 & 30    \\
Fake              &  3388 & 224 & 1729 & 118 & 2312 & 155 & 2072 & 140 & 293 & 32 & 54 & 20  \\
Other backgrounds &  61 & 9 & 74.2 & 11.3 & 116 & 19 & 251 & 14 & 157 & 10 & 53.6 & 5.5  \\
\midrule
ggF, \Htautau     & 54.4 & 9.7 & 23.1 & 4.1 & 112.8 & 20.5 & 109 & 21 & 96.2 & 17.2 & 30 & 7  \\
VBF, \Htautau     & 11.3 & 2.0 & 5.8 & 0.9 & 27.6 & 4.7 & 24.6 & 4.2 & 23.7 & 3.6 & 7.3 & 1.1   \\
$WH$, \Htautau    &  2.1 & 0.6 & 1.5 & 0.3 & 3.8 & 1.1 & 7.0 & 1.1 & 4.6 & 0.7 & 2.5 & 0.7 \\
$ZH$, \Htautau    & 1.4 & 0.3 & 1.1 & 0.3 & 2.7 & 0.9 & 5.3 & 1.0 & 3.7 & 0.5 & 1.5 & 0.2 \\
\ttH, \Htautau    & \multicolumn{2}{c}{$<0.01$} & 0.27 & 0.04 & \multicolumn{2}{c}{$<0.01$} & 1.01 & 0.14 & 0.8 & 0.1 & 0.35 & 0.05  \\
$tH$, \Htautau  & 0.023 & 0.003 & 0.06 & 0.01 & 0.029 & 0.004 & 0.30 & 0.04 & 0.39 & 0.05 & 0.09 & 0.01 \\
$bbH$,  \Htautau & 0.19 & 0.03 & 0.07 & 0.01 & 0.21 & 0.03 & 0.30 & 0.04 & 0.21 & 0.03 & 0.05 & 0.01 \\
\midrule
Total background & 9084 & 244 & 4444 & 132 & 14291 & 199 & 12398 & 188 & 7702 & 96 & 1080 & 32   \\
Total signal     &  69.5 & 11.0 & 32 & 5 & 147 & 23 & 148 & 22 & 130 & 18 & 41.7 & 7.2 \\
\midrule
Total            & 9153.5 & 243.5 & 4476 & 132 & 14438 & 198 & 12546 & 187 & 7832 & 95 & 1122 & 32  \\
\midrule
Data          & \multicolumn{1}{l}{$\quad 9163\phantom{.00}$} &&  \multicolumn{1}{l}{$\quad 4503\phantom{.000}$} && \multicolumn{1}{l}{$\quad 14389\phantom{.0}$} &&  \multicolumn{1}{l}{$\quad 12585\phantom{.00}$} && \multicolumn{1}{l}{$\quad 7800\phantom{.00}$} && \multicolumn{1}{l}{$\quad 1124\phantom{.00}$} & \\
\bottomrule
\end{tabular}}
\end{center}
\end{table}
 
\begin{table}[htbp]
\begin{center}
\caption{
Observed event yields and predictions as computed by the fit in the boost signal regions of the $\tau_{\text{lep}}\tau_{\text{had}}$ channel.
Uncertainties include statistical and systematic components.
The prediction for each sample is determined from the likelihood fit performed to measure the $pp\to\Htautau$ cross-section.
\label{tab:yield:boost:lh}
}
\resizebox{\textwidth}{!}{
\begin{tabular}{
@{}l
S[table-format=5.2]@{$\,\pm\,$}S[table-format=3.2]
S[table-format=4.3]@{$\,\pm\,$}S[table-format=2.3]
S[table-format=6.3]@{$\,\pm\,$}S[table-format=4.3]
S[table-format=5.2]@{$\,\pm\,$}S[table-format=3.2]
S[table-format=4.2]@{$\,\pm\,$}S[table-format=2.2]
S[table-format=4.3]@{$\,\pm\,$}S[table-format=2.3]
@{}
}
 
\toprule
&  \multicolumn{12}{S}{{Boost \tlhad}}\\
\cmidrule(l){2-13}
\pTH[\gev]                       & \multicolumn{4}{S}{{$[100, 120]$}} & \multicolumn{4}{S}{{$[120, 200]$}} & \multicolumn{2}{S}{{$[200, 300]$}} & \multicolumn{2}{S}{{$\left[300, \infty\right[$}} \\
\cmidrule(lr){2-5}\cmidrule(lr){6-9}
\njetsthirty                      & \multicolumn{2}{S}{{$=1$}} & \multicolumn{2}{S}{{$\geq\,2$}} & \multicolumn{2}{S}{{$=1$}} & \multicolumn{2}{S}{{$\geq\,2$}} & \multicolumn{2}{S}{{$\geq\,1$}} & \multicolumn{2}{S}{{$\geq\,1$}} \\
\midrule
 
\Ztautau          & 5583 & 100 & 3228 & 69 & 10927 & 130 & 9546 & 118 & 7195 & 92 & 2413 & 46   \\
Fake              & 1119 & 58 & 832 & 44 & 1324 & 76 & 1434 & 81 & 536 & 35 & 139 & 12   \\
Top               & 65  & 9 & 123 & 16 & 81 & 13 & 317 & 25 & 129 & 12 & 52 & 8   \\
Other backgrounds & 214 & 40 & 177 & 22 & 374 & 40 & 447 & 33 & 300 & 17 & 164 & 6   \\
\midrule
ggF, \Htautau     & 45.4 & 11.6 & 44.8 & 15.2 & 99.5 & 19.7 & 123 & 29 & 91.2 & 24.4 & 33.5 & 8.9   \\
VBF, \Htautau     & 12.1 & 2.0 & 7.2 & 1.1 & 26.9 & 4.1 & 23.6 & 3.7 & 21.6 & 3.2 & 8.6 & 1.4   \\
$WH$, \Htautau    & 1.7 & 0.3 & 2.2 & 0.6 & 3.5 & 2.0 & 6.5 & 1.2 & 4.4 & 0.8 & 3.3 & 1.0  \\
$ZH$, \Htautau    & 1.2 & 0.3 & 1.3 & 0.2 & 2.4 & 0.3 & 4.7 & 0.6 & 3.0 & 0.4 & 1.8 & 0.2  \\
\ttH, \Htautau    &  \multicolumn{2}{c}{$<0.01$} & 0.11 & 0.01 &  \multicolumn{2}{c}{$<0.01$} & 0.55 & 0.08 & 0.36 & 0.05 & 0.22 & 0.03  \\
$tH$,  \Htautau  & \multicolumn{2}{c}{$<0.01$} & 0.09 & 0.01 & 0.018 & 0.002 & 0.28 & 0.04 & 0.17 & 0.02 & 0.042 & 0.006 \\
$bbH$,  \Htautau  & 0.13 & 0.02 & 0.13 & 0.02 & 0.17 & 0.02 & 0.21 & 0.03 & 0.12 & 0.02 & 0.04 & 0.01 \\
 
\midrule
 
Total background & 6981 & 112 & 4360 & 82 & 12706 & 144 & 11743 & 135 & 8160 & 94 & 2768 & 49  \\
Total signal     & 61   & 13  & 56.0 & 15.4 & 133 & 22 & 160 & 31 & 121 & 25 & 47.6 & 9.2 \\
\midrule
Total            & 7042 & 112 & 4416 & 81 & 12839 & 143 & 11903 & 134 & 8281 & 93 & 2816 & 49 \\

\midrule
Data              & \multicolumn{1}{l}{$\quad 7094\phantom{.00}$} && \multicolumn{1}{l}{$4374\phantom{.000}$} && \multicolumn{1}{l}{$\quad 12779\phantom{.00}$} && \multicolumn{1}{l}{$11\,886\phantom{.00}$} && \multicolumn{1}{l}{$8236\phantom{.00}$} && \multicolumn{1}{l}{$2848\phantom{.000}$} \\
\bottomrule
\end{tabular}
}
\end{center}
\end{table}
 
\begin{table}[htbp]
\begin{center}
\caption{
Observed event yields and predictions as computed by the fit in the boost signal regions of the $\tau_{e}\tau_{\mu}$  channel.
Uncertainties include statistical and systematic components.
The prediction for each sample is determined from the likelihood fit performed to measure the $pp\to\Htautau$ cross-section.
\label{tab:yield:boost:emu}
}
\resizebox{\textwidth}{!}{
\begin{tabular}{
@{}l
S[table-format=6.3]@{$\,\pm\,$}S[table-format=4.3]
S[table-format=4.3]@{$\,\pm\,$}S[table-format=2.3]
S[table-format=6.3]@{$\,\pm\,$}S[table-format=4.3]
S[table-format=4.3]@{$\,\pm\,$}S[table-format=2.3]
S[table-format=4.3]@{$\,\pm\,$}S[table-format=2.3]
S[table-format=6.3]@{$\,\pm\,$}S[table-format=4.3]
@{}
}
\toprule
&  \multicolumn{12}{S}{{Boost \temu}}\\
\cmidrule(l){2-13}
\pTH[\gev]        & \multicolumn{4}{S}{{$[100, 120]$}} & \multicolumn{4}{S}{{$[120, 200]$}} & \multicolumn{2}{S}{{$[200, 300]$}} & \multicolumn{2}{S}{{$\left[300, \infty\right[$}} \\
\cmidrule(lr){2-5}\cmidrule(lr){6-9}
\njetsthirty       & \multicolumn{2}{S}{{$=1$}} & \multicolumn{2}{S}{{$\geq\,2$}} & \multicolumn{2}{S}{{$=1$}} & \multicolumn{2}{S}{{$\geq\,2$}} & \multicolumn{2}{S}{{$\geq\,1$}} & \multicolumn{2}{S}{{$\geq\,1$}} \\
\midrule
 
\Ztautau          &  2642 & 64 & 1523 & 42 & 3912 & 69 & 3453 & 68 & 1734 & 37 & 469 & 20  \\
Fake              &  101 & 31 & 85 & 23 & 117 & 36 & 179 & 42 & 88 & 24 & 37 & 13  \\
Top               &  101 & 8 & 187 & 20 & 157 & 11 & 569 & 44 & 258 & 16 & 104 & 11 \\
Other backgrounds &  118 & 17 & 101 & 16 & 273 & 14 & 325 & 30 & 294 & 8 & 173 & 5   \\
\midrule
ggF, \Htautau     & 16.6 & 3.2 & 11 & 2 & 35 & 7 & 36.8 & 7.1 & 25.5 & 4.6 & 8.7 & 2.2   \\
VBF, \Htautau     & 3.4 & 0.6 & 2.2 & 0.3 & 8.7 & 1.6 & 7.8 & 1.1 & 6.2 & 0.9 & 2.1 & 0.3   \\
$WH$, \Htautau    & 0.44 & 0.06 & 0.57 & 0.13 & 1.35 & 0.56 & 2.22 & 0.57 & 1.29 & 0.18 & 0.87 & 0.35  \\
$ZH$, \Htautau    & 0.29 & 0.04 & 0.33 & 0.05 & 0.73 & 0.10 & 1.57 & 0.22 & 0.85 & 0.12 & 0.41 & 0.06  \\
\ttH, \Htautau    &  \multicolumn{2}{c}{$<0.01$} & 0.029 & 0.004 &  \multicolumn{2}{c}{$<0.01$} & 0.20 & 0.03 & 0.08 & 0.01 & 0.07 & 0.01  \\
$tH$,  \Htautau  & \multicolumn{2}{c}{} & 0.08 & 0.01 & 0.025 & 0.003 & 0.13 & 0.02 & 0.11 & 0.02 & 0.033 & 0.005 \\
$bbH$,  \Htautau  & 0.038 & 0.005 & 0.026 & 0.004 & 0.06 & 0.01 & 0.11 & 0.01 & 0.06 & 0.01 & 0.017 & 0.002 \\
 
\midrule
Total background  & 2961 & 65 & 1896 & 47 & 4458 & 70 & 4526 & 75 & 2373 & 43 & 783 & 25   \\
Total signal      & 21   & 3  & 14   & 2  & 46 & 6 & 49 & 7 & 34 & 4 & 12 & 2   \\
\midrule
Total             &  2982 & 65 & 1910 & 47 & 4504 & 70 & 4575 & 75 & 2407 & 42 & 795 & 25 \\
\midrule
 
Data               & \multicolumn{1}{l}{$\quad 2973\phantom{.00}$} && \multicolumn{1}{l}{$1877\phantom{.000}$} && \multicolumn{1}{l}{$\quad 4458\phantom{.00}$} && \multicolumn{1}{l}{$4594\phantom{.000}$} && \multicolumn{1}{l}{$2325\phantom{.000}$} && \multicolumn{1}{l}{$\quad \quad  743\phantom{.0}$} & \\
\bottomrule
\end{tabular}
}
\end{center}
\end{table}
 
\clearpage
 
\begin{table}[!htbp]
\setlength{\extrarowheight}{4pt}
\begin{center}
\caption{
Best-fit values and uncertainties for the $pp\to\Htautau$ cross-section measurement and the measurement in the four dominant production modes.
All measurements include the branching ratio of $\Htautau$ and refers to true Higgs boson rapidity $|y_H|<2.5$.
The SM predictions for each region, computed using the inclusive cross-section calculations and the simulated event samples are also shown.
The contributions to the total uncertainty in the measurements from statistical (Stat.~unc.) or systematic uncertainties (Syst.~unc.) in the signal prediction (Th.~sig.), background prediction (Th.~bkg.), and in experimental performance (Exp.) are given separately.
Each uncertainty is reported as the average of its upward and downward fluctuations.
The total systematic uncertainty, equal to the difference in quadrature between the total uncertainty and the statistical uncertainty, differs from the sum in quadrature of the Th. sig., Th. bkg., and Exp. systematic uncertainties due to correlations.
\label{tab:res:4poi}}
\resizebox{\textwidth}{!}{
\begin{tabular}{
@{}l
S[table-format=1.4]@{$\,\pm\,$}S[table-format=1.4]
S[table-format=1.3]@{$\,\pm\,$}S[table-format=1.3]
|
S[table-format=1.3]
S[table-format=1.3]
S[table-format=1.3]
S[table-format=1.3]
@{}
}
\toprule
Production Mode  & \multicolumn{2}{c}{SM prediction} & \multicolumn{2}{c|}{Result} & \multicolumn{1}{c}{Stat.~unc.} & \multicolumn{3}{c}{Syst.~unc. [pb]}                                                                                                           \\
\cmidrule(l){7-9}
& \multicolumn{2}{c}{[pb]}          & \multicolumn{2}{c|}{[pb]}   & \multicolumn{1}{c}{[pb]}       & \multicolumn{1}{c}{Th.~sig.} & \multicolumn{1}{c}{Th.~bkg.} & \multicolumn{1}{c}{Exp.}                                                        \\
\midrule
\ttH         &  0.0313 & 0.0032 & 0.033  & 0.037  & \pm0.031  & \pm0.010  & \pm0.010  & \pm0.010 \\
$VH$         &  0.1176 & 0.0025 & 0.115  & 0.070  & \pm0.058  & \pm0.016  & \pm0.005  & \pm0.021 \\
ggF          &  2.77   & 0.09   & 2.65   & 0.85   & \pm0.41   & \pm0.56   & \pm0.07   & \pm0.45 \\
VBF          &  0.220  & 0.005  & 0.197  & 0.041  & \pm0.028  & \pm0.024  & \pm0.005  & \pm0.012 \\ \midrule
$pp\,\to\,H$ &  3.17   & 0.09   & 2.94   & 0.41   & \pm0.21   & \pm0.26   & \pm0.05   & \pm0.19 \\
 
\bottomrule
\end{tabular}
}
\end{center}
\end{table}

\begin{table}[!htbp]
\setlength{\extrarowheight}{5pt}
\begin{center}
\caption{
Best-fit values and uncertainties for the $pp\to\Htautau$ cross-sections, in the reduced \emph{stage 1.2} STXS scheme described in the text.
The EW production mode includes vector-boson fusion and $qq\to{V}(\to{qq})H$ processes.
All measurements include the branching ratio of $H\to\tau\tau$ and refers to true Higgs boson rapidity $|y_H|<2.5$.
The SM predictions for each region, computed using the inclusive cross-section calculations and the simulated event samples are also shown.
The contributions to the total uncertainty in the measurements from statistical (Stat.~unc.) or systematic uncertainties (Syst.~unc.) in the signal prediction (Th.~sig.), background prediction (Th.~bkg.), and in experimental performance (Exp.) are given separately.
Each uncertainty is reported as the average of its upward and downward fluctuations.
The total systematic uncertainty, equal to the difference in quadrature between the total uncertainty and the statistical uncertainty, differs from the sum in quadrature of the Th. sig., Th. bkg., and Exp. systematic uncertainties due to correlations.
The spades symbol ($\spadesuit$) indicates that the criteria for \mjj only apply to events with at least two reconstructed jets.
\label{tab:res:9pois}}
\resizebox{\textwidth}{!}{
\begin{tabular}{
@{}
cccc
S[table-format=3.1]@{$\,\pm\,$}S[table-format=2.1]
S[table-format=3.2]@{$\,\pm\,$}S[table-format=3.2]
| l  l  l  l
@{}
}
\toprule
 
\multicolumn{4}{c}{STXS bin}                               & \multicolumn{2}{c}{SM prediction}  & \multicolumn{2}{c|}{Result} & \multicolumn{1}{c}{Stat.~unc.} & \multicolumn{3}{c}{Syst.~unc. [fb]} \\
\cmidrule(r){1-4}\cmidrule(l){10-12}
Process & \mjj[GeV] & \pTH[GeV] & $\text{N}_{\text{jets}}$ & \multicolumn{2}{c}{[fb]}           & \multicolumn{2}{c|}{[fb]}   & \multicolumn{1}{c}{[fb]}       & \multicolumn{1}{c}{Th.~sig.} & \multicolumn{1}{c}{Th.~bkg.} & \multicolumn{1}{c}{Exp.} \\
\midrule
 
\multirow{6}{*}{\rotatebox{90}{ggF + $gg\to Z(\to qq)H$}}
& $[0,\,350]^{\spadesuit}$         & $[60,\,120]$     & $\geq1$ & 394   & 60   & 189      & 390     & $\pm220$  & $\pm59$   & $\pm152$  & $\pm240$ \\
&                                  & $[120,\,200]$    & $=1$    & 47    & 11   & 17       & 30       & $\pm18$   & $\pm4$    & $\pm4$    & $\pm16$  \\
& $[0,\,350]\phantom{^\spadesuit}$ & $[120,\,200]$    & $\geq2$ & 59    & 20   & 33       & 39       & $\pm27$   & $\pm10$   & $\pm10$   & $\pm23$  \\
&                                  & $[200,\,300]$    & $\geq0$ & 30    & 9    & 30.3     & 11.0     & $\pm8.6$  & $\pm2.9$  & $\pm0.8$  & $\pm5.6$ \\
&                                  & $\left[300, \infty\right[$ & $\geq0$ & 7.7   & 3.0  & 9.35     & 3.80   & $\pm3.50$ & $\pm1.00$    & $\pm0.22$   & $\pm1.20$\\
& $\left[350, \infty\right[$                 & $[0,\,200]$      & $\geq2$ & 55    & 13   & 143      & 110 & $\pm54$  & $\pm58$ & $\pm6$   & $\pm71$           \\
\midrule
\multirow{2}{*}{EW} & $[60, 120]$      &   & $\geq2$ & 33.1  & 1.1  & 32       & 20  & $\pm17$  & $\pm4$  & $\pm2$   & $\pm6$             \\
& $\left[350, \infty\right[$ &   & $\geq2$ & 90.1  & 2.2  & 71  & 17  & $\pm13$  & $\pm10$  & $\pm2$   & $\pm4$                 \\
\midrule
\ttH &  &  &                                        & 31.3  & 3.2  & 34  & 37  & $\pm32$  & $\pm7$  & $\pm10$   & $\pm8$\\
\bottomrule
\end{tabular}
}
\end{center}
\end{table}

\clearpage


\FloatBarrier

\section{Conclusion}
\label{sec:conclusion}

Measurements of production cross-sections for Standard Model Higgs bosons decaying into a pair of $\tau$-leptons are presented.
The measurements use data collected by the ATLAS experiment from proton--proton collisions in Run~2 of the LHC.
The data correspond to an integrated luminosity of $139\,\text{fb}^{-1}$.
 
All measurements include the branching ratio of $H\to\tau\tau$ and refer to true Higgs boson rapidity $|y_H|<2.5$.
The $pp\to\Htautau$ cross-section is measured to be $2.94 \pm 0.21 \text{(stat)} ^{+\,0.37}_{-\,0.32} \text{(syst)}$\,pb, in agreement with the SM prediction of \mbox{$3.17\pm0.09$\,pb}.
The observed (expected) uncertainty in the $pp\to\Htautau$ cross-section determination was reduced from $^{+\,28}_{-\,25}\,\%$ ($^{+\,27}_{-\,24}\,\%$) in the measurement reported in Ref.~\autocite{HIGG-2017-07} to $\pm$\SI{13.9}{\percent} ($\pm$\SI{13.2}{\percent}) in this work.
In particular, the impact of the systematic uncertainties was reduced by approximately a factor of two from $\pm$\SI{21.5}{\percent} to $\pm$\SI{12}{\percent}.
 
Cross-sections are determined separately for the four main production modes:
$2.65  \pm 0.41 \text{(stat)}  ^{+\,0.91}_{-\,0.67} \text{(syst)}$\,pb for the gluon--gluon fusion mode,
$0.197 \pm 0.028 \text{(stat)} ^{+\,0.032}_{-\,0.026} \text{(syst)}$\,pb for the vector-boson fusion mode,
$0.115 \pm 0.058 \text{(stat)} ^{+\,0.042}_{-\,0.040} \text{(syst)}$\,pb for production with a vector boson, and
$0.033 \pm 0.031 \text{(stat)} ^{+\,0.022}_{-\,0.017} \text{(syst)}$\,pb for production with a top-quark pair.
 
Measurements are also performed as a function of key kinematic properties of the reconstructed event.
Cross-sections of the production of a Higgs boson decaying into $\tau$-leptons are measured as a function of the Higgs boson transverse momentum, the number of jets produced in association with the Higgs boson, and the invariant mass of the two leading jets when applicable.
They represent the most detailed study of Higgs boson production in the $\tau\tau$ decay channel to date.
The measurements attain an uncertainty of $\pm$\SI{24}{\percent} for electroweak production with two jets of invariant mass greater than \SI{350}{\GeV}.
The ggF production mode is measured with an uncertainty of $\pm$\SI{36}{\percent} and $\pm$\SI{40}{\percent} when the Higgs boson transverse momentum is between 200 and \SI{300}{\GeV} and above \SI{300}{\GeV} respectively.
All measurements are in agreement with the Standard Model predictions.


\section*{Acknowledgements}


We thank CERN for the very successful operation of the LHC, as well as the
support staff from our institutions without whom ATLAS could not be
operated efficiently.
 
We acknowledge the support of
ANPCyT, Argentina;
YerPhI, Armenia;
ARC, Australia;
BMWFW and FWF, Austria;
ANAS, Azerbaijan;
CNPq and FAPESP, Brazil;
NSERC, NRC and CFI, Canada;
CERN;
ANID, Chile;
CAS, MOST and NSFC, China;
Minciencias, Colombia;
MEYS CR, Czech Republic;
DNRF and DNSRC, Denmark;
IN2P3-CNRS and CEA-DRF/IRFU, France;
SRNSFG, Georgia;
BMBF, HGF and MPG, Germany;
GSRI, Greece;
RGC and Hong Kong SAR, China;
ISF and Benoziyo Center, Israel;
INFN, Italy;
MEXT and JSPS, Japan;
CNRST, Morocco;
NWO, Netherlands;
RCN, Norway;
MEiN, Poland;
FCT, Portugal;
MNE/IFA, Romania;
MESTD, Serbia;
MSSR, Slovakia;
ARRS and MIZ\v{S}, Slovenia;
DSI/NRF, South Africa;
MICINN, Spain;
SRC and Wallenberg Foundation, Sweden;
SERI, SNSF and Cantons of Bern and Geneva, Switzerland;
MOST, Taiwan;
TENMAK, T\"urkiye;
STFC, United Kingdom;
DOE and NSF, United States of America.
In addition, individual groups and members have received support from
BCKDF, CANARIE, Compute Canada and CRC, Canada;
PRIMUS 21/SCI/017 and UNCE SCI/013, Czech Republic;
COST, ERC, ERDF, Horizon 2020, ICSC-NextGenerationEU and Marie Sk{\l}odowska-Curie Actions, European Union;
Investissements d'Avenir Labex, Investissements d'Avenir Idex and ANR, France;
DFG and AvH Foundation, Germany;
Herakleitos, Thales and Aristeia programmes co-financed by EU-ESF and the Greek NSRF, Greece;
BSF-NSF and MINERVA, Israel;
Norwegian Financial Mechanism 2014-2021, Norway;
NCN and NAWA, Poland;
La Caixa Banking Foundation, CERCA Programme Generalitat de Catalunya and PROMETEO and GenT Programmes Generalitat Valenciana, Spain;
G\"{o}ran Gustafssons Stiftelse, Sweden;
The Royal Society and Leverhulme Trust, United Kingdom.
 
The crucial computing support from all WLCG partners is acknowledged gratefully, in particular from CERN, the ATLAS Tier-1 facilities at TRIUMF (Canada), NDGF (Denmark, Norway, Sweden), CC-IN2P3 (France), KIT/GridKA (Germany), INFN-CNAF (Italy), NL-T1 (Netherlands), PIC (Spain), ASGC (Taiwan), RAL (UK) and BNL (USA), the Tier-2 facilities worldwide and large non-WLCG resource providers. Major contributors of computing resources are listed in Ref.~\cite{ATL-SOFT-PUB-2023-001}.


\clearpage
\appendix
\part*{Appendix}
 
\begin{figure}[htbp]
\centering
\begin{center}
\includegraphics[height=0.25\textheight, keepaspectratio]{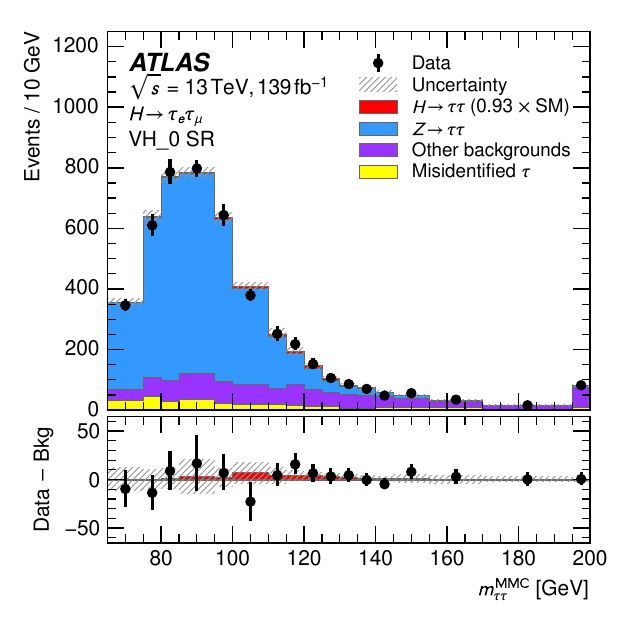}
\includegraphics[height=0.25\textheight, keepaspectratio]{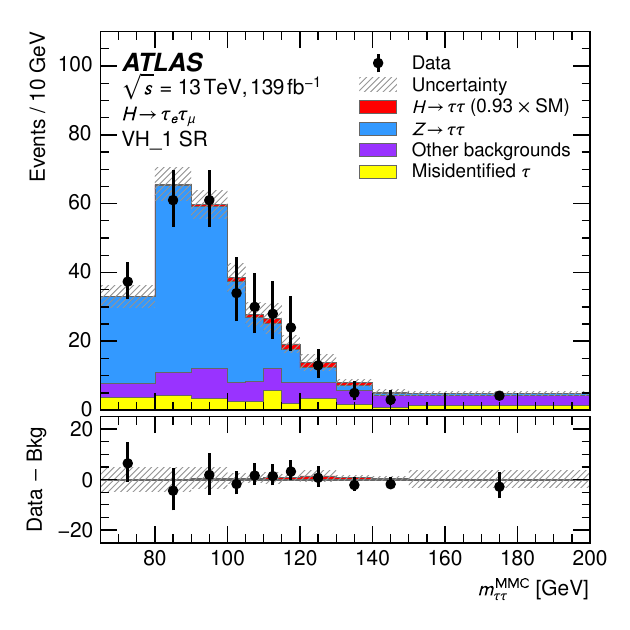}
\includegraphics[height=0.25\textheight, keepaspectratio]{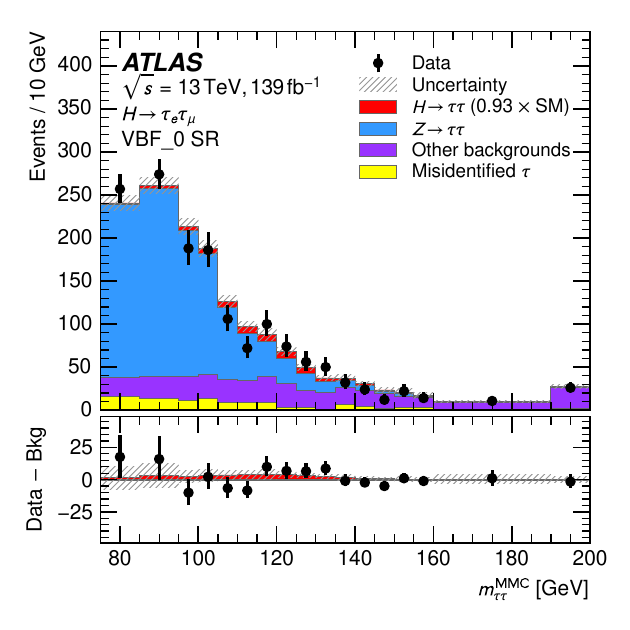}
\includegraphics[height=0.25\textheight, keepaspectratio]{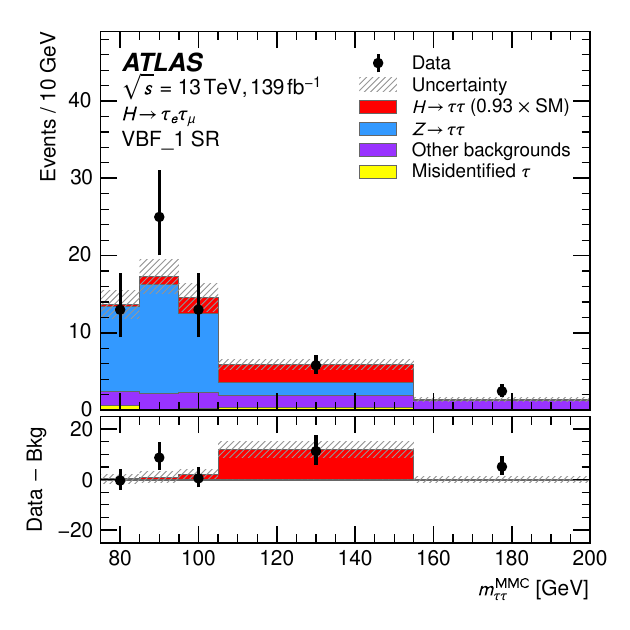}
\end{center}
\caption{
Distribution of the reconstructed $\tau\tau$ invariant mass (\mmmc) for all events in the V(had)H and VBF categories of the $\tau_{e}\tau_{\mu}$ channel.
The bottom panel shows the differences between the numbers of observed data events and expected background events (black points).
The observed Higgs boson signal, corresponding to $(\sigma\times B)/(\sigma\times B)_{\text{SM}}\,=\,0.93$, is shown with a filled red histogram.
Entries with values above the $x$-axis range are shown in the last bin of each distributions.
The dashed band indicates the total uncertainty on the total predicted yields.
The prediction for each sample is determined from the likelihood fit performed to measure the $pp\to\Htautau$ cross-section.
}
\label{fig:app:ll_rest_sr}
\end{figure}

\begin{figure}[htbp]
\centering
\begin{center}
\includegraphics[height=0.25\textheight, keepaspectratio]{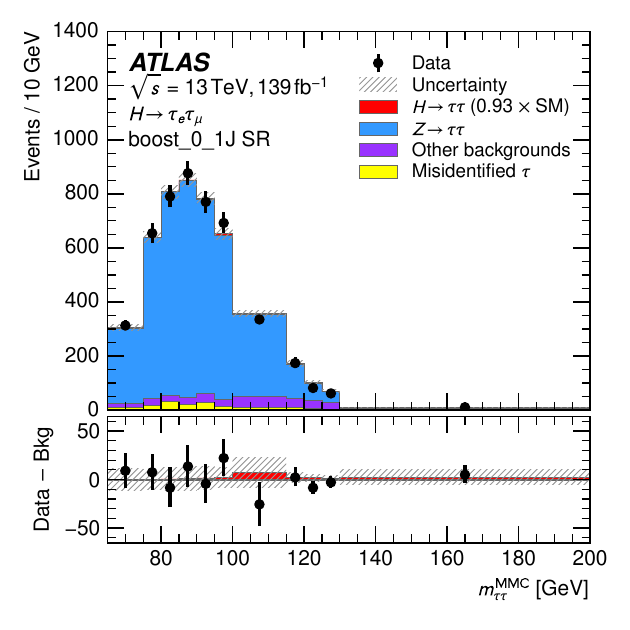}
\includegraphics[height=0.25\textheight, keepaspectratio]{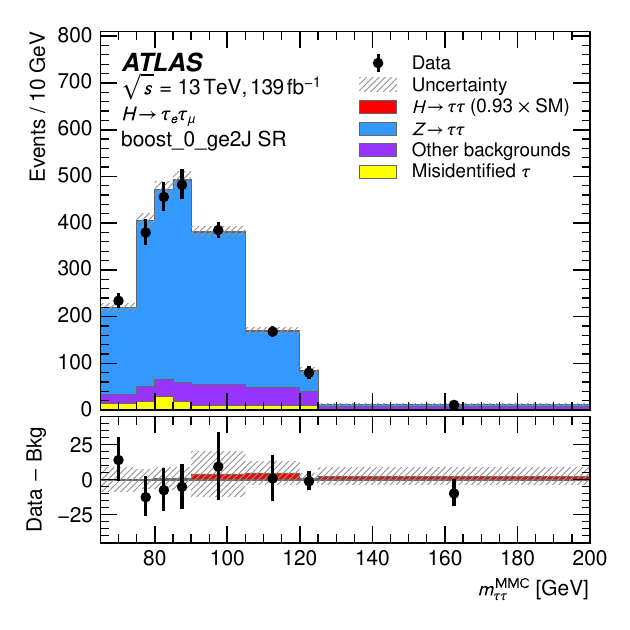}
\includegraphics[height=0.25\textheight, keepaspectratio]{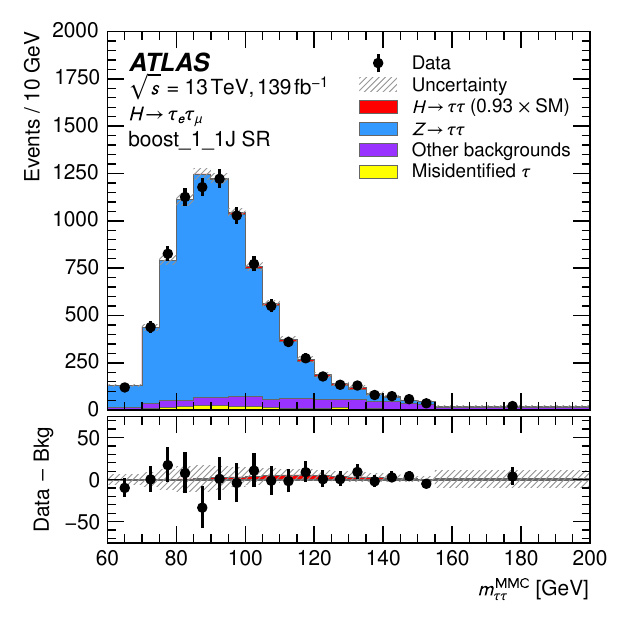}
\includegraphics[height=0.25\textheight, keepaspectratio]{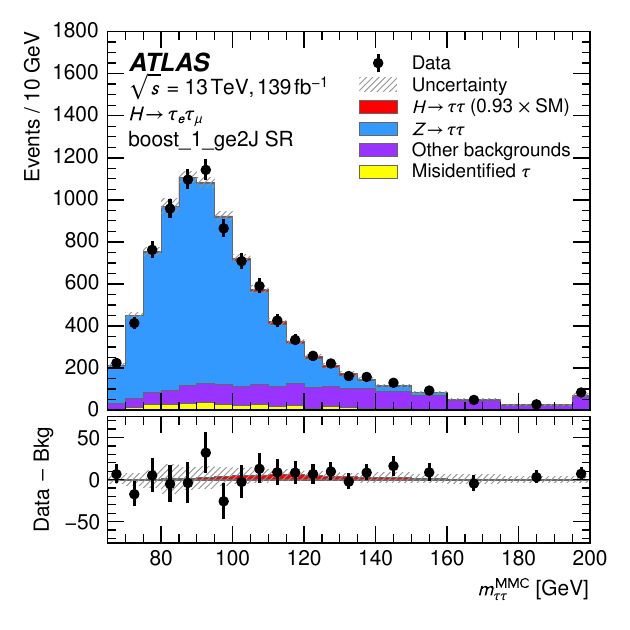}
\includegraphics[height=0.25\textheight, keepaspectratio]{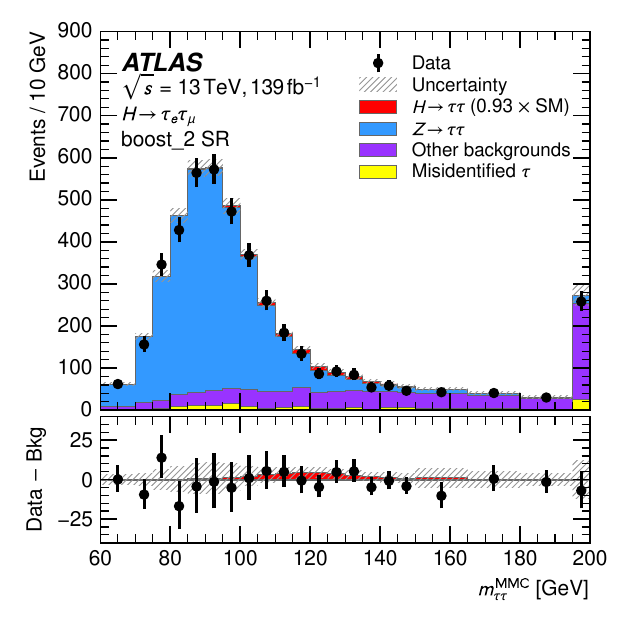}
\includegraphics[height=0.25\textheight, keepaspectratio]{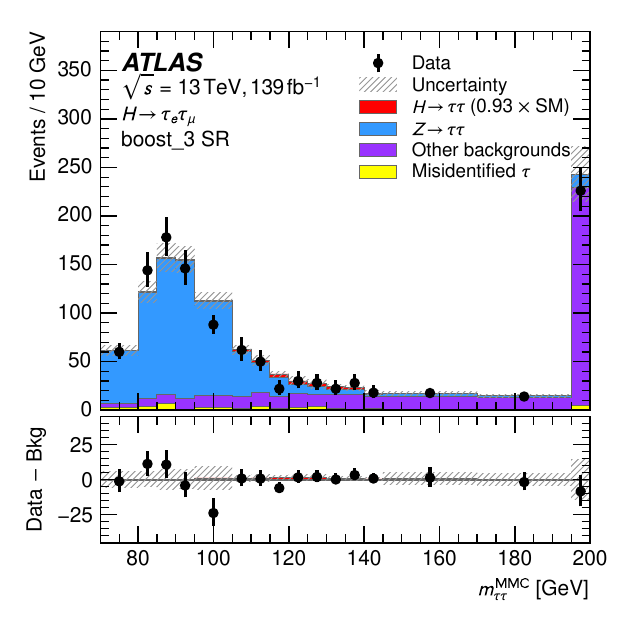}
 
\end{center}
\caption{
Distribution of the reconstructed $\tau\tau$ invariant mass (\mmmc) for all events in the boost categories of the $\tau_{e}\tau_{\mu}$ channel.
The bottom panel shows the differences between the numbers of observed data events and expected background events (black points).
The observed Higgs boson signal, corresponding to $(\sigma\times B)/(\sigma\times B)_{\text{SM}}\,=\,0.93$, is shown with a filled red histogram.
Entries with values above the $x$-axis range are shown in the last bin of each distributions.
The dashed band indicates the total uncertainty on the total predicted yields.
The prediction for each sample is determined from the likelihood fit performed to measure the $pp\to\Htautau$ cross-section.
}
\label{fig:app:ll_boost_sr}
\end{figure}

\begin{figure}[htbp]
\centering
\begin{center}
\includegraphics[height=0.25\textheight, keepaspectratio]{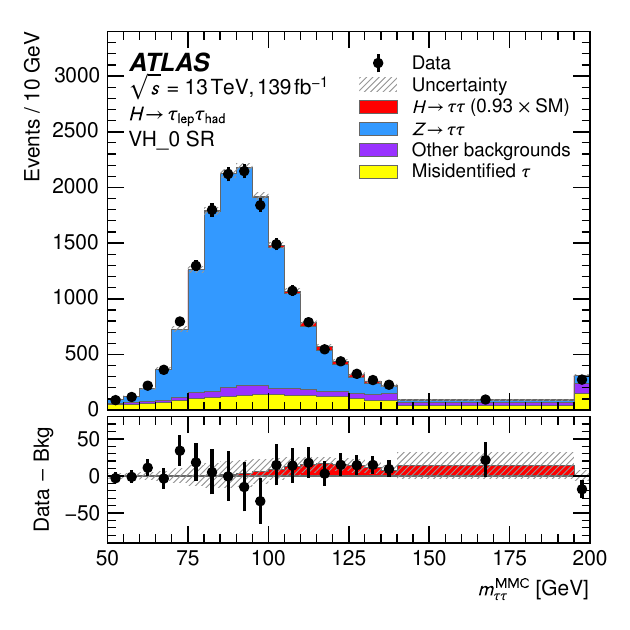}
\includegraphics[height=0.25\textheight, keepaspectratio]{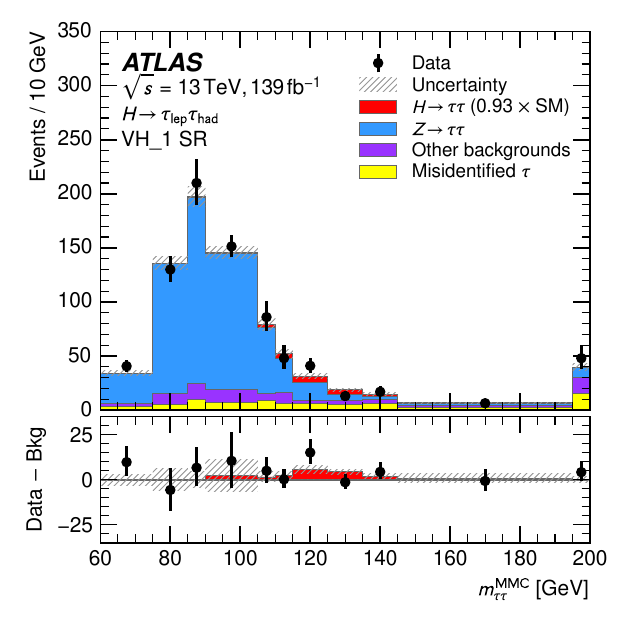}
\includegraphics[height=0.25\textheight, keepaspectratio]{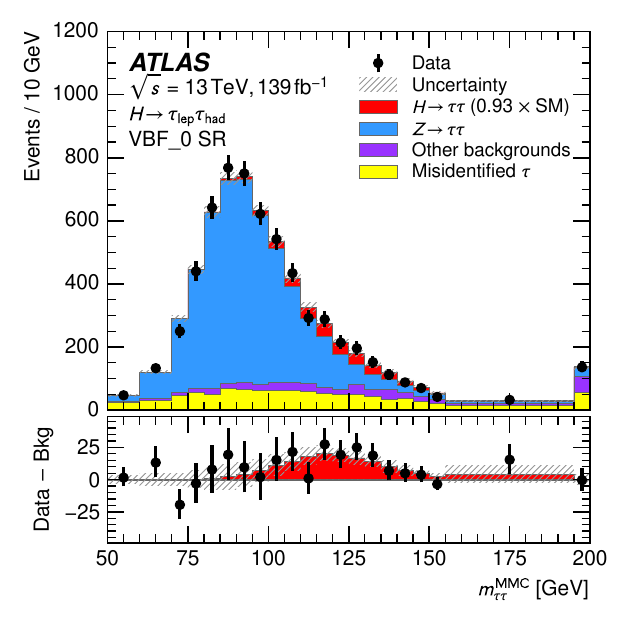}
\includegraphics[height=0.25\textheight, keepaspectratio]{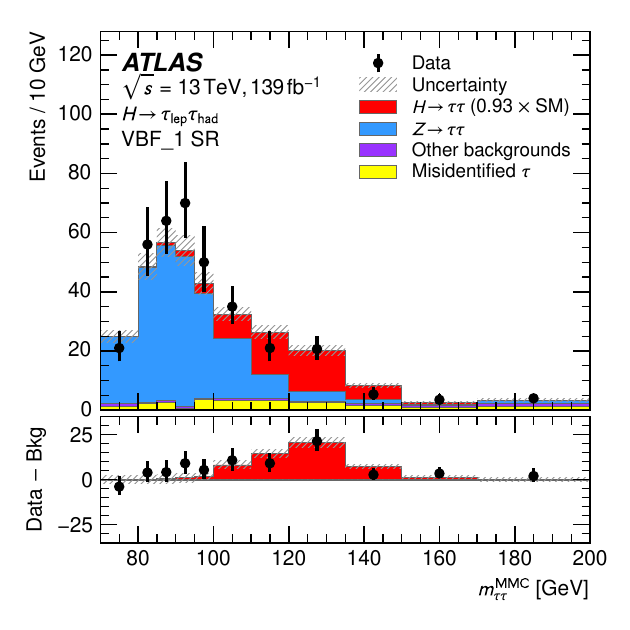}
\end{center}
\caption{
Distribution of the reconstructed $\tau\tau$ invariant mass (\mmmc) for all events in the V(had)H and VBF categories of the \tlhad channel.
The bottom panel shows the differences between the numbers of observed data events and expected background events (black points).
The observed Higgs boson signal, corresponding to $(\sigma\times B)/(\sigma\times B)_{\text{SM}}\,=\,0.93$, is shown with a filled red histogram.
Entries with values above the $x$-axis range are shown in the last bin of each distributions.
The dashed band indicates the total uncertainty on the total predicted yields.
The prediction for each sample is determined from the likelihood fit performed to measure the $pp\to\Htautau$ cross-section.
}
\label{fig:app:lh_rest_sr}
\end{figure}

\begin{figure}[htbp]
\centering
\begin{center}
\includegraphics[height=0.25\textheight, keepaspectratio]{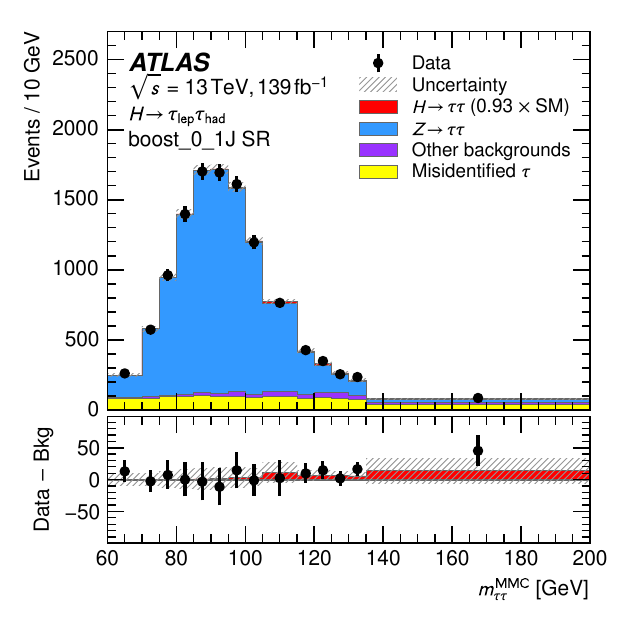}
\includegraphics[height=0.25\textheight, keepaspectratio]{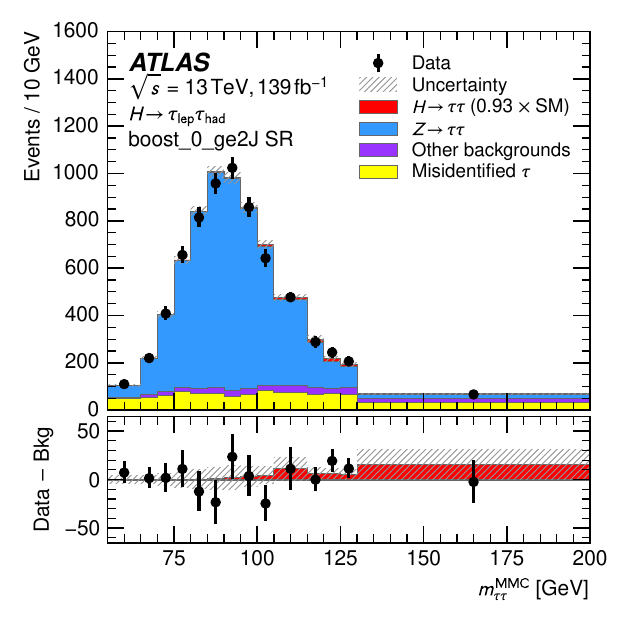}
\includegraphics[height=0.25\textheight, keepaspectratio]{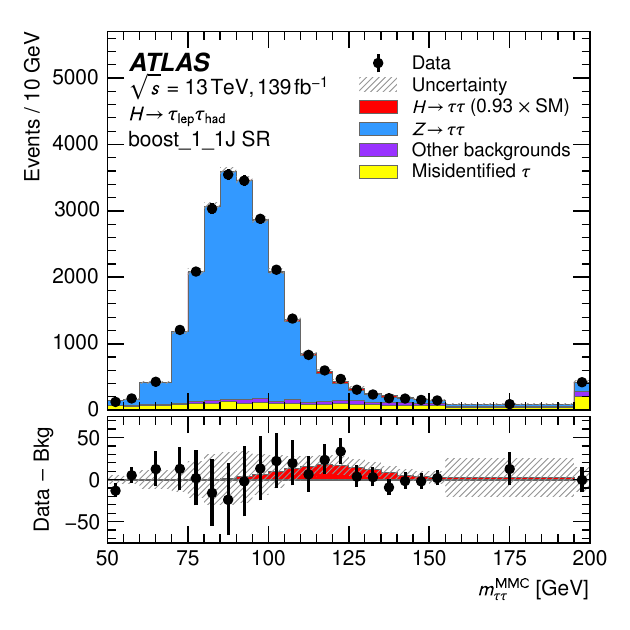}
\includegraphics[height=0.25\textheight, keepaspectratio]{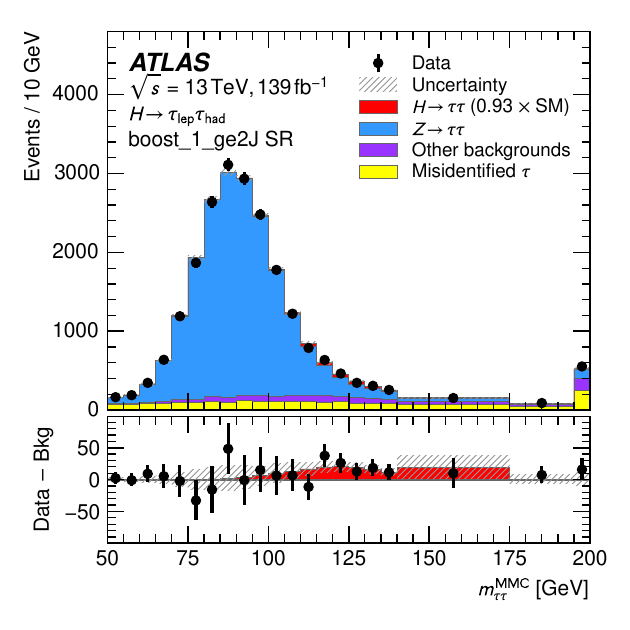}
\includegraphics[height=0.25\textheight, keepaspectratio]{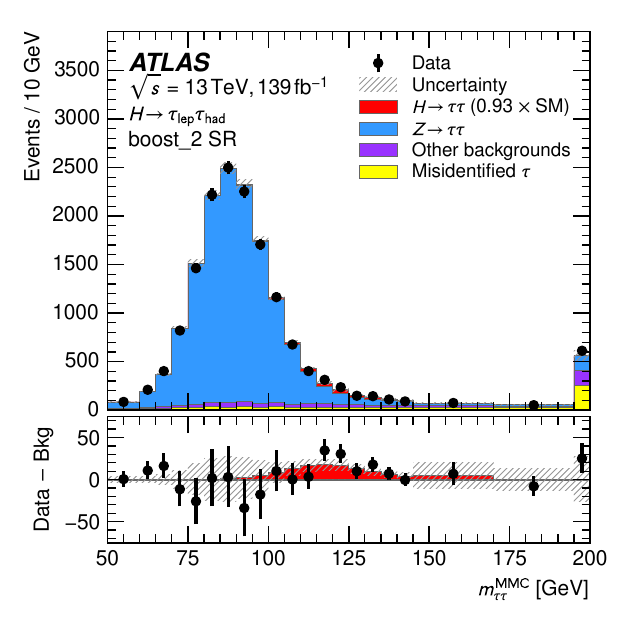}
\includegraphics[height=0.25\textheight, keepaspectratio]{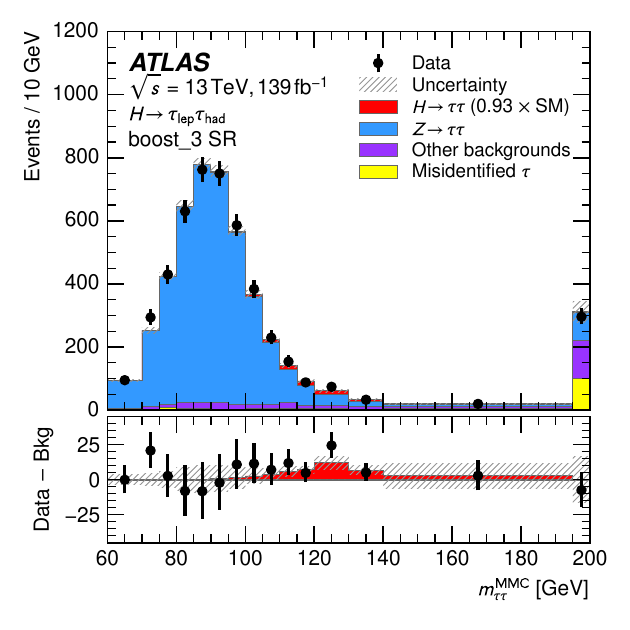}
\end{center}
\caption{
Distribution of the reconstructed $\tau\tau$ invariant mass (\mmmc) for all events in the boost categories of the \tlhad channel.
The bottom panel shows the differences between the numbers of observed data events and expected background events (black points).
The observed Higgs boson signal, corresponding to $(\sigma\times B)/(\sigma\times B)_{\text{SM}}\,=\,0.93$, is shown with a filled red histogram.
Entries with values above the $x$-axis range are shown in the last bin of each distributions.
The dashed band indicates the total uncertainty on the total predicted yields.
The prediction for each sample is determined from the likelihood fit performed to measure the $pp\to\Htautau$ cross-section.
}
\label{fig:app:lh_boost_sr}
\end{figure}

\begin{figure}[htbp]
\centering
\begin{center}
\includegraphics[height=0.25\textheight, keepaspectratio]{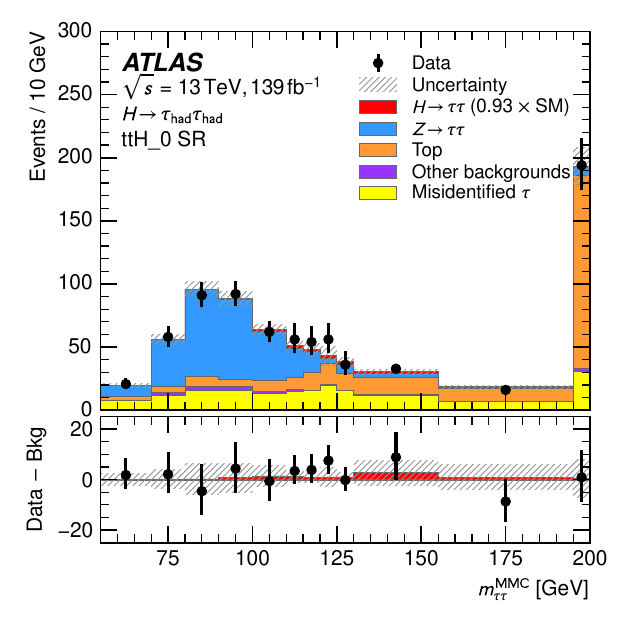}
\includegraphics[height=0.25\textheight, keepaspectratio]{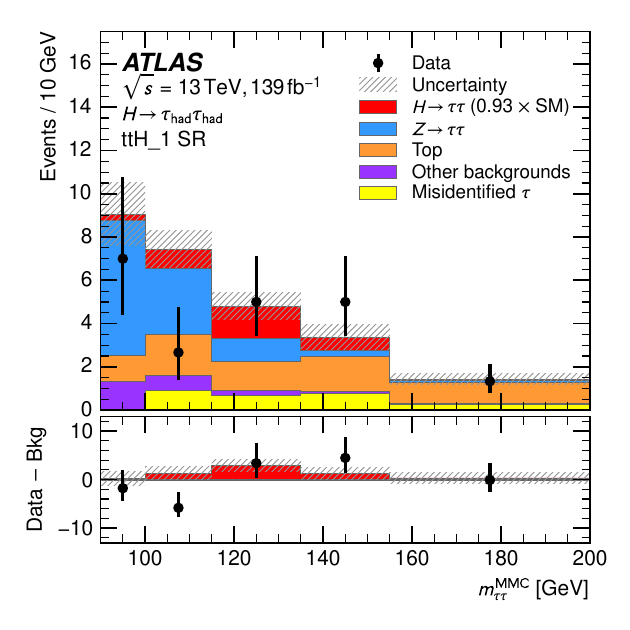}
\includegraphics[height=0.25\textheight, keepaspectratio]{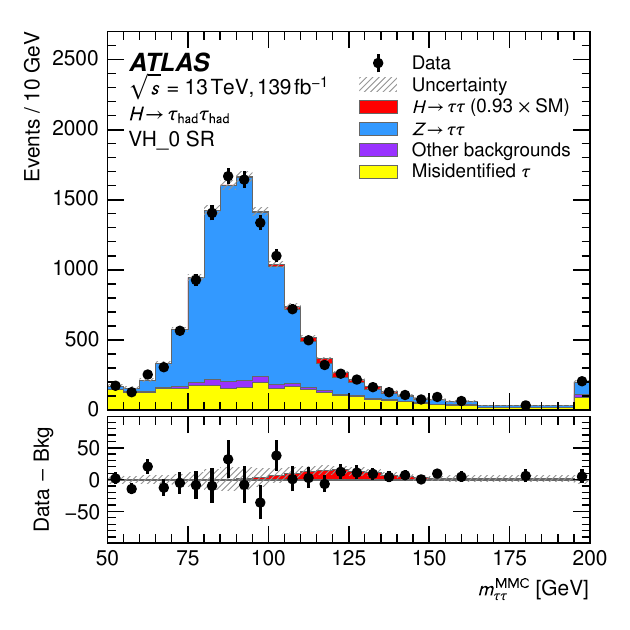}
\includegraphics[height=0.25\textheight, keepaspectratio]{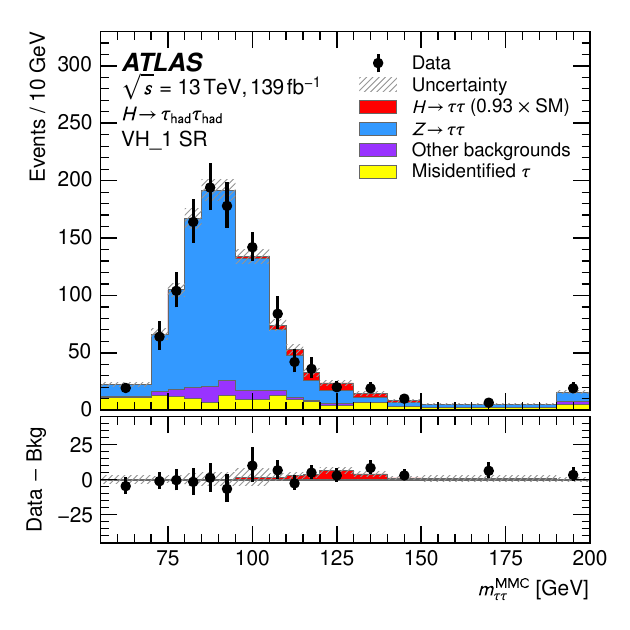}
\includegraphics[height=0.25\textheight, keepaspectratio]{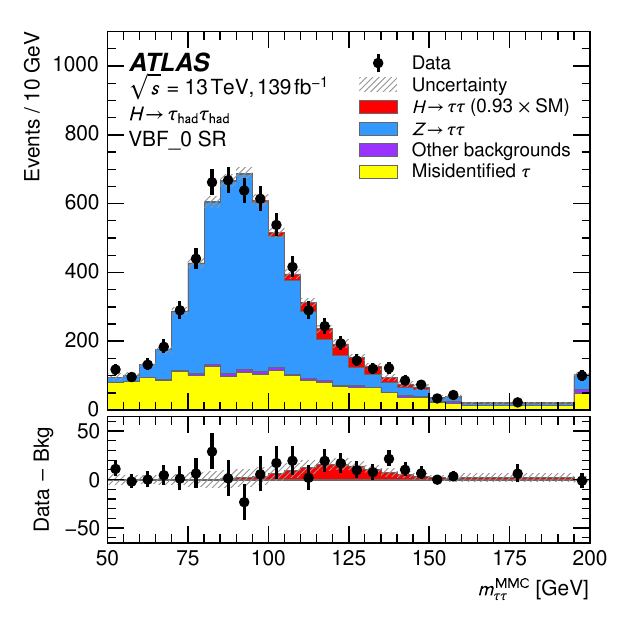}
\includegraphics[height=0.25\textheight, keepaspectratio]{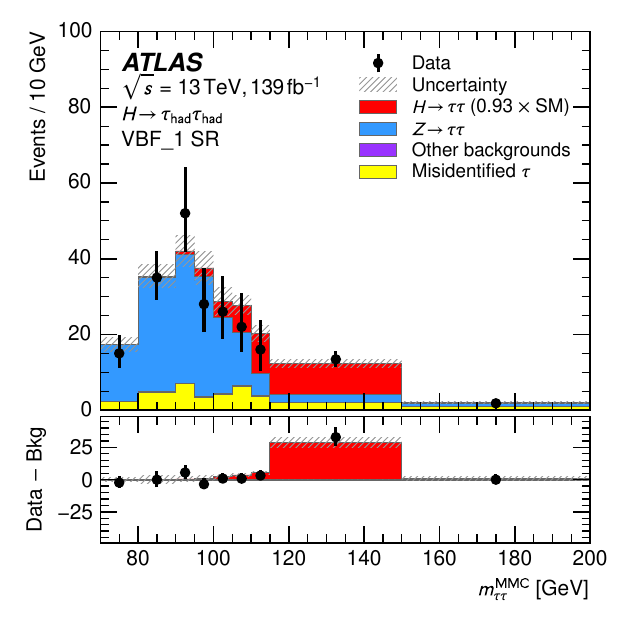}
 
\end{center}
\caption{
Distribution of the reconstructed $\tau\tau$ invariant mass (\mmmc) for all events in the ttH, V(had)H and VBF categories of the \thadhad channel.
The bottom panel shows the differences between the numbers of observed data events and expected background events (black points).
The observed Higgs boson signal, corresponding to $(\sigma\times B)/(\sigma\times B)_{\text{SM}}\,=\,0.93$, is shown with a filled red histogram.
Entries with values above the $x$-axis range are shown in the last bin of each distributions.
The dashed band indicates the total uncertainty on the total predicted yields.
The prediction for each sample is determined from the likelihood fit performed to measure the $pp\to\Htautau$ cross-section.
}
\label{fig:app:hh_rest_sr}
\end{figure}

\begin{figure}[htbp]
\centering
\begin{center}
\includegraphics[height=0.25\textheight, keepaspectratio]{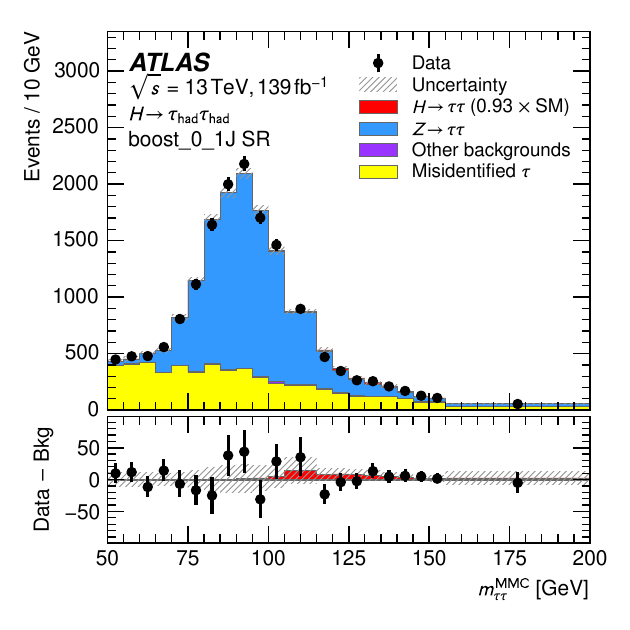}
\includegraphics[height=0.25\textheight, keepaspectratio]{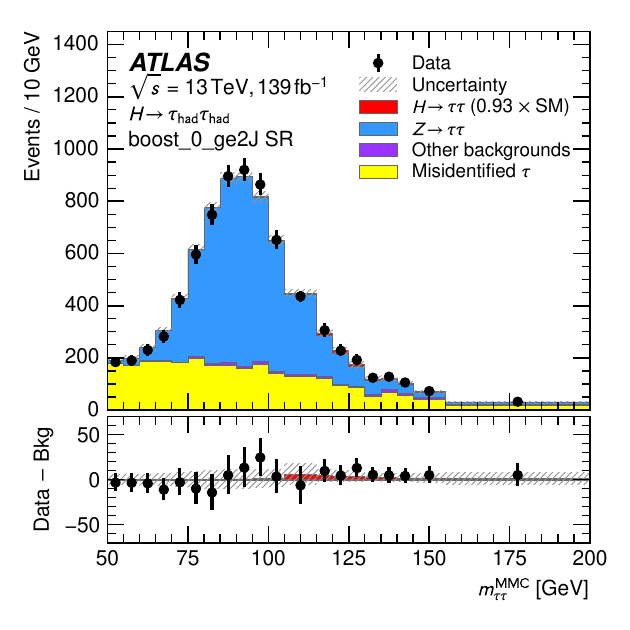}
\includegraphics[height=0.25\textheight, keepaspectratio]{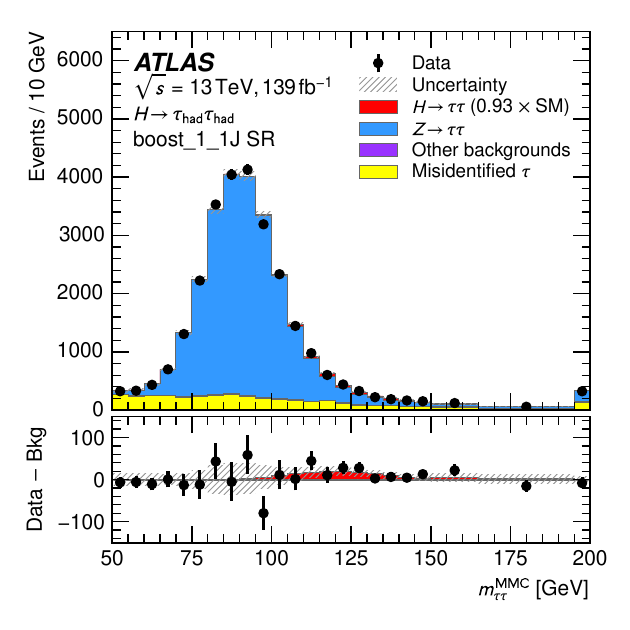}
\includegraphics[height=0.25\textheight, keepaspectratio]{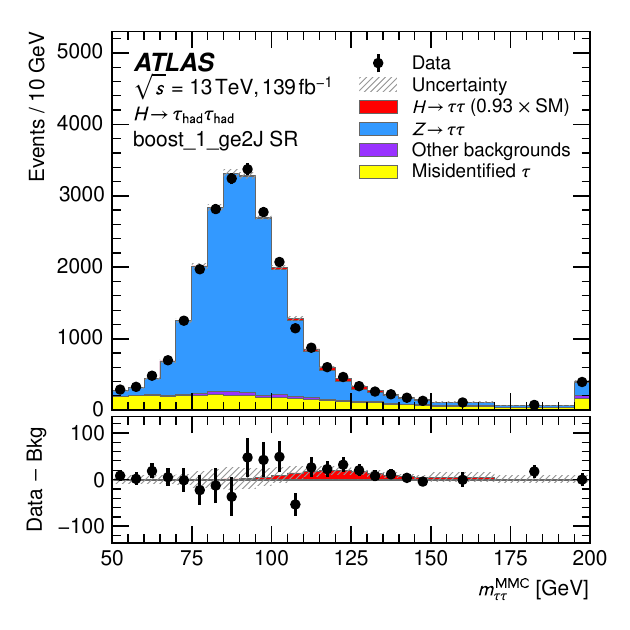}
\includegraphics[height=0.25\textheight, keepaspectratio]{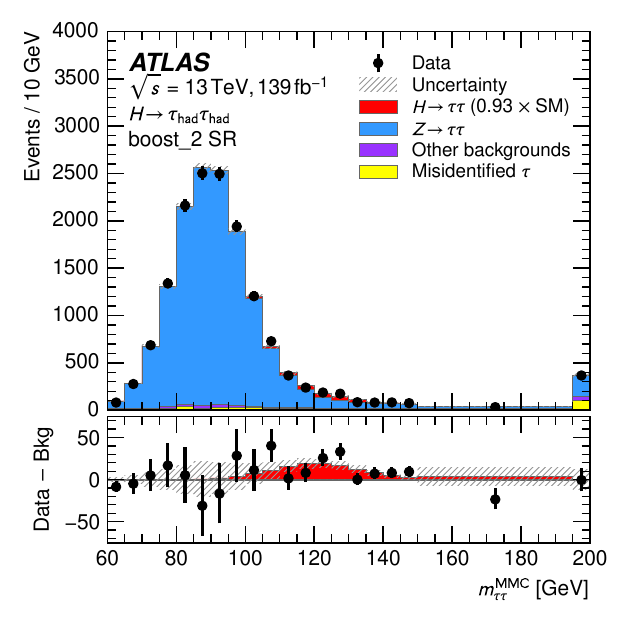}
\includegraphics[height=0.25\textheight, keepaspectratio]{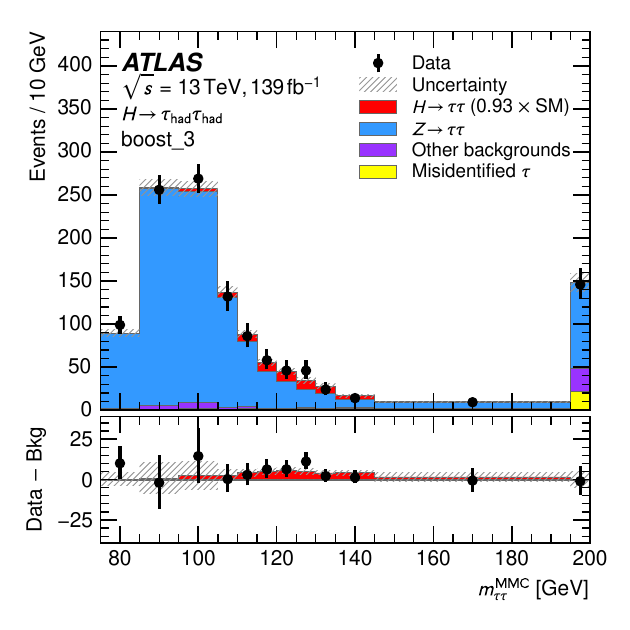}
\end{center}
\caption{
Distribution of the reconstructed $\tau\tau$ invariant mass (\mmmc) for all events in the boost categories of the \thadhad channel.
The bottom panel shows the differences between the numbers of observed data events and expected background events (black points).
The observed Higgs boson signal, corresponding to $(\sigma\times B)/(\sigma\times B)_{\text{SM}}\,=\,0.93$, is shown with a filled red histogram.
Entries with values above the $x$-axis range are shown in the last bin of each distributions.
The dashed band indicates the total uncertainty on the total predicted yields.
The prediction for each sample is determined from the likelihood fit performed to measure the $pp\to\Htautau$ cross-section.
}
\label{fig:app:hh_boost_sr}
\end{figure}

\Cref{fig:app:hh_boost_sr,fig:app:hh_rest_sr,fig:app:lh_boost_sr,fig:app:lh_rest_sr,fig:app:ll_boost_sr,fig:app:ll_rest_sr} show all distributions that enter the likelihood fit with the best-fit parameters derived from the fit with a single parameter of interest (inclusive cross-section measurement).

 
\addcontentsline{toc}{part}{Appendix}
\FloatBarrier
\printbibliography

@article{Englert:1964et,
        author                  = "Englert, F. and Brout, R.",
        title                           = "{Broken Symmetry and the Mass of Gauge Vector Mesons}",
        journal                 = "Phys. Rev. Lett.",
        volume                  = "13",
        pages                   = "321-323",
        doi                             = "10.1103/PhysRevLett.13.321",
        year                    = "1964",
        SLACcitation    = "%%CITATION = PRLTA,13,321;%%",
}

@article{Higgs:1964ia,
        author                  = "Higgs, Peter W.",
        title                           = "{Broken symmetries, massless particles and gauge fields}",
        journal                 = "Phys. Lett.",
        volume                  = "12",
        pages                   = "132-133",
        doi                             = "10.1016/0031-9163(64)91136-9",
        year                    = "1964",
        SLACcitation    = "%%CITATION = PHLTA,12,132;%%",
}

@article{Higgs:1964pj,
        author                  = "Higgs, Peter W.",
        title                           = "{Broken Symmetries and the Masses of Gauge Bosons}",
        journal                 = "Phys. Rev. Lett.",
        volume                  = "13",
        pages                   = "508-509",
        doi                             = "10.1103/PhysRevLett.13.508",
        year                    = "1964",
        SLACcitation    = "%%CITATION = PRLTA,13,508;%%",
}

@article{Guralnik:1964eu,
        author                  = "Guralnik, G.S. and Hagen, C.R. and Kibble, T.W.B.",
        title                           = "{Global Conservation Laws and Massless Particles}",
        journal                 = "Phys. Rev. Lett.",
        volume                  = "13",
        pages                   = "585-587",
        doi                             = "10.1103/PhysRevLett.13.585",
        year                    = "1964",
        SLACcitation    = "%%CITATION = PRLTA,13,585;%%",
}

@article{Higgs:1966ev,
        author                  = "Higgs, Peter W.",
        title                           = "{Spontaneous Symmetry Breakdown without Massless Bosons}",
        journal                 = "Phys. Rev.",
        volume                  = "145",
        pages                   = "1156-1163",
        doi                             = "10.1103/PhysRev.145.1156",
        year                    = "1966",
        SLACcitation    = "%%CITATION = PHRVA,145,1156;%%",
}

@article{Kibble:1967sv,
        author                  = "Kibble, T.W.B.",
        title                           = "{Symmetry Breaking in Non-Abelian Gauge Theories}",
        journal                 = "Phys. Rev.",
        volume                  = "155",
        pages                   = "1554-1561",
        doi                             = "10.1103/PhysRev.155.1554",
        year                    = "1967",
        SLACcitation    = "%%CITATION = PHRVA,155,1554;%%",
}

@article{Stewart:2011cf,
        author                  = "Stewart, Iain W. and Tackmann, Frank J.",
        title                           = "{Theory uncertainties for Higgs mass and other searches using jet bins}",
        journal                 = "Phys. Rev. D",
        volume                  = "85",
        pages                   = "034011",
        doi                             = "10.1103/PhysRevD.85.034011",
        year                    = "2012",
        eprint                  = "1107.2117",
        archivePrefix   = "arXiv",
        primaryClass    = "hep-ph",
        SLACcitation    = "%%CITATION = ARXIV:1107.2117;%%",
}

@article{Beenakker:2002nc,
        author                  = "Beenakker, W. and Dittmaier, S. and Kr{\"a}mer, M. and Plumper, B. and Spira, M. and others",
        title                   = "{NLO QCD corrections to $t\bar{t}H$  production in hadron collisions}",      
        journal                 = "Nucl. Phys. B",
        volume                  = "653",
        pages                   = "151-203",
        doi                             = "10.1016/S0550-3213(03)00044-0",
        year                    = "2003",
        eprint                  = "hep-ph/0211352",
        archivePrefix   = "arXiv",
        primaryClass    = "hep-ph",
        reportNumber = "DESY-02-177, EDINBURGH-2002-18, MPI-PHT-2002-70, PSI-PR-02-22",
        SLACcitation    = "%%CITATION = HEP-PH/0211352;%%",
}

@article{Dawson:2003zu,
        author                  = "Dawson, S. and Jackson, C. and Orr, L.H. and Reina, L. and Wackeroth, D.",
        title                           = "{Associated Higgs boson production with top quarks at the CERN Large Hadron Collider: NLO QCD corrections}",
        journal                 = "Phys. Rev. D",
        volume                  = "68",
        pages                   = "034022",
        doi                             = "10.1103/PhysRevD.68.034022",
        year                    = "2003",
        eprint                  = "hep-ph/0305087",
        archivePrefix   = "arXiv",
        primaryClass    = "hep-ph",
        reportNumber    = "BNL-HET-03-9, FSU-HEP-2003-0503, UB-HET-03-02",
        SLACcitation    = "%%CITATION = HEP-PH/0305087;%%",
}

@article{Yu:2014cka,
      author                    = "Zhang, Yu and Ma, Wen-Gan and Zhang, Ren-You and Chen, Chong and Guo, Lei",
      title                     = "{QCD NLO and EW NLO corrections to $t\bar{t}H$ production with top quark decays at hadron collider}",
      journal                   = "Phys. Lett. B",
      volume            = "738",
      year                      = "2014",
      pages             = "1-5",
      doi                       = "10.1016/j.physletb.2014.09.022",
      eprint                    = "1407.1110",
      archivePrefix     = "arXiv",
      primaryClass      = "hep-ph",
      SLACcitation      = "%%CITATION = ARXIV:1407.1110;%%"
}

@article{Frixione:2014qaa,
      author                    = "Frixione, S. and Hirschi, V. and Pagani, D. and Shao, H. S. and Zaro, M.",
      title                     = "{Weak corrections to Higgs hadroproduction in association with a top-quark pair}",
      journal                   = "JHEP",
      volume            = "09",
      year                      = "2014",
      pages             = "065",
      doi                       = "10.1007/JHEP09(2014)065",
      eprint                    = "1407.0823",
      archivePrefix     = "arXiv",
      primaryClass      = "hep-ph",
      reportNumber      = "CERN-PH-TH-2014-123, CP3-14-49",
      SLACcitation      = "%%CITATION = ARXIV:1407.0823;%%"
}

@Article{MMCpaper,
     author         = "{A. Elagin, P. Murat, A. Pranko and A. Safonov}",
     title          = "{A new mass reconstruction technique for resonances
                        decaying to $\tau\tau$}",
      journal        = "Nucl. Instrum. Meth. ",
      volume         = "A 654",
      pages          = "481-489",
      doi            = "10.1016/j.nima.2011.07.009",
      year           = "2011",
      eprint         = "1012.4686",
      archivePrefix  = "arXiv",
      primaryClass   = "hep-ex",
      SLACcitation   = "%%CITATION = ARXIV:1012.4686;%%",
}

@article{Avoni:2018iuv,
  author         = "{G. Avoni et al.}",
      title          = "{The new LUCID-2 detector for luminosity measurement and
                        monitoring in ATLAS}",
      journal        = "JINST",
      volume         = "13",
      year           = "2018",
      number         = "07",
      pages          = "P07017",
      doi            = "10.1088/1748-0221/13/07/P07017",
      SLACcitation   = "%%CITATION = JINST,13,P07017;%%"
}

@Article{geant,
     author    = "Agostinelli, S. and others",
     collaboration = "GEANT4",
     title     = "{GEANT4: a simulation toolkit}",
     journal   = "Nucl. Instrum. Meth. A",
     volume    = "506",
     year      = "2003",
     pages     = "250-303",
     doi="10.1016/S0168-9002(03)01368-8"
}

@Article{Pythia8,
      author         = "Sj{\"o}strand, Torbjorn and Mrenna, Stephen and Skands, Peter
                        Z.",
      title          = "{A brief introduction to PYTHIA 8.1}",
      journal        = "Comput. Phys. Commun.",
      volume         = 178,
      pages          = "852-867",
      doi            = "10.1016/j.cpc.2008.01.036",
      year           = 2008,
      eprint         = "0710.3820",
      archivePrefix  = "arXiv",
      primaryClass   = "hep-ph",
}

@article{Jager:2015hka,
    author = "Jager, Barbara and Reina, Laura and Wackeroth, Doreen",
    title = "{Higgs boson production in association with b jets in the POWHEG BOX}",
    eprint = "1509.05843",
    archivePrefix = "arXiv",
    primaryClass = "hep-ph",
    doi = "10.1103/PhysRevD.93.014030",
    journal = "Phys. Rev. D",
    volume = "93",
    number = "1",
    pages = "014030",
    year = "2016"
}

@article{GOLONKA2006818,
title = "The tauola-photos-F environment for the TAUOLA and PHOTOS packages, release II",
journal = "Comput. Phys. Commun.",
volume = "174",
number = "10",
pages = "818 - 835",
year = "2006",
%issn = "0010-4655",
doi = "https://doi.org/10.1016/j.cpc.2005.12.018",
%url = "http://www.sciencedirect.com/science/article/pii/S0010465506000403",
author = "P. Golonka and B. Kersevan and T. Pierzchała and E. Richter-Wa̧s and Z. Wa̧s and M. Worek",
keywords = "Particle physics, Monte Carlo methods, Tau decays, TAUOLA, PHOTOS",
}

@article{ILTEN201477,
title = "Tau Decays in Pythia 8",
journal = "Nucl. Phys. Proc. Suppl.",
volume = "253-255",
pages = "77 - 80",
year = "2014",
note = "The Twelfth International Workshop on Tau-Lepton Physics (TAU2012)",
%issn = "0920-5632",
doi = "https://doi.org/10.1016/j.nuclphysbps.2014.09.019",
%url = "http://www.sciencedirect.com/science/article/pii/S0920563214001509",
author = "Philip Ilten",
keywords = "Monte Carlo, tau decays, polarization, hadronic currents",
}

@article{collinear_approximation,
title = "{Higgs decay to \(\tau^+\tau^-\): A possible signature of intermediate mass Higgs bosons at high energy hadron colliders}",
journal = "Nucl. Phys. B",
volume = "297",
number = "2",
pages = "{221-243}",
year = "1988",
doi = "10.1016/0550-3213(88)90019-3",
author = "R.K. Ellis and I. Hinchliffe and M. Soldate and J.J. {van der Bij}",
}

@misc{stxs1.1,
      title={Simplified Template Cross Sections - Stage 1.1}, 
      author={Nicolas Berger and Claudia Bertella and Thomas P. Calvet and Milene Calvetti and Valerio Dao and Marco Delmastro and Michael Duehrssen-Debling and Paolo Francavilla and Yacine Haddad and Oleh Kivernyk and Jonathon M. Langford and Changqiao Li and Giovanni Marchiori and Predrag Milenovic and Carlo E. Pandini and Edward Scott and Frank J. Tackmann and Kerstin Tackmann and Lorenzo Viliani and Meng Xiao},
      year={2019},
      eprint={1906.02754},
      archivePrefix={arXiv},
      primaryClass={hep-ph}
}

@article{demartin2015higgs,
      title={Higgs production in association with a single top quark at the LHC}, 
      author={Federico Demartin and Fabio Maltoni and Kentarou Mawatari and Marco Zaro},
      journal="Eur. Phys. J. C",
      volume="75",
      pages="267",
      year={2015},
      eprint={1504.00611},
      archivePrefix={arXiv},
      primaryClass={hep-ph},
      doi="10.1140/epjc/s10052-015-3475-9"
}

@misc{bendavid2018les,
      title={Les Houches 2017: Physics at TeV Colliders Standard Model Working Group Report}, 
      author={J. Bendavid and F. Caola and V. Ciulli and R. Harlander and G. Heinrich and J. Huston and S. Kallweit and S. Prestel and E. Re and K. Tackmann and J. Thaler and K. Theofilatos and J. R. Andersen and J. Bellm and N. Berger and D. Bhatia and B. Biedermann and S. Bräuer and D. Britzger and A. G. Buckley and R. Camacho and G. Chachamis and S. Chatterjee and X. Chen and M. Chiesa and J. R. Currie and A. Denner and F. Dreyer and F. Driencourt-Mangin and S. Forte and M. V. Garzelli and T. Gehrmann and S. Gieseke and E. W. N. Glover and P. Gras and N. Greiner and C. Gütschow and C. Gwenlan and M. Heil and M. Herndon and V. Hirschi and A. H. Hoang and S. Höche and A. Huss and S. P. Jones and D. Kar and A. Karlberg and Z. Kassabov and M. Kerner and J. Klappert and S. Kuttimalai and J. -N. Lang and A. Larkoski and J. M. Lindert and P. Loch and K. Long and L. Lönnblad and G. Luisoni and A. Maier and P. Maierhöfer and D. Maître and S. Marzani and J. A. McFayden and I. Moult and M. Mozer and S. Mrenna and B. Nachman and D. Napoletano and C. Pandini and A. Papaefstathiou and M. Pellen and L. Perrozzi and J. Pires and S. Plätzer and S. Pozzorini and S. Quackenbush and K. Rabbertz and M. Rauch and C. Reuschle and P. Richardson and A. Gehrmann-De Ridder and G. Rodrigo and J. Rojo and R. Röntsch and L. Rottoli and D. Samitz and T. Samui and G. Sborlini and M. Schönherr and S. Schumann and L. Scyboz and S. Seth and H. -S. Shao and A. Siódmok and P. Z. Skands and J. M. Smillie and G. Soyez and P. Sun and M. R. Sutton and F. J. Tackmann and S. Uccirati and S. Weinzierl and E. Yazgan and C. -P. Yuan and F. Yuan},
      year={2018},
      eprint={1803.07977},
      archivePrefix={arXiv},
      primaryClass={hep-ph}
}

@article{PhysRevD.87.093008,
  title = {Next-to-leading-order uncertainties in $\mathrm{\text{Higgs}}\mathbf{+}2$ jets from gluon fusion},
  author = {Gangal, Shireen and Tackmann, Frank J.},
  journal = {Phys. Rev. D},
  volume = {87},
%  issue = {9},
  pages = {093008},
  numpages = {14},
  year = {2013},
  month = {5},
  publisher = {American Physical Society},
  doi = {10.1103/PhysRevD.87.093008},
%  url = {https://link.aps.org/doi/10.1103/PhysRevD.87.093008},
  eprint        = "1302.5437",
  archivePrefix = "arXiv",
  primaryClass = "hep-ph",
}

@article{article,
author = {Cascioli, Fabio and Hoeche, Stefan and Krauss, Frank and Maierhofer, Philipp and Pozzorini, Stefano and Siegert, Frank},
year = {2014},
month = {09},
pages = {},
title = {Precise Higgs-background predictions: Merging NLO QCD and squared quark-loop corrections to four-lepton + 0,1 jet production},
volume = {046},
journal = {JHEP},
doi = {10.1007/JHEP01(2014)046}
}

@article{id_trigger,
    author         = "{ATLAS Collaboration}",
    title = "{The ATLAS Inner Detector Trigger performance in pp collisions at 13 TeV during LHC Run 2}",
    eprint = "2107.02485",
    archivePrefix = "arXiv",
    primaryClass = "hep-ex",
    reportNumber = "CERN-EP-2021-076",
    month = "7",
    year = "2021"
}

@article{l1topo_trigger,
    author         = "{ATLAS Collaboration}",
    title = "{Performance of the ATLAS Level-1 topological trigger in Run 2}",
    eprint = "2105.01416",
    archivePrefix = "arXiv",
    primaryClass = "hep-ex",
    reportNumber = "CERN-EP-2021-040",
    month = "5",
    year = "2021"
}

@Article{CMS-HIG-20-015,
    author         = "{CMS Collaboration}",
    title          = "{Measurement of the inclusive and differential Higgs boson production cross sections in the decay mode to a pair of $\tau$ leptons in pp collisions at $\sqrt{s} = $ 13 TeV}",
    year           = "2021",
    reportNumber   = "CERN-EP-2021-134",
    eprint         = "2107.11486",
    archivePrefix  = "arXiv",
    primaryClass   = "hep-ex",
}

@Booklet{hepdata,
    author         = "{ATLAS Collaboration}",
    title          = "{Measurements of Higgs boson production cross-sections in the $H\to\tau^{+}\tau^{-}$ decay channel in $pp$ collisions at $\sqrt{s}=13\,\text{TeV}$ with the ATLAS detector}",
    howpublished   = "{HEPData}",
    url            = "https://www.hepdata.net/record/115994",
    year           = "2022",
}

@Article{Evans:2008zzb,
      author         = "Evans, Lyndon and Bryant, Philip",
      title          = "{LHC Machine}",
      journal        = "JINST",
      volume         = "3",
      pages          = "S08001",
      doi            = "10.1088/1748-0221/3/08/S08001",
      year           = "2008",
      SLACcitation   = "%%CITATION = JINST,3,S08001;%%",
}

@Article{Cacciari:2008gp,
     author    = "Cacciari, Matteo and Salam, Gavin P. and Soyez, Gregory",
     title     = "{The anti-\(k_{t}\) jet clustering algorithm}",
     journal   = "JHEP",
     volume    = "04",
     year      = "2008",
     pages     = "063",
     eprint    = "0802.1189",
     archivePrefix = "arXiv",
     primaryClass  =  "hep-ph",
     doi       = "10.1088/1126-6708/2008/04/063",
     SLACcitation  = "%%CITATION = 0802.1189;%%"
}

@Article{Butterworth:2015oua,
      author         = "Butterworth, Jon and others",
      title          = "{PDF4LHC recommendations for LHC Run II}",
      journal        = "J. Phys. G",
      volume         = "43",
      year           = "2016",
      pages          = "023001",
      doi            = "10.1088/0954-3899/43/2/023001",
      eprint         = "1510.03865",
      archivePrefix  = "arXiv",
      primaryClass   = "hep-ph",
      reportNumber   = "OUTP-15-17P, SMU-HEP-15-12, TIF-UNIMI-2015-14,
                        LCTS-2015-27, CERN-PH-TH-2015-249",
      SLACcitation   = "%%CITATION = ARXIV:1510.03865;%%"
}

@article{Lai:2010vv,
      author         = "Lai, H.-L. and others",
      title          = "{New parton distributions for collider physics}",
      journal        = "Phys. Rev. D",
      volume         = "82",
      pages          = "074024",
      doi            = "10.1103/PhysRevD.82.074024",
      year           = "2010",
      eprint         = "1007.2241",
      archivePrefix  = "arXiv",
      primaryClass   = "hep-ph",
      reportNumber   = "MSUHEP-100707, SMU-HEP-10-10",
      SLACcitation   = "%%CITATION = ARXIV:1007.2241;%%",
}

@article{Dulat:2015mca,
      author         = "Dulat, Sayipjamal and Hou, Tie-Jiun and Gao, Jun and
                        Guzzi, Marco and Huston, Joey and Nadolsky, Pavel and
                        Pumplin, Jon and Schmidt, Carl and Stump, Daniel and Yuan,
                        C. P.",
      title          = "{New parton distribution functions from a global analysis
                        of quantum chromodynamics}",
      journal        = "Phys. Rev. D",
      volume         = "93",
      year           = "2016",
      number         = "3",
      pages          = "033006",
      doi            = "10.1103/PhysRevD.93.033006",
      eprint         = "1506.07443",
      archivePrefix  = "arXiv",
      primaryClass   = "hep-ph",
      SLACcitation   = "%%CITATION = ARXIV:1506.07443;%%"
}

@article{Harland-Lang:2014zoa,
      author         = "Harland-Lang, L. A. and Martin, A. D. and Motylinski, P.
                        and Thorne, R. S.",
      title          = "{Parton distributions in the LHC era: MMHT 2014 PDFs}",
      journal        = "Eur. Phys. J. C",
      number         = "5",
      volume         = "75",
      pages          = "204",
      doi            = "10.1140/epjc/s10052-015-3397-6",
      year           = "2015",
      eprint         = "1412.3989",
      archivePrefix  = "arXiv",
      primaryClass   = "hep-ph",
      reportNumber   = "LCTS-2014-47, IPPP-14-97, DCPT-14-194",
      SLACcitation   = "%%CITATION = ARXIV:1412.3989;%%",
}

@article{Ball:2012cx,
      author         = "{NNPDF Collaboration} and Ball, Richard D. and others",
      title          = "{Parton distributions with LHC data}",
      journal        = "Nucl. Phys. B",
      volume         = "867",
      year           = "2013",
      pages          = "244",
      doi            = "10.1016/j.nuclphysb.2012.10.003",
      eprint         = "1207.1303",
      archivePrefix  = "arXiv",
      primaryClass   = "hep-ph",
      reportNumber   = "EDINBURGH-2012-08, IFUM-FT-997, FR-PHENO-2012-014,
                        RWTH-TTK-12-25, CERN-PH-TH-2012-037, SFB-CPP-12-47\,
                        --CERN-PH-TH-2012-037",
      SLACcitation   = "%%CITATION = ARXIV:1207.1303;%%"
}

@article{Ball:2014uwa,
      author         = "{The NNPDF Collaboration} and Ball, Richard D. and others",
      title          = "{Parton distributions for the LHC run II}",
      journal        = "JHEP",
      volume         = "04",
      year           = "2015",
      pages          = "040",
      doi            = "10.1007/JHEP04(2015)040",
      eprint         = "1410.8849",
      archivePrefix  = "arXiv",
      primaryClass   = "hep-ph",
      reportNumber   = "EDINBURGH-2014-15, IFUM-1034-FT, CERN-PH-TH-2013-253,
                        OUTP-14-11P, CAVENDISH-HEP-14-11",
      SLACcitation   = "%%CITATION = ARXIV:1410.8849;%%"
}

@article{Czakon:2012pz,
      author         = "Czakon, Michal and Mitov, Alexander",
      title          = "{NNLO corrections to top pair production at hadron
                        colliders: the quark-gluon reaction}",
      journal        = "JHEP",
      volume         = "01",
      pages          = "080",
      doi            = "10.1007/JHEP01(2013)080",
      year           = "2013",
      eprint         = "1210.6832",
      archivePrefix  = "arXiv",
      primaryClass   = "hep-ph",
      SLACcitation   = "%%CITATION = ARXIV:1210.6832;%%",
}

@article{Sjostrand:2014zea,
      author         = "Sj{\"o}strand, Torbj{\"o}rn and Ask, Stefan and Christiansen,
                        Jesper R. and Corke, Richard and Desai, Nishita and Ilten,
                        Philip and Mrenna, Stephen and Prestel, Stefan and
                        Rasmussen, Christine O. and Skands, Peter Z.",
      title          = "{An introduction to PYTHIA 8.2}",
      journal        = "Comput. Phys. Commun.",
      volume         = "191",
      year           = "2015",
      pages          = "159",
      doi            = "10.1016/j.cpc.2015.01.024",
      eprint         = "1410.3012",
      archivePrefix  = "arXiv",
      primaryClass   = "hep-ph",
      reportNumber   = "LU-TP-14-36, MCNET-14-22, CERN-PH-TH-2014-190,
                        FERMILAB-PUB-14-316-CD, DESY-14-178, SLAC-PUB-16122,
                        --FERMILAB-PUB-14-316-CD",
      SLACcitation   = "%%CITATION = ARXIV:1410.3012;%%"
}

@Article{Lange:2001uf,
      author         = "Lange, D. J.",
      title          = "{The EvtGen particle decay simulation package}",
      booktitle      = "{Proceedings, 7th International Conference on B physics
                        at hadron machines (BEAUTY 2000)}",
      journal        = "Nucl. Instrum. Meth. A",
      volume         = "462",
      year           = "2001",
      pages          = "152",
      doi            = "10.1016/S0168-9002(01)00089-4",
      SLACcitation   = "%%CITATION = NUIMA,A462,152;%%"
}

@Article{Alwall:2014hca,
      author         = "Alwall, J. and Frederix, R. and Frixione, S. and Hirschi,
                        V. and Maltoni, F. and Mattelaer, O. and Shao, H. -S. and
                        Stelzer, T. and Torrielli, P. and Zaro, M.",
      title          = "{The automated computation of tree-level and
                        next-to-leading order differential cross sections, and
                        their matching to parton shower simulations}",
      journal        = "JHEP",
      volume         = "07",
      year           = "2014",
      pages          = "079",
      doi            = "10.1007/JHEP07(2014)079",
      eprint         = "1405.0301",
      archivePrefix  = "arXiv",
      primaryClass   = "hep-ph",
      reportNumber   = "CERN-PH-TH-2014-064, CP3-14-18, LPN14-066, MCNET-14-09,
                        ZU-TH-14-14",
      SLACcitation   = "%%CITATION = ARXIV:1405.0301;%%"
}

@article{Frederix:2012ps,
      author         = "Frederix, Rikkert and Frixione, Stefano",
      title          = "{Merging meets matching in MC@NLO}",
      journal        = "JHEP",
      volume         = "12",
      year           = "2012",
      pages          = "061",
      doi            = "10.1007/JHEP12(2012)061",
      eprint         = "1209.6215",
      archivePrefix  = "arXiv",
      primaryClass   = "hep-ph",
      reportNumber   = "CERN-PH-TH-2012-247, ZU-TH-21-12",
      SLACcitation   = "%%CITATION = ARXIV:1209.6215;%%"
}

@Article{Bahr:2008pv,
      author         = "B{\"a}hr, M. and others",
      title          = "{Herwig++ physics and manual}",
      journal        = "Eur. Phys. J. C",
      volume         = "58",
      year           = "2008",
      pages          = "639",
      doi            = "10.1140/epjc/s10052-008-0798-9",
      eprint         = "0803.0883",
      archivePrefix  = "arXiv",
      primaryClass   = "hep-ph",
      reportNumber   = "CERN-PH-TH-2008-038, CAVENDISH-HEP-08-03, KA-TP-05-2008,
                        DCPT-08-22, IPPP-08-11, CP3-08-05",
      SLACcitation   = "%%CITATION = ARXIV:0803.0883;%%"
}

@Article{Bellm:2015jjp,
      author         = "Bellm, Johannes and others",
      title          = "{Herwig 7.0/Herwig++ 3.0 release note}",
      journal        = "Eur. Phys. J. C",
      volume         = "76",
      year           = "2016",
      number         = "4",
      pages          = "196",
      doi            = "10.1140/epjc/s10052-016-4018-8",
      eprint         = "1512.01178",
      archivePrefix  = "arXiv",
      primaryClass   = "hep-ph",
      reportNumber   = "CERN-PH-TH-2015-289, MAN-HEP-2015-15, IFJPAN-IV-2015-13,
                        HERWIG-2015-01, KA-TP-18-2015, DCPT-15-142, MCNET-15-28,
                        IPPP-15-71, --HERWIG-2015-01",
      SLACcitation   = "%%CITATION = ARXIV:1512.01178;%%"
}

@Article{Nason:2004rx,
      author         = "Nason, Paolo",
      title          = "{A new method for combining NLO QCD with shower Monte Carlo algorithms}",
      journal        = "JHEP",
      volume         = "11",
      pages          = "040",
      doi            = "10.1088/1126-6708/2004/11/040",
      year           = "2004",
      eprint         = "hep-ph/0409146",
      archivePrefix  = "arXiv",
}

@Article{Frixione:2007vw,
      author         = "Frixione, Stefano and Nason, Paolo and Oleari, Carlo",
      title          = "{Matching NLO QCD computations with parton shower
                        simulations: the POWHEG method}",
      journal        = "JHEP",
      volume         = "11",
      pages          = "070",
      doi            = "10.1088/1126-6708/2007/11/070",
      year           = "2007",
      eprint         = "0709.2092",
      archivePrefix  = "arXiv",
      primaryClass   = "hep-ph",
}

@Article{Alioli:2010xd,
      author         = "Alioli, Simone and Nason, Paolo and Oleari, Carlo and Re,
                        Emanuele",
      title          = "{A general framework for implementing NLO calculations in
                        shower Monte Carlo programs: the POWHEG BOX}",
      journal        = "JHEP",
      volume         = "06",
      pages          = "043",
      doi            = "10.1007/JHEP06(2010)043",
      year           = "2010",
      eprint         = "1002.2581",
      archivePrefix  = "arXiv",
      primaryClass   = "hep-ph",
}

@article{Hamilton:2012rf,
      author         = "Hamilton, Keith and Nason, Paolo and Oleari, Carlo and
                        Zanderighi, Giulia",
      title          = "{Merging H/W/Z + 0 and 1 jet at NLO with no merging
                        scale: a path to parton shower + NNLO matching}",
      journal        = "JHEP",
      volume         = "05",
      year           = "2013",
      pages          = "082",
      doi            = "10.1007/JHEP05(2013)082",
      eprint         = "1212.4504",
      archivePrefix  = "arXiv",
      primaryClass   = "hep-ph",
      reportNumber   = "CERN-PH-TH-2012-356",
      SLACcitation   = "%%CITATION = ARXIV:1212.4504;%%"
}

@article{Hamilton:2012np,
      author         = "Hamilton, Keith and Nason, Paolo and Zanderighi, Giulia",
      title          = "{MINLO: multi-scale improved NLO}",
      journal        = "JHEP",
      volume         = "10",
      year           = "2012",
      pages          = "155",
      doi            = "10.1007/JHEP10(2012)155",
      eprint         = "1206.3572",
      archivePrefix  = "arXiv",
      primaryClass   = "hep-ph",
      reportNumber   = "CERN-PH-TH-2012-166, OUTP-12-11P, MCNET-12-07",
      SLACcitation   = "%%CITATION = ARXIV:1206.3572;%%"
}

@article{Hartanto:2015uka,
      author         = "Hartanto, Heribertus B. and J{\"a}ger, Barbara and Reina,
                        Laura and Wackeroth, Doreen",
      title          = "{Higgs boson production in association with top quarks in
                        the POWHEG BOX}",
      journal        = "Phys. Rev. D",
      volume         = "91",
      year           = "2015",
      number         = "9",
      pages          = "094003",
      doi            = "10.1103/PhysRevD.91.094003",
      eprint         = "1501.04498",
      archivePrefix  = "arXiv",
      primaryClass   = "hep-ph",
      SLACcitation   = "%%CITATION = ARXIV:1501.04498;%%"
}

@article{Aliev:2010zk,
      author         = "Aliev, M. and Lacker, H. and Langenfeld, U. and Moch, S.
                        and Uwer, P. and Wiedermann, M.",
      title          = "{HATHOR -- HAdronic Top and Heavy quarks crOss section
                        calculatoR}",
      journal        = "Comput. Phys. Commun.",
      volume         = "182",
      year           = "2011",
      pages          = "1034-1046",
      doi            = "10.1016/j.cpc.2010.12.040",
      eprint         = "1007.1327",
      archivePrefix  = "arXiv",
      primaryClass   = "hep-ph",
      reportNumber   = "DESY-10-091, HU-EP-10-33, SFB-CPP-10-60",
      SLACcitation   = "%%CITATION = ARXIV:1007.1327;%%"
}

@article{Kant:2014oha,
      author         = "Kant, P. and Kind, O. M. and Kintscher, T. and Lohse, T.
                        and Martini, T. and Mölbitz, S. and Rieck, P. and Uwer, P.",
      title          = "{HatHor for single top-quark production: Updated
                        predictions and uncertainty estimates for single top-quark
                        production in hadronic collisions}",
      journal        = "Comput. Phys. Commun.",
      volume         = "191",
      year           = "2015",
      pages          = "74-89",
      doi            = "10.1016/j.cpc.2015.02.001",
      eprint         = "1406.4403",
      archivePrefix  = "arXiv",
      primaryClass   = "hep-ph",
      reportNumber   = "HU-EP-14-22",
      SLACcitation   = "%%CITATION = ARXIV:1406.4403;%%"
}

@article{Beneke:2011mq,
      author         = "Beneke, M. and Falgari, P. and Klein, S. and Schwinn, C.",
      title          = "{Hadronic top-quark pair production with NNLL threshold
                        resummation}",
      journal        = "Nucl. Phys. B",
      volume         = "855",
      year           = "2012",
      pages          = "695-741",
      doi            = "10.1016/j.nuclphysb.2011.10.021",
      eprint         = "1109.1536",
      archivePrefix  = "arXiv",
      primaryClass   = "hep-ph",
      reportNumber   = "TTK-11-38, ITP-UU-11-26, SPIN-11-19, FR-PHENO-2011-015,
                        SFB-CPP-11-49",
      SLACcitation   = "%%CITATION = ARXIV:1109.1536;%%"
}

@article{Anastasiou:2003ds,
      author         = "Anastasiou, Charalampos and Dixon, Lance and Melnikov,
                        Kirill and Petriello, Frank",
      title          = "{High-precision QCD at hadron colliders: Electroweak
                        gauge boson rapidity distributions at next-to-next-to leading order}",
      journal        = "Phys. Rev. D",
      volume         = "69",
      year           = "2004",
      pages          = "094008",
      doi            = "10.1103/PhysRevD.69.094008",
      eprint         = "hep-ph/0312266",
      archivePrefix  = "arXiv",
      reportNumber   = "SLAC-PUB-10288, UH-511-1042-03",
      SLACcitation   = "%%CITATION = HEP-PH/0312266;%%"
}

@article{Cacciari:2011hy,
      author         = "Cacciari, Matteo and Czakon, Michal and Mangano,
                        Michelangelo and Mitov, Alexander and Nason, Paolo",
      title          = "{Top-pair production at hadron colliders with
                        next-to-next-to-leading logarithmic soft-gluon
                        resummation}",
      journal        = "Phys. Lett. B",
      volume         = "710",
      pages          = "612-622",
      doi            = "10.1016/j.physletb.2012.03.013",
      year           = "2012",
      eprint         = "1111.5869",
      primaryClass   = "hep-ph",
      archivePrefix  = "arXiv",
      reportNumber   = "CERN-PH-TH-2011-277, TTK-11-54",
      SLACcitation   = "%%CITATION = ARXIV:1111.5869;%%",
}

@article{Czakon:2012zr,
      author         = "Czakon, Michal and Mitov, Alexander",
      title          = "{NNLO corrections to top-pair production at hadron
                        colliders: the all-fermionic scattering channels}",
      journal        = "JHEP",
      volume         = "12",
      pages          = "054",
      doi            = "10.1007/JHEP12(2012)054",
      year           = "2012",
      eprint         = "1207.0236",
      archivePrefix  = "arXiv",
      primaryClass   = "hep-ph",
      SLACcitation   = "%%CITATION = ARXIV:1207.0236;%%",
}

@article{Czakon:2013goa,
      author         = "Czakon, Michal and Fiedler, Paul and Mitov, Alexander",
      title          = "{Total Top-Quark Pair-Production Cross Section at
                        Hadron Colliders Through \(O(\alpha_S^4)\)}",
      journal        = "Phys. Rev. Lett.",
      volume         = "110",
      pages          = "252004",
      doi            = "10.1103/PhysRevLett.110.252004",
      year           = "2013",
      eprint         = "1303.6254",
      archivePrefix  = "arXiv",
      primaryClass   = "hep-ph",
      reportNumber   = "CERN-PH-TH-2013-056, TTK-13-08",
      SLACcitation   = "%%CITATION = ARXIV:1303.6254;%%",
}

@article{Baernreuther:2012ws,
      author         = "B{\"a}rnreuther, Peter and Czakon, Michal and Mitov,
                        Alexander",
      title          = "{Percent-Level-Precision Physics at the
                  Tevatron: Next-to-Next-to-Leading Order QCD
                  Corrections to \(q \bar{q} \to t \bar{t} + X\)}",
      journal        = "Phys. Rev. Lett.",
      volume         = "109",
      pages          = "132001",
      doi            = "10.1103/PhysRevLett.109.132001",
      year           = "2012",
      eprint         = "1204.5201",
      archivePrefix  = "arXiv",
      primaryClass   = "hep-ph",
      SLACcitation   = "%%CITATION = ARXIV:1204.5201;%%",
}

@article{Frixione:2007nw,
      author         = "Frixione, Stefano and Ridolfi, Giovanni and Nason, Paolo",
      title          = "{A positive-weight next-to-leading-order Monte Carlo for
                        heavy flavour hadroproduction}",
      journal        = "JHEP",
      volume         = "09",
      pages          = "126",
      doi            = "10.1088/1126-6708/2007/09/126",
      year           = "2007",
      eprint         = "0707.3088",
      archivePrefix  = "arXiv",
      primaryClass   = "hep-ph",
      SLACcitation   = "%%CITATION = ARXIV:0707.3088;%%",
}

@article{deFlorian:2016spz,
      author         = "de Florian, D. and others",
      title          = "{Handbook of LHC Higgs Cross Sections: 4. Deciphering the
                        Nature of the Higgs Sector}",
      collaboration  = "LHC Higgs Cross Section Working Group",
      doi            = "10.23731/CYRM-2017-002",
      year           = "2016",
      eprint         = "1610.07922",
      archivePrefix  = "arXiv",
      primaryClass   = "hep-ph",
      reportNumber   = "FERMILAB-FN-1025-T, CERN-2017-002-M",
      SLACcitation   = "%%CITATION = ARXIV:1610.07922;%%"
}

@Article{Hoeche:2009rj,
      author         = "H{\"o}che, Stefan and Krauss, Frank and Schumann, Steffen and Siegert, Frank",
      title          = "{QCD matrix elements and truncated showers}",
      journal        = "JHEP",
      volume         = "05",
      pages          = "053",
      doi            = "10.1088/1126-6708/2009/05/053",
      year           = "2009",
      eprint         = "0903.1219",
      archivePrefix  = "arXiv",
      primaryClass   = "hep-ph",
}

@Article{Gleisberg:2008fv,
      author         = "Gleisberg, Tanju and H{\"o}che, Stefan",
      title          = "{Comix, a new matrix element generator}",
      journal        = "JHEP",
      volume         = "12",
      pages          = "039",
      doi            = "10.1088/1126-6708/2008/12/039",
      year           = "2008",
      eprint         = "0808.3674",
      archivePrefix  = "arXiv",
      primaryClass   = "hep-ph",
}

@Article{Schumann:2007mg,
      author         = "Schumann, Steffen and Krauss, Frank",
      title          = "{A parton shower algorithm based on Catani--Seymour dipole factorisation}",
      journal        = "JHEP",
      volume         = "03",
      pages          = "038",
      doi            = "10.1088/1126-6708/2008/03/038",
      year           = "2008",
      eprint         = "0709.1027",
      archivePrefix  = "arXiv",
      primaryClass   = "hep-ph",
}

@article{Hoeche:2012yf,
      author         = "H{\"o}che, Stefan and Krauss, Frank and Sch{\"o}nherr, Marek and
                        Siegert, Frank",
      title          = "{QCD matrix elements + parton showers. The NLO case}",
      journal        = "JHEP",
      volume         = "04",
      year           = "2013",
      pages          = "027",
      doi            = "10.1007/JHEP04(2013)027",
      eprint         = "1207.5030",
      archivePrefix  = "arXiv",
      primaryClass   = "hep-ph",
      reportNumber   = "SLAC-PUB-15191, IPPP-12-52, DCPT-12-104, LPN12-081,
                        FR-PHENO-2012-017, MCNET-12-09, --FR-PHENO-2012-017",
      SLACcitation   = "%%CITATION = ARXIV:1207.5030;%%"
}

@article{Bothmann:2016nao,
      author         = "Bothmann, Enrico and Sch{\"o}nherr, Marek and Schumann, Steffen",
      title          = "{Reweighting QCD matrix-element and parton-shower calculations}",
      journal        = "Eur. Phys. J. C",
      volume         = "76",
      year           = "2016",
      number         = "11",
      pages          = "590",
      doi            = "10.1140/epjc/s10052-016-4430-0",
      eprint         = "1606.08753",
      archivePrefix  = "arXiv",
      primaryClass   = "hep-ph",
      reportNumber   = "MCNET-16-22, ZU-TH-21-16, ZU--TH--21-16",
      SLACcitation   = "%%CITATION = ARXIV:1606.08753;%%"
}

@article{Bothmann:2019yzt,
      author = "Bothmann, Enrico and others",
      title = "{Event generation with Sherpa 2.2}",
      journal = "SciPost Phys.",
      volume = "7",
      year = "2019",
      number = "3",
      pages = "034",
      doi = "10.21468/SciPostPhys.7.3.034",
      reportNumber = "FERMILAB-PUB-19-218-T, SLAC-PUB-17433, IPPP/19/42, MCNET-19-11",
      eprint = "1905.09127",
      archivePrefix = "arXiv",
      primaryClass = "hep-ph",
}

@article{Hoeche:2011fd,
      author         = "H{\"o}che, Stefan and Krauss, Frank and Sch{\"o}nherr, Marek and
                        Siegert, Frank",
      title          = "{A critical appraisal of NLO+PS matching methods}",
      journal        = "JHEP",
      volume         = "09",
      year           = "2012",
      pages          = "049",
      doi            = "10.1007/JHEP09(2012)049",
      eprint         = "1111.1220",
      archivePrefix  = "arXiv",
      primaryClass   = "hep-ph",
      reportNumber   = "SLAC-PUB-14661, IPPP-11-67, DCPT-11-134, LPN11-58,
                        FR-PHENO-2011-019, MCNET-11-24",
      SLACcitation   = "%%CITATION = ARXIV:1111.1220;%%"
}

@article{Catani:2001cc,
      author         = "Catani, S. and Krauss, F. and Webber, B. R. and Kuhn, R.",
      title          = "{QCD Matrix Elements + Parton Showers}",
      journal        = "JHEP",
      volume         = "11",
      year           = "2001",
      pages          = "063",
      doi            = "10.1088/1126-6708/2001/11/063",
      eprint         = "hep-ph/0109231",
      archivePrefix  = "arXiv",
      reportNumber   = "CERN-TH-2000-367, CAVENDISH-HEP-00-03",
      SLACcitation   = "%%CITATION = HEP-PH/0109231;%%"
}

@article{Cascioli:2011va,
      author         = "Cascioli, Fabio and Maierh{\"o}fer, Philipp and Pozzorini, Stefano",
      title          = "{Scattering Amplitudes with Open Loops}",
      journal        = "Phys. Rev. Lett.",
      volume         = "108",
      year           = "2012",
      pages          = "111601",
      doi            = "10.1103/PhysRevLett.108.111601",
      eprint         = "1111.5206",
      archivePrefix  = "arXiv",
      primaryClass   = "hep-ph",
      reportNumber   = "ZU-TH-23-11, LPN11-66",
      SLACcitation   = "%%CITATION = ARXIV:1111.5206;%%"
}

@article{Buccioni:2019sur,
    author = "Buccioni, Federico and Lang, Jean-Nicolas and Lindert, Jonas M. and Maierh{\"o}fer, Philipp and Pozzorini, Stefano and Zhang, Hantian and Zoller, Max F.",
    title = "{OpenLoops 2}",
    eprint = "1907.13071",
    archivePrefix = "arXiv",
    primaryClass = "hep-ph",
    reportNumber = "IPPP/19/62, FR-PHENO-2019-12, PSI-PR-19-15, ZU-TH 37/19",
    doi = "10.1140/epjc/s10052-019-7306-2",
    journal = "Eur. Phys. J. C",
    volume = "79",
    number = "10",
    pages = "866",
    year = "2019"
}

@article{Denner:2016kdg,
      author         = "Denner, Ansgar and Dittmaier, Stefan and Hofer, Lars",
      title          = "{\textsc{Collier}: A fortran-based complex one-loop library in
                        extended regularizations}",
      journal        = "Comput. Phys. Commun.",
      volume         = "212",
      year           = "2017",
      pages          = "220-238",
      doi            = "10.1016/j.cpc.2016.10.013",
      eprint         = "1604.06792",
      archivePrefix  = "arXiv",
      primaryClass   = "hep-ph",
      reportNumber   = "FR-PHENO-2016-003, ICCUB-16-016",
      SLACcitation   = "%%CITATION = ARXIV:1604.06792;%%"
}

@article{Czakon:2011xx,
  author         = "Czakon, Michal and Mitov, Alexander",
  title          = "{Top++: A program for the calculation of the top-pair cross-section at hadron colliders}",
  journal        = "Comput. Phys. Commun.",
  volume         = "185",
  year           = "2014",
  pages          = "2930",
  doi            = "10.1016/j.cpc.2014.06.021",
  eprint         = "1112.5675",
  archivePrefix  = "arXiv",
  primaryClass   = "hep-ph",
  reportNumber   = "CERN-PH-TH-2011-303, TTK-11-58",
}

@article{Campbell:2012am,
      author         = "Campbell, John M. and Ellis, R. Keith and Frederix,
                        Rikkert and Nason, Paolo and Oleari, Carlo and Williams,
                        Ciaran",
      title          = "{NLO Higgs boson production plus one and two jets using
                        the POWHEG BOX, MadGraph4 and MCFM}",
      journal        = "JHEP",
      volume         = "07",
      year           = "2012",
      pages          = "092",
      doi            = "10.1007/JHEP07(2012)092",
      eprint         = "1202.5475",
      archivePrefix  = "arXiv",
      primaryClass   = "hep-ph",
      reportNumber   = "FERMILAB-PUB-12-040-T, CERN-PH-TH-2012-048",
      SLACcitation   = "%%CITATION = ARXIV:1202.5475;%%"
}

@article{Catani:2007vq,
      author         = "Catani, Stefano and Grazzini, Massimiliano",
      title          = "{Next-to-Next-to-Leading-Order Subtraction Formalism in Hadron Collisions and
                        its Application to Higgs-boson Production at the Large Hadron Collider}",
      journal        = "Phys. Rev. Lett.",
      volume         = "98",
      year           = "2007",
      pages          = "222002",
      doi            = "10.1103/PhysRevLett.98.222002",
      eprint         = "hep-ph/0703012",
      archivePrefix  = "arXiv",
      primaryClass   = "hep-ph",
      SLACcitation   = "%%CITATION = HEP-PH/0703012;%%"
}

@article{Anastasiou:2016cez,
      author         = "Anastasiou, Charalampos and Duhr, Claude and Dulat, Falko
                        and Furlan, Elisabetta and Gehrmann, Thomas and Herzog,
                        Franz and Lazopoulos, Achilleas and Mistlberger, Bernhard",
      title          = "{High precision determination of the gluon fusion Higgs
                        boson cross-section at the LHC}",
      journal        = "JHEP",
      volume         = "05",
      year           = "2016",
      pages          = "058",
      doi            = "10.1007/JHEP05(2016)058",
      eprint         = "1602.00695",
      archivePrefix  = "arXiv",
      primaryClass   = "hep-ph",
      reportNumber   = "CP3-16-01, ZU-TH-27-15, NIKHEF-2016-004,
                        CERN-TH-2016-006",
      SLACcitation   = "%%CITATION = ARXIV:1602.00695;%%"
}

@article{Anastasiou:2015ema,
      author         = "Anastasiou, Charalampos and Duhr, Claude and Dulat, Falko
                        and Herzog, Franz and Mistlberger, Bernhard",
      title          = "{Higgs Boson Gluon-Fusion Production in QCD at Three
                        Loops}",
      journal        = "Phys. Rev. Lett.",
      volume         = "114",
      year           = "2015",
      pages          = "212001",
      doi            = "10.1103/PhysRevLett.114.212001",
      eprint         = "1503.06056",
      archivePrefix  = "arXiv",
      primaryClass   = "hep-ph",
      reportNumber   = "CERN-PH-TH-2015-055, CP3-15-07",
      SLACcitation   = "%%CITATION = ARXIV:1503.06056;%%"
}

@article{Dulat:2018rbf,
      author         = "Dulat, Falko and Lazopoulos, Achilleas and Mistlberger,
                        Bernhard",
      title          = "{iHixs 2 -- Inclusive Higgs cross sections}",
      journal        = "Comput. Phys. Commun.",
      volume         = "233",
      year           = "2018",
      pages          = "243-260",
      doi            = "10.1016/j.cpc.2018.06.025",
      eprint         = "1802.00827",
      archivePrefix  = "arXiv",
      primaryClass   = "hep-ph",
      reportNumber   = "CERN-TH-2018-019, SLAC-PUB-17222",
      SLACcitation   = "%%CITATION = ARXIV:1802.00827;%%"
}

@article{Harlander:2009mq,
      author         = "Harlander, Robert V. and Ozeren, Kemal J.",
      title          = "{Finite top mass effects for hadronic Higgs production at
                        next-to-next-to-leading order}",
      journal        = "JHEP",
      volume         = "11",
      year           = "2009",
      pages          = "088",
      doi            = "10.1088/1126-6708/2009/11/088",
      eprint         = "0909.3420",
      archivePrefix  = "arXiv",
      primaryClass   = "hep-ph",
      reportNumber   = "WUB-09-10",
      SLACcitation   = "%%CITATION = ARXIV:0909.3420;%%"
}

@article{Harlander:2009bw,
      author         = "Harlander, Robert V. and Ozeren, Kemal J.",
      title          = "{Top mass effects in Higgs production at
                        next-to-next-to-leading order QCD: Virtual corrections}",
      journal        = "Phys. Lett. B",
      volume         = "679",
      year           = "2009",
      pages          = "467-472",
      doi            = "10.1016/j.physletb.2009.08.012",
      eprint         = "0907.2997",
      archivePrefix  = "arXiv",
      primaryClass   = "hep-ph",
      reportNumber   = "WUB-09-08",
      SLACcitation   = "%%CITATION = ARXIV:0907.2997;%%"
}

@article{Harlander:2009my,
      author         = "Harlander, Robert V. and Mantler, Hendrik and Marzani,
                        Simone and Ozeren, Kemal J.",
      title          = "{Higgs production in gluon fusion at
                        next-to-next-to-leading order QCD for finite top mass}",
      journal        = "Eur. Phys. J. C",
      volume         = "66",
      year           = "2010",
      pages          = "359-372",
      doi            = "10.1140/epjc/s10052-010-1258-x",
      eprint         = "0912.2104",
      archivePrefix  = "arXiv",
      primaryClass   = "hep-ph",
      reportNumber   = "WUB-09-18, MAN-HEP-2009-44",
      SLACcitation   = "%%CITATION = ARXIV:0912.2104;%%"
}

@article{Pak:2009dg,
      author         = "Pak, Alexey and Rogal, Mikhail and Steinhauser, Matthias",
      title          = "{Finite top quark mass effects in NNLO Higgs boson
                        production at LHC}",
      journal        = "JHEP",
      volume         = "02",
      year           = "2010",
      pages          = "025",
      doi            = "10.1007/JHEP02(2010)025",
      eprint         = "0911.4662",
      archivePrefix  = "arXiv",
      primaryClass   = "hep-ph",
      reportNumber   = "SFB-CPP-09-116, TTP09-43",
      SLACcitation   = "%%CITATION = ARXIV:0911.4662;%%"
}

@article{Actis:2008ug,
      author         = "Actis, Stefano and Passarino, Giampiero and Sturm,
                        Christian and Uccirati, Sandro",
      title          = "{NLO electroweak corrections to Higgs boson production at
                        hadron colliders}",
      journal        = "Phys. Lett. B",
      volume         = "670",
      year           = "2008",
      pages          = "12-17",
      doi            = "10.1016/j.physletb.2008.10.018",
      eprint         = "0809.1301",
      archivePrefix  = "arXiv",
      primaryClass   = "hep-ph",
      reportNumber   = "PITHA-08-20, SFB-CPP-08-62, TTP08-38",
      SLACcitation   = "%%CITATION = ARXIV:0809.1301;%%"
}

@article{Actis:2008ts,
      author         = "Actis, Stefano and Passarino, Giampiero and Sturm,
                        Christian and Uccirati, Sandro",
      title          = "{NNLO computational techniques: The cases \(H \to \gamma\gamma\) and \(H \to gg\)}",
      journal        = "Nucl. Phys. B",
      volume         = "811",
      year           = "2009",
      pages          = "182-273",
      doi            = "10.1016/j.nuclphysb.2008.11.024",
      eprint         = "0809.3667",
      archivePrefix  = "arXiv",
      primaryClass   = "hep-ph",
      reportNumber   = "PITHA-08-24, SFB-CPP-08-73, TTP08-42",
      SLACcitation   = "%%CITATION = ARXIV:0809.3667;%%"
}

@article{Bonetti:2018ukf,
      author         = "Bonetti, Marco and Melnikov, Kirill and Tancredi,
                        Lorenzo",
      title          = "{Higher order corrections to mixed QCD-EW contributions
                        to Higgs boson production in gluon fusion}",
      journal        = "Phys. Rev. D",
      volume         = "97",
      year           = "2018",
      number         = "5",
      pages          = "056017",
      doi            = "10.1103/PhysRevD.97.056017",
      eprint         = "1801.10403",
      archivePrefix  = "arXiv",
      primaryClass   = "hep-ph",
      reportNumber   = "CERN-TH-2018-011, TTP18-004",
      SLACcitation   = "%%CITATION = ARXIV:1801.10403;%%",
      related = "Bonetti:2018ukf-err",
      relatedstring  = "Erratum:",
}

@article{Ciccolini:2007jr,
     author         = "Ciccolini, M. and Denner, Ansgar and Dittmaier, S.",
     title          = "{Strong and Electroweak Corrections to the Production of a Higgs Boson + 2 Jets via Weak Interactions at the Large Hadron Collider}",
     journal        = "Phys. Rev. Lett.",
     volume         = "99",
     pages          = "161803",
     doi            = "10.1103/PhysRevLett.99.161803",
     year           = "2007",
     eprint         = "0707.0381",
     archivePrefix  = "arXiv",
     primaryClass   = "hep-ph",
}

@Article{Ciccolini:2007ec,
    author    = "Ciccolini, Mariano and Denner, Ansgar and Dittmaier, Stefan",
    title     = "{Electroweak and QCD corrections to Higgs production via
                 vector-boson fusion at the CERN LHC}",
    journal   = "Phys. Rev. D",
    volume    = "77",
    year      = "2008",
    pages     = "013002",
    eprint    = "0710.4749",
    archivePrefix = "arXiv",
    primaryClass  =  "hep-ph",
    doi       = "10.1103/PhysRevD.77.013002",
    SLACcitation  = "%%CITATION = 0710.4749;%%"
}

@Article{Bolzoni:2010xr,
    author    = "Bolzoni, Paolo and Maltoni, Fabio and Moch, Sven-Olaf and
                 Zaro, Marco",
    title     = "{Higgs Boson Production via Vector-Boson Fusion at Next-to-Next-to-Leading Order in QCD}",
    journal   = "Phys. Rev. Lett.",
    volume    = "105",
    year      = "2010",
    pages     = "011801",
    eprint    = "1003.4451",
    archivePrefix = "arXiv",
    primaryClass  =  "hep-ph",
    doi       = "10.1103/PhysRevLett.105.011801",
    SLACcitation  = "%%CITATION = 1003.4451;%%"
}

@article{Djouadi:1997yw,
    author    = "Djouadi, A. and Kalinowski, J. and Spira, M.",
    title     = "{HDECAY: A program for Higgs boson decays in the Standard
                 Model and its  supersymmetric extension}",
    journal   = "Comput. Phys. Commun.",
    volume    = "108",
    year      = "1998",
    pages     = "56",
    eprint    = "hep-ph/9704448",
    archivePrefix = "arXiv",
    doi       = "10.1016/S0010-4655(97)00123-9",
    reportNumber   = "DESY-97-079, IFT-96-29, PM-97-04",
    SLACcitation  = "%%CITATION = HEP-PH/9704448;%%"
}

@article{Spira:1997dg,
      author         = "Spira, Michael",
      title          = "{QCD Effects in Higgs Physics}",
      journal        = "Fortsch. Phys.",
      volume         = "46",
      year           = "1998",
      pages          = "203-284",
      doi            = "10.1002/(SICI)1521-3978(199804)46:3<203::AID-PROP203>3.0.CO;2-4",
      eprint         = "hep-ph/9705337",
      archivePrefix  = "arXiv",
      reportNumber   = "CERN-TH-97-068, CERN-TH-97-68",
      SLACcitation   = "%%CITATION = HEP-PH/9705337;%%"
}

@article{Djouadi:2006bz,
      author         = "Djouadi, A. and M{\"u}hlleitner, M. M. and Spira, M.",
      title          = "{Decays of Supersymmetric particles: The Program SUSY-HIT
                        (SUspect-SdecaY-Hdecay-InTerface)}",
      booktitle      = "{Proceedings, Physics at LHC, 3rd Conference}",
      venue          = "Cracow, Poland",
      eventdate      = "2006-07-03/2006-07-08",
      journal        = "Acta Phys. Polon. B",
      volume         = "38",
      year           = "2007",
      pages          = "635-644",
      eprint         = "hep-ph/0609292",
      archivePrefix  = "arXiv",
      XprimaryClass   = "hep-ph",
      reportNumber   = "CERN-PH-TH-2006-200",
      SLACcitation   = "%%CITATION = HEP-PH/0609292;%%"
}

@article{Bredenstein:2006ha,
      author         = "Bredenstein, A. and Denner, Ansgar and Dittmaier, S. and Weber, M. M.",
      title          = "{Radiative corrections to the semileptonic and hadronic
                        Higgs-boson decays \(H \to WW / ZZ \to 4\) fermions}",
      journal        = "JHEP",
      volume         = "02",
      year           = "2007",
      pages          = "080",
      doi            = "10.1088/1126-6708/2007/02/080",
      eprint         = "hep-ph/0611234",
      archivePrefix  = "arXiv",
      xprimaryClass   = "hep-ph",
      reportNumber   = "KEK-CP-187, KEK-PREPRINT-2006-47, MPP-2006-138,
                        PSI-PR-06-11, WUB-06-08",
      SLACcitation   = "%%CITATION = HEP-PH/0611234;%%"
}

@article{Bredenstein:2006rh,
      author         = "Bredenstein, A. and Denner, Ansgar and Dittmaier, S. and Weber, M. M.",
      title          = "{Precise predictions for the Higgs-boson decay $H \to WW/ZZ \to 4$ leptons}",
      journal        = "Phys. Rev. D",
      volume         = "74",
      year           = "2006",
      pages          = "013004",
      doi            = "10.1103/PhysRevD.74.013004",
      eprint         = "hep-ph/0604011",
      archivePrefix  = "arXiv",
      primaryClass   = "hep-ph",
      reportNumber   = "MPP-2005-24, PSI-PR-06-05",
      SLACcitation   = "%%CITATION = HEP-PH/0604011;%%"
}

@article{Bredenstein:2006nk,
      author         = "Bredenstein, A. and Denner, Ansgar and Dittmaier, S. and Weber, M. M.",
      title          = "{Precision calculations for the Higgs decays \(H \to ZZ/WW \to 4\) leptons}",
      booktitle      = "{Proceedings, 8th DESY Workshop on Elementary Particle
                        Theory: Loops and Legs in Quantum Field Theory}",
      venue          = "Eisenach, Germany",
      eventdate      = "2006-04-23/2006-04-28",
      journal        = "Nucl. Phys. Proc. Suppl.",
      volume         = "160",
      year           = "2006",
      pages          = "131-135",
      doi            = "10.1016/j.nuclphysbps.2006.09.104",
      eprint         = "hep-ph/0607060",
      archivePrefix  = "arXiv",
      primaryClass   = "hep-ph",
      reportNumber   = "PSI-PR-06-09",
      SLACcitation   = "%%CITATION = HEP-PH/0607060;%%"
}

@article{Ciccolini:2003jy,
      author         = "Ciccolini, M. L. and Dittmaier, S. and Kr{\"a}mer, M.",
      title          = "{Electroweak radiative corrections to associated \(WH\) and
                        \(ZH\) production at hadron colliders}",
      journal        = "Phys. Rev. D",
      volume         = "68",
      year           = "2003",
      pages          = "073003",
      doi            = "10.1103/PhysRevD.68.073003",
      eprint         = "hep-ph/0306234",
      archivePrefix  = "arXiv",
      primaryClass   = "hep-ph",
      reportNumber   = "EDINBURGH-2003-05, MPI-PHT-2003-24",
      SLACcitation   = "%%CITATION = HEP-PH/0306234;%%"
}

@article{Brein:2003wg,
      author         = "Brein, Oliver and Djouadi, Abdelhak and Harlander, Robert",
      title          = "{NNLO QCD corrections to the Higgs-strahlung processes at
                        hadron colliders}",
      journal        = "Phys. Lett. B",
      volume         = "579",
      year           = "2004",
      pages          = "149-156",
      doi            = "10.1016/j.physletb.2003.10.112",
      eprint         = "hep-ph/0307206",
      archivePrefix  = "arXiv",
      XprimaryClass   = "hep-ph",
      reportNumber   = "MPP-2003-35, CERN-TH-2003-161, PM-03-16",
      SLACcitation   = "%%CITATION = HEP-PH/0307206;%%"
}

@article{Brein:2011vx,
      author         = "Brein, Oliver and Harlander, Robert V. and Wiesemann, Marius and Zirke, Tom",
      title          = "{Top-quark mediated effects in hadronic Higgs-Strahlung}",
      journal        = "Eur. Phys. J. C",
      volume         = "72",
      year           = "2012",
      pages          = "1868",
      doi            = "10.1140/epjc/s10052-012-1868-6",
      eprint         = "1111.0761",
      archivePrefix  = "arXiv",
      primaryClass   = "hep-ph",
      reportNumber   = "CERN-PH-TH-2011-268, FR-PHENO-2011-016, WUB-11-15",
      SLACcitation   = "%%CITATION = ARXIV:1111.0761;%%"
}

@article{Altenkamp:2012sx,
      author         = "Altenkamp, Lukas and Dittmaier, Stefan and Harlander,
                        Robert V. and Rzehak, Heidi and Zirke, Tom J. E.",
      title          = "{Gluon-induced Higgs-strahlung at next-to-leading order
                        QCD}",
      journal        = "JHEP",
      volume         = "02",
      year           = "2013",
      pages          = "078",
      doi            = "10.1007/JHEP02(2013)078",
      eprint         = "1211.5015",
      archivePrefix  = "arXiv",
      primaryClass   = "hep-ph",
      reportNumber   = "CERN-PH-TH-2012-312, FR-PHENO-2012-023, WUB-12-21",
      SLACcitation   = "%%CITATION = ARXIV:1211.5015;%%"
}

@article{Denner:2014cla,
      author         = "Denner, Ansgar and Dittmaier, Stefan and Kallweit, Stefan
                        and M{\"u}ck, Alexander",
      title          = "{HAWK 2.0: A Monte Carlo program for Higgs production
                        in vector-boson fusion and Higgs strahlung at hadron
                        colliders}",
      journal        = "Comput. Phys. Commun.",
      volume         = "195",
      year           = "2015",
      pages          = "161-171",
      doi            = "10.1016/j.cpc.2015.04.021",
      eprint         = "1412.5390",
      archivePrefix  = "arXiv",
      primaryClass   = "hep-ph",
      reportNumber   = "FR-PHENO-2014-013, MITP-14-101, TTK-14-36",
      SLACcitation   = "%%CITATION = ARXIV:1412.5390;%%"
}

@article{Brein:2012ne,
      author         = "Brein, Oliver and Harlander, Robert V. and Zirke, Tom J.
                        E.",
      title          = "{vh@nnlo -- Higgs Strahlung at hadron colliders}",
      journal        = "Comput. Phys. Commun.",
      volume         = "184",
      year           = "2013",
      pages          = "998-1003",
      doi            = "10.1016/j.cpc.2012.11.002",
      eprint         = "1210.5347",
      archivePrefix  = "arXiv",
      primaryClass   = "hep-ph",
      SLACcitation   = "%%CITATION = ARXIV:1210.5347;%%"
}

@article{Harlander:2014wda,
      author         = "Harlander, Robert V. and Kulesza, Anna and Theeuwes,
                        Vincent and Zirke, Tom",
      title          = "{Soft gluon resummation for gluon-induced Higgs
                        Strahlung}",
      journal        = "JHEP",
      volume         = "11",
      year           = "2014",
      pages          = "082",
      doi            = "10.1007/JHEP11(2014)082",
      eprint         = "1410.0217",
      archivePrefix  = "arXiv",
      primaryClass   = "hep-ph",
      reportNumber   = "MS-TP-14-17, WUB-14-10, LPN14-112",
      SLACcitation   = "%%CITATION = ARXIV:1410.0217;%%"
}

@article{Demartin:2016axk,
    author = "Demartin, Federico and Maier, Benedikt and Maltoni, Fabio and Mawatari, Kentarou and Zaro, Marco",
    title = "{tWH associated production at the LHC}",
    eprint = "1607.05862",
    archivePrefix = "arXiv",
    primaryClass = "hep-ph",
    reportNumber = "MCNET-16-30, CP3-16-40, LPSC16158",
    doi = "10.1140/epjc/s10052-017-4601-7",
    journal = "Eur. Phys. J. C",
    volume = "77",
    number = "1",
    pages = "34",
    year = "2017"
}

@Article{Cowan:2010js,
   Author = {Cowan, Glen and Cranmer, Kyle and Gross, Eilam and Vitells, Ofer},
   Title = {{Asymptotic formulae for likelihood-based tests of new physics}},
   Journal = {Eur. Phys. J. C},
   Volume = {71},
   Year = {2011},
   Pages = {1554},
   doi = "10.1140/epjc/s10052-011-1554-0",
   Eprint = "1007.1727",
   Archiveprefix = {arXiv},
   Primaryclass = {physics.data-an},
   related = "Cowan:2010js-err",
   relatedstring  = "Erratum:",
}

@Article{PERF-2007-01,
    author         = "{ATLAS Collaboration}",
    title          = "{The ATLAS Experiment at the CERN Large Hadron Collider}",
    journal        = "JINST",
    volume         = "3",
    year           = "2008",
    pages          = "S08003",
    doi            = "10.1088/1748-0221/3/08/S08003",
    primaryClass   = "hep-ex",
}

@Article{SOFT-2010-01,
    author         = "{ATLAS Collaboration}",
    title          = "{The ATLAS Simulation Infrastructure}",
    journal        = "Eur. Phys. J. C",
    volume         = "70",
    year           = "2010",
    pages          = "823",
    doi            = "10.1140/epjc/s10052-010-1429-9",
    eprint         = "1005.4568",
    archivePrefix  = "arXiv",
    primaryClass   = "physics.ins-det",
}

@Article{HIGG-2012-27,
    author         = "{ATLAS Collaboration}",
    title          = "{Observation of a new particle in the search for the Standard Model Higgs boson with the ATLAS detector at the LHC}",
    journal        = "Phys. Lett. B",
    volume         = "716",
    year           = "2012",
    pages          = "1",
    doi            = "10.1016/j.physletb.2012.08.020",
    reportNumber   = "CERN-PH-EP-2012-218",
    eprint         = "1207.7214",
    archivePrefix  = "arXiv",
    primaryClass   = "hep-ex",
}

@Article{STDM-2012-23,
    author         = "{ATLAS Collaboration}",
    title          = "{Measurement of the \(Z/\gamma^*\) boson transverse momentum distribution in \(pp\) collisions at \(\sqrt{s} = 7\,\text{TeV}\) with the ATLAS detector}",
    journal        = "JHEP",
    volume         = "09",
    year           = "2014",
    pages          = "145",
    doi            = "10.1007/JHEP09(2014)145",
    reportNumber   = "CERN-PH-EP-2014-075",
    eprint         = "1406.3660",
    archivePrefix  = "arXiv",
    primaryClass   = "hep-ex",
}

@Article{HIGG-2013-21,
    author         = "{ATLAS Collaboration}",
    title          = "{Measurements of Higgs boson production and couplings in the four-lepton channel in \(pp\) collisions at center-of-mass energies of \(7\) and \(8\,\text{TeV}\) with the ATLAS detector}",
    journal        = "Phys. Rev. D",
    volume         = "91",
    year           = "2015",
    pages          = "012006",
    doi            = "10.1103/PhysRevD.91.012006",
    reportNumber   = "CERN-PH-EP-2014-170",
    eprint         = "1408.5191",
    archivePrefix  = "arXiv",
    primaryClass   = "hep-ex",
}

@Article{HIGG-2013-26,
    author         = "{ATLAS Collaboration}",
    title          = "{Search for the associated production of the Higgs boson with a top quark pair in multilepton final states with the ATLAS detector}",
    journal        = "Phys. Lett. B",
    volume         = "749",
    year           = "2015",
    pages          = "519",
    doi            = "10.1016/j.physletb.2015.07.079",
    reportNumber   = "CERN-PH-EP-2015-109",
    eprint         = "1506.05988",
    archivePrefix  = "arXiv",
    primaryClass   = "hep-ex",
}

@Article{HIGG-2013-32,
    author         = "{ATLAS Collaboration}",
    title          = "{Evidence for the Higgs-boson Yukawa coupling to tau leptons with the ATLAS detector}",
    journal        = "JHEP",
    volume         = "04",
    year           = "2015",
    pages          = "117",
    doi            = "10.1007/JHEP04(2015)117",
    reportNumber   = "CERN-PH-EP-2014-262",
    eprint         = "1501.04943",
    archivePrefix  = "arXiv",
    primaryClass   = "hep-ex",
}

@Article{HIGG-2014-01,
    author         = "{ATLAS Collaboration}",
    title          = "{Search for the standard model Higgs boson produced in association with a vector boson and decaying into a tau pair in \(pp\) collisions at \(\sqrt{s} = 8\,\text{TeV}\) with the ATLAS detector}",
    journal        = "Phys. Rev. D",
    volume         = "93",
    year           = "2016",
    pages          = "092005",
    doi            = "10.1103/PhysRevD.93.092005",
    eprint         = "1511.08352",
    archivePrefix  = "arXiv",
    primaryClass   = "hep-ex",
}

@Article{HIGG-2014-09,
    author         = "{ATLAS Collaboration}",
    title          = "{Modelling \(Z\rightarrow\tau\tau\) processes in ATLAS with \(\tau\)-embedded \(Z\rightarrow\mu\mu\) data}",
    journal        = "JINST",
    volume         = "10",
    year           = "2015",
    pages          = "P09018",
    doi            = "10.1088/1748-0221/10/09/P09018",
    reportNumber   = "CERN-PH-EP-2015-130",
    eprint         = "1506.05623",
    archivePrefix  = "arXiv",
    primaryClass   = "hep-ex",
}

@Article{HIGG-2014-14,
    author         = "{ATLAS and CMS Collaborations}",
    title          = "{Combined Measurement of the Higgs Boson Mass in \(pp\) Collisions at \(\sqrt{s} = 7\) and \(8\,\text{TeV}\) with the ATLAS and CMS Experiments}",
    journal        = "Phys. Rev. Lett.",
    volume         = "114",
    year           = "2015",
    pages          = "191803",
    doi            = "10.1103/PhysRevLett.114.191803",
    eprint         = "1503.07589",
    archivePrefix  = "arXiv",
    primaryClass   = "hep-ex",
}

@Article{PERF-2014-03,
    author         = "{ATLAS Collaboration}",
    title          = "{Performance of pile-up mitigation techniques for jets in \(pp\) collisions at \(\sqrt{s} = 8\,\text{TeV}\) using the ATLAS detector}",
    journal        = "Eur. Phys. J. C",
    volume         = "76",
    year           = "2016",
    pages          = "581",
    doi            = "10.1140/epjc/s10052-016-4395-z",
    reportNumber   = "CERN-PH-EP-2015-206",
    eprint         = "1510.03823",
    archivePrefix  = "arXiv",
    primaryClass   = "hep-ex",
}

@Article{PERF-2014-06,
    author         = "{ATLAS Collaboration}",
    title          = "{Reconstruction of hadronic decay products of tau leptons with the ATLAS experiment}",
    journal        = "Eur. Phys. J. C",
    volume         = "76",
    year           = "2016",
    pages          = "295",
    doi            = "10.1140/epjc/s10052-016-4110-0",
    reportNumber   = "CERN-PH-EP-2015-294",
    eprint         = "1512.05955",
    archivePrefix  = "arXiv",
    primaryClass   = "hep-ex",
}

@Article{PERF-2014-07,
    author         = "{ATLAS Collaboration}",
    title          = "{Topological cell clustering in the ATLAS calorimeters and its performance in LHC Run~1}",
    journal        = "Eur. Phys. J. C",
    volume         = "77",
    year           = "2017",
    pages          = "490",
    doi            = "10.1140/epjc/s10052-017-5004-5",
    reportNumber   = "CERN-PH-EP-2015-304",
    eprint         = "1603.02934",
    archivePrefix  = "arXiv",
    primaryClass   = "hep-ex",
}

@Article{HIGG-2015-07,
    author         = "{ATLAS and CMS Collaborations}",
    title          = "{Measurements of the Higgs boson production and decay rates and constraints on its couplings from a combined ATLAS and CMS analysis of the LHC \(pp\) collision data at \(\sqrt{s} = 7\) and \(8\,\text{TeV}\)}",
    journal        = "JHEP",
    volume         = "08",
    year           = "2016",
    pages          = "045",
    doi            = "10.1007/JHEP08(2016)045",
    eprint         = "1606.02266",
    archivePrefix  = "arXiv",
    primaryClass   = "hep-ex",
}

@Article{PERF-2015-01,
    author         = "{ATLAS Collaboration}",
    title          = "{Reconstruction of primary vertices at the ATLAS experiment in Run~1 proton--proton collisions at the LHC}",
    journal        = "Eur. Phys. J. C",
    volume         = "77",
    year           = "2017",
    pages          = "332",
    doi            = "10.1140/epjc/s10052-017-4887-5",
    reportNumber   = "CERN-EP-2016-150",
    eprint         = "1611.10235",
    archivePrefix  = "arXiv",
    primaryClass   = "hep-ex",
}

@Article{PERF-2015-09,
    author         = "{ATLAS Collaboration}",
    title          = "{Jet reconstruction and performance using particle flow with the ATLAS Detector}",
    journal        = "Eur. Phys. J. C",
    volume         = "77",
    year           = "2017",
    pages          = "466",
    doi            = "10.1140/epjc/s10052-017-5031-2",
    reportNumber   = "CERN-EP-2017-024",
    eprint         = "1703.10485",
    archivePrefix  = "arXiv",
    primaryClass   = "hep-ex",
}

@Article{HIGG-2016-07,
    author         = "{ATLAS Collaboration}",
    title          = "{Measurements of gluon-gluon fusion and vector-boson fusion Higgs boson production cross-sections in the \(H \rightarrow WW^{*} \rightarrow e\nu\mu\nu\) decay channel in \(pp\) collisions at \(\sqrt{s} = 13\,\text{TeV}\) with the ATLAS detector}",
    journal        = "Phys. Lett. B",
    volume         = "789",
    year           = "2019",
    pages          = "508",
    doi            = "10.1016/j.physletb.2018.11.064",
    reportNumber   = "CERN-EP-2018-212",
    eprint         = "1808.09054",
    archivePrefix  = "arXiv",
    primaryClass   = "hep-ex",
}

@Article{HIGG-2016-21,
    author         = "{ATLAS Collaboration}",
    title          = "{Measurements of Higgs boson properties in the diphoton decay channel with \(36\,\text{fb}^{-1}\) of \(pp\) collision data at \(\sqrt{s} = 13\,\text{TeV}\) with the ATLAS detector}",
    journal        = "Phys. Rev. D",
    volume         = "98",
    year           = "2018",
    pages          = "052005",
    doi            = "10.1103/PhysRevD.98.052005",
    reportNumber   = "CERN-EP-2017-288",
    eprint         = "1802.04146",
    archivePrefix  = "arXiv",
    primaryClass   = "hep-ex",
}

@Article{HIGG-2016-22,
    author         = "{ATLAS Collaboration}",
    title          = "{Measurement of the Higgs boson coupling properties in the \(H \rightarrow ZZ^{*} \rightarrow 4\ell\) decay channel at \(\sqrt{s} = 13\,\text{TeV}\) with the ATLAS detector}",
    journal        = "JHEP",
    volume         = "03",
    year           = "2018",
    pages          = "095",
    doi            = "10.1007/JHEP03(2018)095",
    reportNumber   = "CERN-EP-2017-206",
    eprint         = "1712.02304",
    archivePrefix  = "arXiv",
    primaryClass   = "hep-ex",
}

@Article{HIGG-2016-25,
    author         = "{ATLAS Collaboration}",
    title          = "{Measurement of inclusive and differential cross sections in the \(H \rightarrow ZZ^* \rightarrow 4\ell\) decay channel in \(pp\) collisions at \(\sqrt{s} = 13\,\text{TeV}\) with the ATLAS detector}",
    journal        = "JHEP",
    volume         = "10",
    year           = "2017",
    pages          = "132",
    doi            = "10.1007/JHEP10(2017)132",
    reportNumber   = "CERN-EP-2017-139",
    eprint         = "1708.02810",
    archivePrefix  = "arXiv",
    primaryClass   = "hep-ex",
}

@Article{HIGG-2016-33,
    author         = "{ATLAS Collaboration}",
    title          = "{Measurement of the Higgs boson mass in the \(H \rightarrow ZZ^* \rightarrow 4\ell\) and \(H \rightarrow \gamma\gamma\) channels with \(\sqrt{s} = 13\,\text{TeV}\) \(pp\) collisions using the ATLAS detector}",
    journal        = "Phys. Lett. B",
    volume         = "784",
    year           = "2018",
    pages          = "345",
    doi            = "10.1016/j.physletb.2018.07.050",
    reportNumber   = "CERN-EP-2018-085",
    eprint         = "1806.00242",
    archivePrefix  = "arXiv",
    primaryClass   = "hep-ex",
}

@Article{PERF-2016-06,
    author         = "{ATLAS Collaboration}",
    title          = "{Identification and rejection of pile-up jets at high pseudorapidity with the ATLAS detector}",
    journal        = "Eur. Phys. J. C",
    volume         = "77",
    year           = "2017",
    pages          = "580",
    doi            = "10.1140/epjc/s10052-017-5081-5",
    reportNumber   = "CERN-EP-2017-055",
    eprint         = "1705.02211",
    archivePrefix  = "arXiv",
    primaryClass   = "hep-ex",
    related        = "PERF-2016-06-err",
    relatedstring  = "Erratum:",
}

@Article{PERF-2016-07,
    author         = "{ATLAS Collaboration}",
    title          = "{Performance of missing transverse momentum reconstruction with the ATLAS detector using proton--proton collisions at \(\sqrt{s} = 13\,\text{TeV}\)}",
    journal        = "Eur. Phys. J. C",
    volume         = "78",
    year           = "2018",
    pages          = "903",
    doi            = "10.1140/epjc/s10052-018-6288-9",
    reportNumber   = "CERN-EP-2017-274",
    eprint         = "1802.08168",
    archivePrefix  = "arXiv",
    primaryClass   = "hep-ex",
}

@Article{STDM-2016-01,
    author         = "{ATLAS Collaboration}",
    title          = "{Measurements of the production cross section of a \(Z\) boson in association with jets in \(pp\) collisions at \(\sqrt{s} = 13\,\text{TeV}\) with the ATLAS detector}",
    journal        = "Eur. Phys. J. C",
    volume         = "77",
    year           = "2017",
    pages          = "361",
    doi            = "10.1140/epjc/s10052-017-4900-z",
    reportNumber   = "CERN-EP-2016-297",
    eprint         = "1702.05725",
    archivePrefix  = "arXiv",
    primaryClass   = "hep-ex",
}

@Article{STDM-2016-09,
    author         = "{ATLAS Collaboration}",
    title          = "{Measurement of the cross-section for electroweak production of dijets in association with a \(Z\) boson in \(pp\) collisions at \(\sqrt{s} = 13\,\text{TeV}\) with the ATLAS detector}",
    journal        = "Phys. Lett. B",
    volume         = "775",
    year           = "2017",
    pages          = "206",
    doi            = "10.1016/j.physletb.2017.10.040",
    reportNumber   = "CERN-EP-2017-115",
    eprint         = "1709.10264",
    archivePrefix  = "arXiv",
    primaryClass   = "hep-ex",
}

@Article{TRIG-2016-01,
    author         = "{ATLAS Collaboration}",
    title          = "{Performance of the ATLAS trigger system in 2015}",
    journal        = "Eur. Phys. J. C",
    volume         = "77",
    year           = "2017",
    pages          = "317",
    doi            = "10.1140/epjc/s10052-017-4852-3",
    reportNumber   = "CERN-EP-2016-241",
    eprint         = "1611.09661",
    archivePrefix  = "arXiv",
    primaryClass   = "hep-ex",
}

@Article{HIGG-2017-07,
    author         = "{ATLAS Collaboration}",
    title          = "{Cross-section measurements of the Higgs boson decaying into a pair of \(\tau\)-leptons in proton--proton collisions at \(\sqrt{s} = 13\,\text{TeV}\) with the ATLAS detector}",
    journal        = "Phys. Rev. D",
    volume         = "99",
    year           = "2019",
    pages          = "072001",
    doi            = "10.1103/PhysRevD.99.072001",
    reportNumber   = "CERN-EP-2018-232",
    eprint         = "1811.08856",
    archivePrefix  = "arXiv",
    primaryClass   = "hep-ex",
}

@Article{HIGG-2017-11,
    author         = "{ATLAS Collaboration}",
    title          = "{Combined measurement of differential and total cross sections in the \(H \rightarrow \gamma \gamma\) and the \(H \rightarrow ZZ^* \rightarrow 4\ell\) decay channels at \(\sqrt{s} = 13\,\text{TeV}\) with the ATLAS detector}",
    journal        = "Phys. Lett. B",
    volume         = "786",
    year           = "2018",
    pages          = "114",
    doi            = "10.1016/j.physletb.2018.09.019",
    reportNumber   = "CERN-EP-2018-080",
    eprint         = "1805.10197",
    archivePrefix  = "arXiv",
    primaryClass   = "hep-ex",
}

@Article{DAPR-2018-01,
    author         = "{ATLAS Collaboration}",
    title          = "{ATLAS data quality operations and performance for 2015--2018 data-taking}",
    journal        = "JINST",
    volume         = "15",
    year           = "2020",
    pages          = "P04003",
    doi            = "10.1088/1748-0221/15/04/P04003",
    reportNumber   = "CERN-EP-2019-207",
    eprint         = "1911.04632",
    archivePrefix  = "arXiv",
    primaryClass   = "physics.ins-det",
}

@Article{EGAM-2018-01,
    author         = "{ATLAS Collaboration}",
    title          = "{Electron and photon performance measurements with the ATLAS detector using the 2015--2017 LHC proton--proton collision data}",
    journal        = "JINST",
    volume         = "14",
    year           = "2019",
    pages          = "P12006",
    doi            = "10.1088/1748-0221/14/12/P12006",
    reportNumber   = "CERN-EP-2019-145",
    eprint         = "1908.00005",
    archivePrefix  = "arXiv",
    primaryClass   = "hep-ex",
}

@Article{FTAG-2018-01,
    author         = "{ATLAS Collaboration}",
    title          = "{ATLAS \(b\)-jet identification performance and efficiency measurement with \(t\bar{t}\) events in \(pp\) collisions at \(\sqrt{s} = 13\,\text{TeV}\)}",
    journal        = "Eur. Phys. J. C",
    volume         = "79",
    year           = "2019",
    pages          = "970",
    doi            = "10.1140/epjc/s10052-019-7450-8",
    reportNumber   = "CERN-EP-2019-132",
    eprint         = "1907.05120",
    archivePrefix  = "arXiv",
    primaryClass   = "hep-ex",
}

@Article{HIGG-2018-13,
    author         = "{ATLAS Collaboration}",
    title          = "{Observation of Higgs boson production in association with a top quark pair at the LHC with the ATLAS detector}",
    journal        = "Phys. Lett. B",
    volume         = "784",
    year           = "2018",
    pages          = "173",
    doi            = "10.1016/j.physletb.2018.07.035",
    reportNumber   = "CERN-EP-2018-138",
    eprint         = "1806.00425",
    archivePrefix  = "arXiv",
    primaryClass   = "hep-ex",
}

@Article{HIGG-2018-14,
    author         = "{ATLAS Collaboration}",
    title          = "{Test of CP invariance in vector-boson fusion production of the Higgs boson in the \(H\to\tau\tau\) channel in proton--proton collisions at \(\sqrt{s} = 13\,\text{TeV}\) with the ATLAS detector}",
    journal        = "Phys. Lett. B",
    volume         = "805",
    year           = "2020",
    pages          = "135426",
    doi            = "10.1016/j.physletb.2020.135426",
    reportNumber   = "CERN-EP-2020-009",
    eprint         = "2002.05315",
    archivePrefix  = "arXiv",
    primaryClass   = "hep-ex",
}

@Article{JETM-2018-05,
    author         = "{ATLAS Collaboration}",
    title          = "{Jet energy scale and resolution measured in proton--proton collisions at \(\sqrt{s} = 13\,\text{TeV}\) with the ATLAS detector}",
    journal        = "Eur. Phys. J. C",
    volume         = "81",
    year           = "2020",
    pages          = "689",
    doi            = "10.1140/epjc/s10052-021-09402-3",
    reportNumber   = "CERN-EP-2020-083",
    eprint         = "2007.02645",
    archivePrefix  = "arXiv",
    primaryClass   = "hep-ex",
}

@Article{MUON-2018-03,
    author         = "{ATLAS Collaboration}",
    title          = "{Muon reconstruction and identification efficiency in ATLAS using the full Run~2 \(pp\) collision data set at \(\sqrt{s} = 13\,\text{TeV}\)}",
    journal        = "Eur. Phys. J. C",
    volume         = "81",
    year           = "2021",
    pages          = "578",
    doi            = "10.1140/epjc/s10052-021-09233-2",
    reportNumber   = "CERN-EP-2020-199",
    eprint         = "2012.00578",
    archivePrefix  = "arXiv",
    primaryClass   = "hep-ex",
}

@Article{PIX-2018-001,
    author         = "Abbott, B. and others",
    title          = "{Production and integration of the ATLAS Insertable B-Layer}",
    journal        = "JINST",
    volume         = "13",
    year           = "2018",
    pages          = "T05008",
    doi            = "10.1088/1748-0221/13/05/T05008",
    eprint         = "1803.00844",
    archivePrefix  = "arXiv",
    primaryClass   = "physics.ins-det",
}

@Article{TRIG-2018-01,
    author         = "{ATLAS Collaboration}",
    title          = "{Performance of the ATLAS muon triggers in Run~2}",
    journal        = "JINST",
    volume         = "15",
    year           = "2020",
    pages          = "P09015",
    doi            = "10.1088/1748-0221/15/09/p09015",
    reportNumber   = "CERN-EP-2020-031",
    eprint         = "2004.13447",
    archivePrefix  = "arXiv",
    primaryClass   = "physics.ins-det",
}

@Article{TRIG-2018-05,
    author         = "{ATLAS Collaboration}",
    title          = "{Performance of electron and photon triggers in ATLAS during LHC Run~2}",
    journal        = "Eur. Phys. J. C",
    volume         = "80",
    year           = "2020",
    pages          = "47",
    doi            = "10.1140/epjc/s10052-019-7500-2",
    reportNumber   = "CERN-EP-2019-169",
    eprint         = "1909.00761",
    archivePrefix  = "arXiv",
    primaryClass   = "hep-ex",
}

@Article{HIGG-2019-09,
    author         = "{ATLAS Collaboration}",
    title          = "{Measurements of Higgs boson production cross-sections in the \(H\to\tau^{+}\tau^{-}\) decay channel in \(pp\) collisions at \(\sqrt{s} = 13\,\text{TeV}\) with the ATLAS detector}",
    journal        = "JHEP",
    volume         = "08",
    year           = "2022",
    pages          = "175",
    doi            = "10.1007/JHEP08(2022)175",
    reportNumber   = "CERN-EP-2021-217",
    eprint         = "2201.08269",
    archivePrefix  = "arXiv",
    primaryClass   = "hep-ex",
}

@Booklet{ATL-SOFT-PUB-2021-001,
    author         = "{ATLAS Collaboration}",
    title          = "{The ATLAS Collaboration Software and Firmware}",
    howpublished   = "{ATL-SOFT-PUB-2021-001}",
    url            = "https://cds.cern.ch/record/2767187",
    year           = "2021",
}

@Booklet{ATL-SOFT-PUB-2023-001,
    author         = "{ATLAS Collaboration}",
    title          = "{ATLAS Computing Acknowledgements}",
    howpublished   = "{ATL-SOFT-PUB-2023-001}",
    url            = "https://cds.cern.ch/record/2869272",
    year           = "2023",
}

@Report{ATLAS-TDR-19,
    author         = "{ATLAS Collaboration}",
    title          = "{ATLAS Insertable B-Layer: Technical Design Report}",
    type           = "ATLAS-TDR-19; CERN-LHCC-2010-013",
    year           = "2010",
    url            = "https://cds.cern.ch/record/1291633",
    related        = "ATLAS-TDR-19-addm",
    relatedstring  = "Addendum:",
}

@Article{CMS-HIG-12-028,
    author         = "{CMS Collaboration}",
    title          = "{Observation of a new boson at a mass of 125 GeV with the CMS experiment at the LHC}",
    journal        = "Phys. Lett. B",
    volume         = "716",
    year           = "2012",
    pages          = "30",
    doi            = "10.1016/j.physletb.2012.08.021",
    reportNumber   = "CERN-PH-EP-2012-220",
    eprint         = "1207.7235",
    archivePrefix  = "arXiv",
    primaryClass   = "hep-ex",
}

@Article{CMS-HIG-13-004,
    author         = "{CMS Collaboration}",
    title          = "{Evidence for the \(125\,\text{GeV}\) Higgs boson decaying to a pair of \(\tau\) leptons}",
    journal        = "JHEP",
    volume         = "05",
    year           = "2014",
    pages          = "104",
    doi            = "10.1007/JHEP05(2014)104",
    reportNumber   = "CERN-PH-EP-2014-001",
    eprint         = "1401.5041",
    archivePrefix  = "arXiv",
    primaryClass   = "hep-ex",
}

@Article{CMS-HIG-16-040,
    author         = "{CMS Collaboration}",
    title          = "{Measurements of Higgs boson properties in the diphoton decay channel in proton--proton collisions at \(\sqrt{s} = 13\,\text{TeV}\)}",
    journal        = "JHEP",
    volume         = "11",
    year           = "2018",
    pages          = "185",
    doi            = "10.1007/JHEP11(2018)185",
    reportNumber   = "CERN-EP-2018-060",
    eprint         = "1804.02716",
    archivePrefix  = "arXiv",
    primaryClass   = "hep-ex",
}

@Article{CMS-HIG-16-041,
    author         = "{CMS Collaboration}",
    title          = "{Measurements of properties of the Higgs boson decaying into the four-lepton final state in \(pp\) collisions at \(\sqrt{s} = 13\,\text{TeV}\)}",
    journal        = "JHEP",
    volume         = "11",
    year           = "2017",
    pages          = "047",
    doi            = "10.1007/JHEP11(2017)047",
    reportNumber   = "CERN-EP-2017-123",
    eprint         = "1706.09936",
    archivePrefix  = "arXiv",
    primaryClass   = "hep-ex",
}

@Article{CMS-HIG-16-042,
    author         = "{CMS Collaboration}",
    title          = "{Measurements of properties of the Higgs boson decaying to a \(W\) boson pair in \(pp\) collisions at \(\sqrt{s} = 13\,\text{TeV}\)}",
    journal        = "Phys. Lett. B",
    volume         = "791",
    year           = "2019",
    pages          = "96",
    doi            = "10.1016/j.physletb.2018.12.073",
    reportNumber   = "CERN-EP-2018-141",
    eprint         = "1806.05246",
    archivePrefix  = "arXiv",
    primaryClass   = "hep-ex",
}

@Article{CMS-HIG-16-043,
    author         = "{CMS Collaboration}",
    title          = "{Observation of the Higgs boson decay to a pair of \(\tau\) leptons with the CMS detector}",
    journal        = "Phys. Lett. B",
    volume         = "779",
    year           = "2018",
    pages          = "283",
    doi            = "10.1016/j.physletb.2018.02.004",
    reportNumber   = "CERN-EP-2017-181",
    eprint         = "1708.00373",
    archivePrefix  = "arXiv",
    primaryClass   = "hep-ex",
}

@Article{CMS-HIG-17-025,
    author         = "{CMS Collaboration}",
    title          = "{Measurement of inclusive and differential Higgs boson production cross sections in the diphoton decay channel in proton--proton collisions at \(\sqrt{s} = 13\,\text{TeV}\)}",
    journal        = "JHEP",
    volume         = "01",
    year           = "2019",
    pages          = "183",
    doi            = "10.1007/JHEP01(2019)183",
    reportNumber   = "CERN-EP-2018-166",
    eprint         = "1807.03825",
    archivePrefix  = "arXiv",
    primaryClass   = "hep-ex",
}

@Article{CMS-HIG-17-035,
    author         = "{CMS Collaboration}",
    title          = "{Observation of \(t\bar{t}H\) Production}",
    journal        = "Phys. Rev. Lett.",
    volume         = "120",
    year           = "2018",
    pages          = "231801",
    doi            = "10.1103/PhysRevLett.120.231801",
    reportNumber   = "CERN-EP-2018-064",
    eprint         = "1804.02610",
    archivePrefix  = "arXiv",
    primaryClass   = "hep-ex",
}

@Article{CMS-HIG-18-007,
    author         = "{CMS Collaboration}",
    title          = "{Search for the associated production of the Higgs boson and a vector boson in proton--proton collisions at \(\sqrt{s} = 13\,\text{TeV}\) via Higgs boson decays to \(\tau\) leptons}",
    journal        = "JHEP",
    volume         = "06",
    year           = "2019",
    pages          = "093",
    doi            = "10.1007/JHEP06(2019)093",
    reportNumber   = "CERN-EP-2018-221",
    eprint         = "1809.03590",
    archivePrefix  = "arXiv",
    primaryClass   = "hep-ex",
}

@Article{CMS-TAU-18-001,
    author         = "{CMS Collaboration}",
    title          = "{An embedding technique to determine \(\tau\tau\) backgrounds in proton--proton collision data}",
    journal        = "JINST",
    volume         = "14",
    year           = "2019",
    pages          = "P06032",
    doi            = "10.1088/1748-0221/14/06/P06032",
    reportNumber   = "CERN-EP-2019-012",
    eprint         = "1903.01216",
    archivePrefix  = "arXiv",
    primaryClass   = "hep-ex",
}

@Booklet{ATLAS-CONF-2014-058,
    author         = "{ATLAS Collaboration}",
    title          = "{Estimation of non-prompt and fake lepton backgrounds in final states with top quarks produced in proton--proton collisions at \(\sqrt{s} = 8~\text{TeV}\) with the ATLAS Detector}",
    howpublished   = "{ATLAS-CONF-2014-058}",
    url            = "https://cds.cern.ch/record/1951336",
    year           = "2014",
}

@Booklet{ATLAS-CONF-2015-029,
    author         = "{ATLAS Collaboration}",
    title          = "{Selection of jets produced in \(13~\text{TeV}\) proton--proton collisions with the ATLAS detector}",
    howpublished   = "{ATLAS-CONF-2015-029}",
    url            = "https://cds.cern.ch/record/2037702",
    year           = "2015",
}

@Booklet{ATLAS-CONF-2017-029,
    author         = "{ATLAS Collaboration}",
    title          = "{Measurement of the tau lepton reconstruction and identification performance in the ATLAS experiment using \(pp\) collisions at \(\sqrt{s} = 13~\text{TeV}\)}",
    howpublished   = "{ATLAS-CONF-2017-029}",
    url            = "https://cds.cern.ch/record/2261772",
    year           = "2017",
}

@Booklet{ATLAS-CONF-2019-021,
    author         = "{ATLAS Collaboration}",
    title          = "{Luminosity determination in \(pp\) collisions at \(\sqrt{s} = 13\,\text{TeV}\) using the ATLAS detector at the LHC}",
    howpublished   = "{ATLAS-CONF-2019-021}",
    url            = "https://cds.cern.ch/record/2677054",
    year           = "2019",
}

@Booklet{ATLAS-CONF-2021-014,
    author         = "{ATLAS Collaboration}",
    title          = "{Measurements of gluon fusion and vector-boson-fusion production of the Higgs boson in \(H \to WW^{*} \to e\nu \mu\nu\) decays using \(pp\) collisions at \(\sqrt{s} = 13\,\text{TeV}\) with the ATLAS detector}",
    howpublished   = "{ATLAS-CONF-2021-014}",
    url            = "https://cds.cern.ch/record/2759651",
    year           = "2021",
}

@Booklet{ATL-PHYS-PUB-2014-021,
    author         = "{ATLAS Collaboration}",
    title          = "{ATLAS Pythia~8 tunes to \(7~\text{TeV}\) data}",
    howpublished   = "{ATL-PHYS-PUB-2014-021}",
    url            = "https://cds.cern.ch/record/1966419",
    year           = "2014",
}

@Booklet{ATL-DAQ-PUB-2016-001,
    author         = "{ATLAS Collaboration}",
    title          = "{2015 start-up trigger menu and initial performance assessment of the ATLAS trigger using Run-2 data}",
    howpublished   = "{ATL-DAQ-PUB-2016-001}",
    url            = "https://cds.cern.ch/record/2136007",
    year           = "2016",
}

@Booklet{ATL-PHYS-PUB-2016-017,
    author         = "{ATLAS Collaboration}",
    title          = "{The Pythia~8 A3 tune description of ATLAS minimum bias and inelastic measurements incorporating the Donnachie--Landshoff diffractive model}",
    howpublished   = "{ATL-PHYS-PUB-2016-017}",
    url            = "https://cds.cern.ch/record/2206965",
    year           = "2016",
}

@Booklet{ATL-PHYS-PUB-2016-020,
    author         = "{ATLAS Collaboration}",
    title          = "{Studies on top-quark Monte Carlo modelling for Top2016}",
    howpublished   = "{ATL-PHYS-PUB-2016-020}",
    url            = "https://cds.cern.ch/record/2216168",
    year           = "2016",
}

@Booklet{ATL-DAQ-PUB-2017-001,
    author         = "{ATLAS Collaboration}",
    title          = "{Trigger Menu in 2016}",
    howpublished   = "{ATL-DAQ-PUB-2017-001}",
    url            = "https://cds.cern.ch/record/2242069",
    year           = "2017",
}

@Booklet{ATL-PHYS-PUB-2017-007,
    author         = "{ATLAS Collaboration}",
    title          = "{Studies on top-quark Monte Carlo modelling with Sherpa and MG5\_aMC@NLO}",
    howpublished   = "{ATL-PHYS-PUB-2017-007}",
    url            = "https://cds.cern.ch/record/2261938",
    year           = "2017",
}

@Booklet{ATL-PHYS-PUB-2017-013,
    author         = "{ATLAS Collaboration}",
    title          = "{Optimisation and performance studies of the ATLAS \(b\)-tagging algorithms for the 2017-18 LHC run}",
    howpublished   = "{ATL-PHYS-PUB-2017-013}",
    url            = "https://cds.cern.ch/record/2273281",
    year           = "2017",
}

@Booklet{ATL-DAQ-PUB-2018-002,
    author         = "{ATLAS Collaboration}",
    title          = "{Trigger Menu in 2017}",
    howpublished   = "{ATL-DAQ-PUB-2018-002}",
    url            = "https://cds.cern.ch/record/2625986",
    year           = "2018",
}

@Booklet{ATL-DAQ-PUB-2019-001,
    author         = "{ATLAS Collaboration}",
    title          = "{Trigger Menu in 2018}",
    howpublished   = "{ATL-DAQ-PUB-2019-001}",
    url            = "https://cds.cern.ch/record/2693402",
    year           = "2019",
}

@Booklet{ATL-PHYS-PUB-2019-033,
    author         = "{ATLAS Collaboration}",
    title          = "{Identification of hadronic tau lepton decays using neural networks in the ATLAS experiment}",
    howpublished   = "{ATL-PHYS-PUB-2019-033}",
    url            = "https://cds.cern.ch/record/2688062",
    year           = "2019",
}

@Booklet{ATL-PHYS-PUB-2020-025,
    author         = "{ATLAS Collaboration}",
    title          = "{Formulae for Estimating Significance}",
    howpublished   = "{ATL-PHYS-PUB-2020-025}",
    url            = "https://cds.cern.ch/record/2736148",
    year           = "2020",
}
 
\clearpage
 
\begin{flushleft}
\hypersetup{urlcolor=black}
{\Large The ATLAS Collaboration}

\bigskip

\AtlasOrcid[0000-0002-6665-4934]{G.~Aad}$^\textrm{\scriptsize 99}$,
\AtlasOrcid[0000-0002-5888-2734]{B.~Abbott}$^\textrm{\scriptsize 117}$,
\AtlasOrcid[0000-0002-7248-3203]{D.C.~Abbott}$^\textrm{\scriptsize 100}$,
\AtlasOrcid[0000-0002-2788-3822]{A.~Abed~Abud}$^\textrm{\scriptsize 34}$,
\AtlasOrcid[0000-0002-1002-1652]{K.~Abeling}$^\textrm{\scriptsize 53}$,
\AtlasOrcid[0000-0002-2987-4006]{D.K.~Abhayasinghe}$^\textrm{\scriptsize 92}$,
\AtlasOrcid[0000-0002-8496-9294]{S.H.~Abidi}$^\textrm{\scriptsize 27}$,
\AtlasOrcid[0000-0002-9987-2292]{A.~Aboulhorma}$^\textrm{\scriptsize 33e}$,
\AtlasOrcid[0000-0001-5329-6640]{H.~Abramowicz}$^\textrm{\scriptsize 148}$,
\AtlasOrcid[0000-0002-1599-2896]{H.~Abreu}$^\textrm{\scriptsize 147}$
% ...
\bigskip
\\

% ...
$^{27}$Physics Department, Brookhaven National Laboratory, Upton NY; United States of America.\\
% ...
$^{33}$$^{(a)}$Facult\'e des Sciences Ain Chock, R\'eseau Universitaire de Physique des Hautes Energies - Universit\'e Hassan II, Casablanca;$^{(b)}$Facult\'{e} des Sciences, Universit\'{e} Ibn-Tofail, K\'{e}nitra;$^{(c)}$Facult\'e des Sciences Semlalia, Universit\'e Cadi Ayyad, LPHEA-Marrakech;$^{(d)}$LPMR, Facult\'e des Sciences, Universit\'e Mohamed Premier, Oujda;$^{(e)}$Facult\'e des sciences, Universit\'e Mohammed V, Rabat;$^{(f)}$Institute of Applied Physics, Mohammed VI Polytechnic University, Ben Guerir; Morocco.\\
$^{34}$CERN, Geneva; Switzerland.\\
% ...
$^{53}$II. Physikalisches Institut, Georg-August-Universit\"{a}t G\"ottingen, G\"ottingen; Germany.\\
% ...
$^{92}$Department of Physics, Royal Holloway University of London, Egham; United Kingdom.\\
% ...
$^{99}$CPPM, Aix-Marseille Universit\'e, CNRS/IN2P3, Marseille; France.\\
$^{100}$Department of Physics, University of Massachusetts, Amherst MA; United States of America.\\
% ...
$^{117}$Homer L. Dodge Department of Physics and Astronomy, University of Oklahoma, Norman OK; United States of America.\\
% ...
$^{147}$Department of Physics, Technion, Israel Institute of Technology, Haifa; Israel.\\
$^{148}$Raymond and Beverly Sackler School of Physics and Astronomy, Tel Aviv University, Tel Aviv; Israel.\\
% ...
$^{a}$ Also Affiliated with an institute covered by a cooperation agreement with CERN.\\
$^{b}$ Also at Borough of Manhattan Community College, City University of New York, New York NY; United States of America.\\
$^{c}$ Also at Bruno Kessler Foundation, Trento; Italy.\\
$^{d}$ Also at Center for High Energy Physics, Peking University; China.\\
$^{e}$ Also at Centro Studi e Ricerche Enrico Fermi; Italy.\\
$^{f}$ Also at CERN, Geneva; Switzerland.\\
% ...

\end{flushleft}


\end{document}